\documentclass[manuscript,screen]{acmart}

\usepackage[printonlyused,nolist]{acronym}
\usepackage{rotating}
\usepackage{wasysym}
\usepackage{adjustbox}
\usepackage{graphicx}
\usepackage{multirow}
\usepackage{booktabs}
\usepackage{longtable}
\usepackage{lscape}
\usepackage{diagbox}
\usepackage{enumerate}

\usepackage{colortbl}
\usepackage{enumitem}
\usepackage{makecell}
\usepackage{wrapfig}

\newcommand\rot[1]{\rlap{\rotatebox{45}{#1}}}

\AtBeginDocument{%
  \providecommand\BibTeX{{%
    \normalfont B\kern-0.5em{\scshape i\kern-0.25em b}\kern-0.8em\TeX}}}

\begin{document}

\title{A Survey on Blockchain Interoperability: Past, Present, and Future
Trends}

\author{Rafael Belchior}
\affiliation{%
  \institution{INESC-ID, Instituto Superior T\'ecnico, Universidade de Lisboa, Portugal}
  \streetaddress{Rua Alves Redol, 9}
  \city{Lisboa}
  \postcode{1000-029}}
\email{rafael.belchior@tecnico.ulisboa.pt}

\author{Andr\'e Vasconcelos}
\affiliation{%
  \institution{INESC-ID, Instituto Superior T\'ecnico, Universidade de Lisboa, Portugal}
  \streetaddress{Rua Alves Redol, 9}
  \city{Lisboa}
  \postcode{1000-029}}
\email{andre.vasconcelos@tecnico.ulisboa.pt}

\author{S\'ergio Guerreiro}
\affiliation{%
  \institution{INESC-ID, Instituto Superior T\'ecnico, Universidade de Lisboa, Portugal}
  \streetaddress{Rua Alves Redol, 9}
  \city{Lisboa}
  \postcode{1000-029}}
\email{sergio.guerreiro@tecnico.ulisboa.pt}

\author{Miguel Correia}
\affiliation{%
  \institution{INESC-ID, Instituto Superior T\'ecnico, Universidade de Lisboa, Portugal}
  \streetaddress{Rua Alves Redol, 9}
  \city{Lisboa}
  \postcode{1000-029}}
\email{miguel.p.correia@tecnico.ulisboa.pt}

\renewcommand{\shortauthors}{Belchior et al.}

\begin{abstract}
Blockchain interoperability is emerging as one of the crucial features of blockchain technology, but the knowledge necessary for achieving it is fragmented. This fact makes it challenging for academics and the industry to achieve interoperability among blockchains seamlessly. Given this new domain's novelty and potential, we conduct a literature review on blockchain interoperability by collecting 284 papers and 120 grey literature documents, constituting a corpus of 404 documents. From those 404 documents, we systematically analyzed and discussed 102 documents, including peer-reviewed papers and grey literature. Our review classifies studies in three categories: Public Connectors, Blockchain of Blockchains, and Hybrid Connectors. Each category is further divided into sub-categories based on defined criteria. We classify 67 existing solutions in one subcategory using the Blockchain Interoperability Framework, providing a holistic overview of blockchain interoperability.
Our findings show that blockchain interoperability has a much broader spectrum than cryptocurrencies and cross-chain asset transfers. Finally, this paper discusses supporting technologies, standards, use cases, open challenges, and
future research directions, paving the way for research in the area.





\end{abstract}



\maketitle
\newcommand{\STAB}[1]{\begin{tabular}{@{}c@{}}#1\end{tabular}}

\section{Introduction}

\begin{wrapfigure}{r}{0.4\textwidth} 

    \centering
    \includegraphics[width=\linewidth]{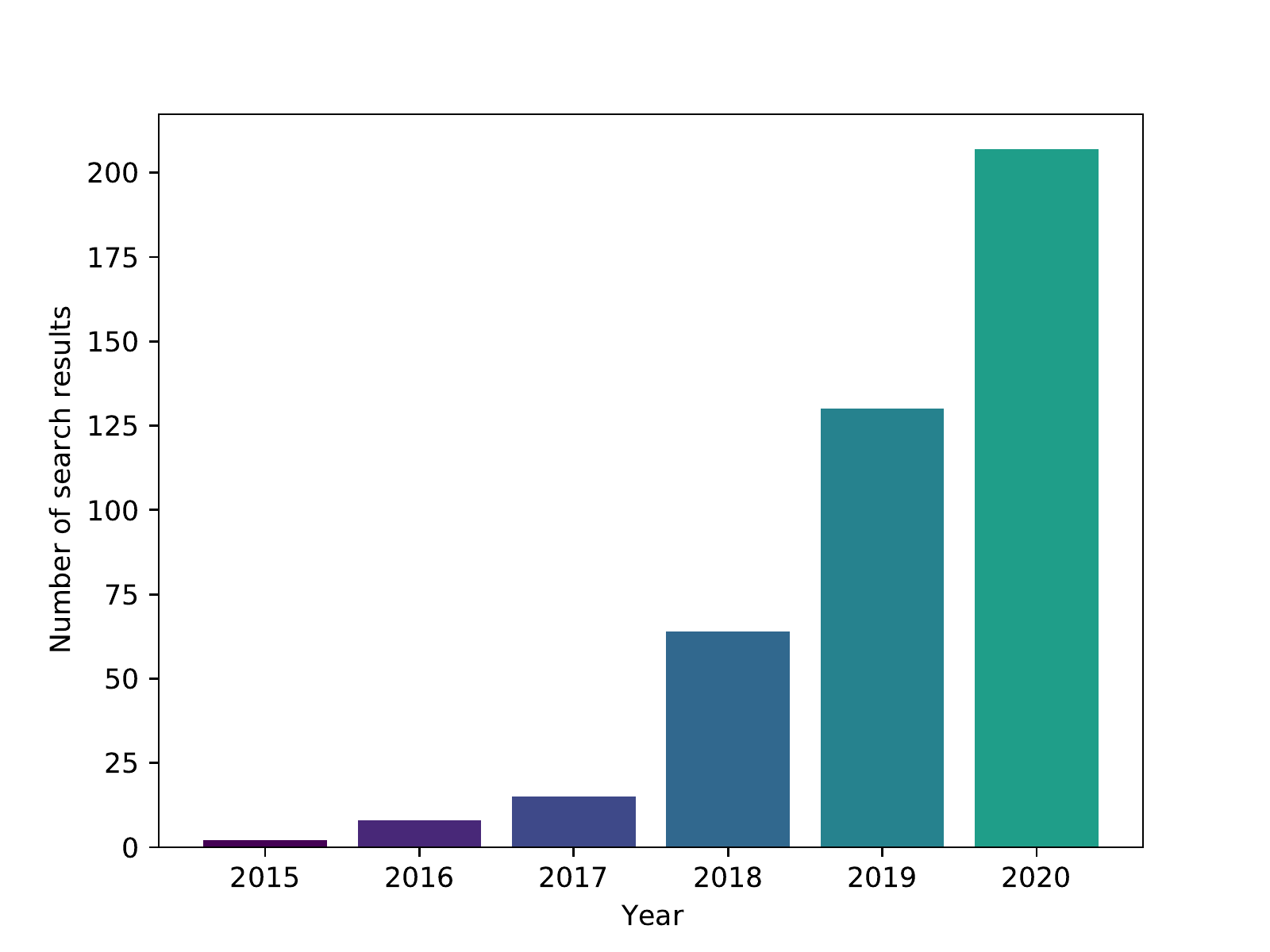}
        \caption{Research trends on blockchain interoperability}
    \label{fig:stats}
\end{wrapfigure}

Blockchain technology is maturing at a fast pace. The development of real-world applications shows real interest from both industry and academia \cite{zohar2015,UlHassan2019}. For instance, applications have been developed in the areas of public administration \cite{Belchior2019, Belchior2019_Audits}, access control \cite{rouhani2019,ssibac}, and others \cite{casino2019}. Additionally, blockchain is progressing towards the performance of centralized systems: for example, the Hyperledger Fabric blockchain is predicted to achieve 50,000 \emph{transactions per second} \cite{Gorenflo2019,Gorenflo2020}. Figure \ref{fig:stats} depicts the number of search results per year for ``blockchain interoperability'' that Google Scholar returned. In 2015, only two documents were related to blockchain interoperability. In 2016, 2017, 2018, 2019, and 2020, the results were 8, 15, 64, 130, and 207, respectively, showing a steep increase regarding interest in this research area.

Serving multiple use cases and stakeholders requires various blockchain features and capabilities \cite{wef2020}. The need for adaptability is a motivating factor for creating different blockchains, leading to a heterogeneous ecosystem \cite{Xu2017,hardjono2021,pilai2020}. Choosing new blockchains allows researchers and developers to implement new use case scenarios and keep up with recent endeavors. However, each blockchain has its security risks, as the technology is still maturing, the user base is limited (e.g., in comparison to the web or databases), and there are uncovered bugs, and security flaws \cite{sok_attacks}. Therefore, developers and researchers have to choose between novelty and stability, leading to a vast diversity of choices \cite{Anceaume2018}. This diversity leads to \emph{fragmentation}: there are many \emph{immature} blockchain solutions (e.g., without extensive testing). Until recently, blockchains did not consider the need for interoperability, as each one focused on resolving specific challenges, leading to \emph{data and value silos} \cite{jin2018, Abebe2019,tasca2017}. 

Moreover, what if the blockchain in which a particular service is running becomes obsolete, vulnerable, or is shutdown? If the user requirements or circumstances change over time, a different blockchain might be more appropriate for a specific use case \cite{agileLikoebe}. What if the service to serve is so crucial that it requires seamless dependability? Furthermore, if we want to reproduce our use case to another blockchain, how can we increase \emph{portability}? 
 
In 1996, Wegner stated that ``interoperability is the ability of two or more software components to cooperate despite differences in language, interface, and execution platform'' \cite{wegner96}. In that context, Wegner established a bridge between the concept of interoperability and existing standards. As authors were influenced by the standards existing at that time, authors nowadays are influenced by the Internet architecture and concepts, in what concerns blockchain interoperability \cite{Hardjono2019,vo2018}. Thus, reflecting on the Internet's architecture seems like a good starting point to understand how blockchains can interoperate. Thus, it is important to solve the \emph{blockchain interoperability} challenge, i.e., to provide interoperability between blockchains in order to explore synergies between different solutions, scale the existing ones, and create new use cases (see Section \ref{subsec:bid}). For example, a user should be able to transfer their assets from a blockchain to another or build \emph{cross-blockchain decentralized applications}. 

While information systems evolve, so do the meaning and scope of interoperability. According to the National Interoperability Framework Observatory (NIFO), endorsed by the European Commission, there are several interoperability layers \cite{NIFO_inter}: \emph{technical interoperability}, \emph{semantic interoperability}, \emph{organizational interoperability}, \emph{legal interoperability}, \emph{integrated public service governance}, and \emph{interoperability governance}. For instance, technical interoperability regards the technical mechanisms that enable integration among blockchains, while semantic interoperability concerns whether the application-specific semantics can be conserved across blockchains. Despite interoperability having an extensive scope, we mainly focus on \emph{technical interoperability}, and \emph{semantic interoperability} as most blockchain interoperability work is concentrated. We leave the study of other interoperability layers for future work. 


Interoperability does not only conflate flexibility and application portability. It also has the potential to solve some of the biggest blockchain research challenges. In particular, interoperability promotes blockchain \emph{scalability}, as it provides a way to offload transactions to other blockchains, e.g., via sharding \cite{scotm,bcsharding}, it can promote privacy (by allowing the end-user to use different blockchain for data objects with different privacy requirements), and creates new business opportunities. Given the complexity of this research area, we attempt to answer three research questions:

 \textbf{RQ1}: What is the current landscape concerning blockchain interoperability, both in industry and academia?

 \textbf{RQ2}: Are technological requirements for blockchain interoperability currently satisfied?
 
\textbf{RQ3}: Are there real use cases requiring blockchain interoperability? 

\subsection{Contributions}

As a systematization of knowledge on blockchain interoperability, this paper yields four-fold contributions:
\begin{itemize}
    \item Introduce the blockchain interoperability research area, presenting the necessary background and highlighting definitions tailored both for industry and academia. We define blockchain interoperability and discuss different blockchain interoperability architectures and standards.
    \item Propose the Blockchain Interoperability Framework (BIF), a framework defining criteria to assess blockchain interoperability solutions.
    \item Present a \emph{systematic literature review}, where we identify and discuss blockchain interoperability solutions, accordingly to BIF, in three categories: \emph{Public Connectors}, \emph{Blockchain of Blockchains}, and \emph{Hybrid Connectors}. In particular, our analysis is based on several sources (e.g., peer-reviewed papers, whitepapers, blog posts, technical reports), enabling an in-depth understanding of each solution's current state and its \emph{roadmap}, i.e., its creator's plans. To achieve this, \emph{we systematically contacted the authors of grey literature papers and industrial solutions}: this is our innovative attempt to provide the reader with high-quality information in this rapidly emerging research area. This method allows us to obtain up-to-date, reliable information that often is cumbersome to obtain.
    \item We identify and propose use cases that benefit from a multiple blockchain approach, pinpoint challenges and obstacles to the development of blockchain interoperability solutions and standards, and propose future research directions, paving the way for systematic research in this area.
\end{itemize}

\subsection{Organization}
Section \ref{sec:back} provides background on blockchain consensus algorithms, previous results on blockchain interoperability, and blockchain interoperability definitions and architecture. Next, Section \ref{sec:related_literature_reviews} presents and discusses related literature reviews, while Section \ref{sec:bif} introduces the Blockchain Interoperability Framework.
Next, a systematic review and analysis of blockchain interoperability categories is conducted, distributed across three categories, in Section \ref{sec:solutions}: Public Connectors (Section \ref{sec_crypto}), Blockchain of Blockchains (Section \ref{sec:be}), and Hybrid Connectors (Section \ref{subsec:blockchain_connectors}). For each category, we provide a detailed analysis and discussion.
To provide a holistic view of the blockchain interoperability landscape, we promote a general discussion in Section \ref{sec:discussion_sec}. This discussion compares solutions across categories (Section \ref{sec:discussion}), presents standardization efforts (Section \ref{sec:technologies_standards}), informs readers regarding use case scenarios with multiple blockchains  (Section \ref{sec:use_cases}), answers to the research questions (Section \ref{sec:res_q_a}),  and indicates challenges related to interoperability (Section \ref{sec:issues}). We present research directions (Section \ref{sec:research_directions}), and, finally, we conclude the paper (Section \ref{sec:concl}).
Six appendices complement this survey. Appendix \ref{a:method} presents the methodology employed. 
Appendix \ref{a:architecture} presents an architecture for blockchain interoperability, reviewing the various efforts on that topic. Appendix \ref{a:crypto}, \ref{a:be} and \ref{a:connectors} presents a description of the surveyed public connectors, blockchain of blockchains, and hybrid connector approaches, respectively. Finally, Appendix \ref{a:use_cases} complements the use case section, by presenting more cross-blockchain use cases.

\section{Background}
\label{sec:back}
In this section, we provide the necessary background to the understanding of this survey. 

\subsection{A Primer on Blockchain Technology}
\label{sec:primer}
The term \emph{blockchain} has at least two different meanings: a type of system and a type of data structure. In this paper, we use the term blockchain to denominate a class of distributed systems. A blockchain maintains a shared state, specifically a replicated data structure that we denominate \emph{distributed ledger}. This ledger is maintained by a set of machines with computational and storage resources, called nodes (or peers or participants). Nodes are not trusted individually to maintain the distributed ledger; they are trusted as a group due to their number and diversity \cite{vukolic_bcpitw}. A blockchain can also be considered a \emph{deterministic state machine} that provides a certain service, given existing incentives that the network can reward. The first blockchain was part of the Bitcoin system and provided as service transactions of a cryptocurrency, a digital currency, also designated Bitcoin \cite{nakamoto2008}. The service provided by Bitcoin is the execution of transactions of bitcoins. 

Most blockchains are programmable, i.e., their state machine is extensible with user programs. These programs are often designated \emph{smart contracts} \cite{szabo1997paper,ethereum-white-paper} and their execution is caused by calls also designated \emph{transactions}. Smart contracts are executed in a virtual machine, e.g., in the Ethereum Virtual Machine (EVM) in Ethereum and other blockchains that adopted the EVM for compatibility (that we designate \emph{EVM-based blockchains}). Smart contracts are often used to implement \emph{tokens}, i.e., blockchain-based abstractions that can be owned and represent currency, resources, assets, access, equity, identity, collectibles, etc.~\cite{antonopoulos2018mastering}. There are several standard token formats, e.g., ERC-20 and ERC-721. These tokens are fungible and non-fungible assets, respectively. A fungible asset is interchangeable with another asset of the same type. Conversely, a non-fungible asset is an asset that is unique and has specific properties. 

In many blockchains, transactions are aggregated in \emph{blocks}, linked by the previous block's cryptographic hash. Hence those data structures are also called blockchains - often viewed as deterministic state machines.

Blockchain systems ought to be resilient to faults (e.g., \emph{crash fault-tolerant} or \emph{Byzantine fault-tolerant}), as there may be crashes or malicious nodes on the network \cite{correia2019byzantine}. They run a consensus algorithm to create agreement on a global ledger state in the presence of Byzantine faults. Consensus algorithms are important because they define the behavior of blockchain nodes and their interaction \cite{correia2019byzantine,zheng2017}, and the security assumptions of each blockchain. They, therefore, affect how blockchain peers communicate and operate with each other: in Bitcoin's Proof-of-Work (PoW), peers have to compute a cryptographic challenge to validate transactions, competing with each other. Another blockchain, Tendermint, uses a Byzantine fault-tolerant state machine replication (BFT) algorithm \cite{Kwon2016}, supporting up to a third less one of faulty participants. In Hyperledger Fabric, a widely-used private blockchain platform, a consensus algorithm allows higher transaction throughput than PoW by allowing a subset of nodes to execute and endorse transactions (called endorser peers) and by typically using a weaker consensus (only crash fault-tolerant). The variety of blockchain infrastructures makes it challenging to categorize blockchains, and their interoperability solutions, as there are no \emph{de facto} blockchain interoperability or blockchain architecture standards.

Apart from differences in the consensus, blockchains can be deemed \emph{public} (also called permissionless) or \emph{private} (also called permissioned).
Permissionless blockchains do not require authentication for participants to access the ledger. \emph{Bitcoin} \cite{nakamoto2008} and \emph{Ethereum} \cite{ethereum-white-paper,ethereum_yellow_paper} are examples of such blockchains. Permissioned blockchains are blockchains in which users are authenticated and can be held accountable according to a governance model suitable for enterprise and governmental needs. Hyperledger Fabric \cite{fabric}, Corda \cite{r3}, Quorum \cite{quorum}, Tendermint \cite{Kwon2016}, and Multichain \cite{multichain} are examples of permissioned blockchains.  

\begin{wrapfigure}{r}{0.6\textwidth} 

    \centering
    \includegraphics[scale=0.32]{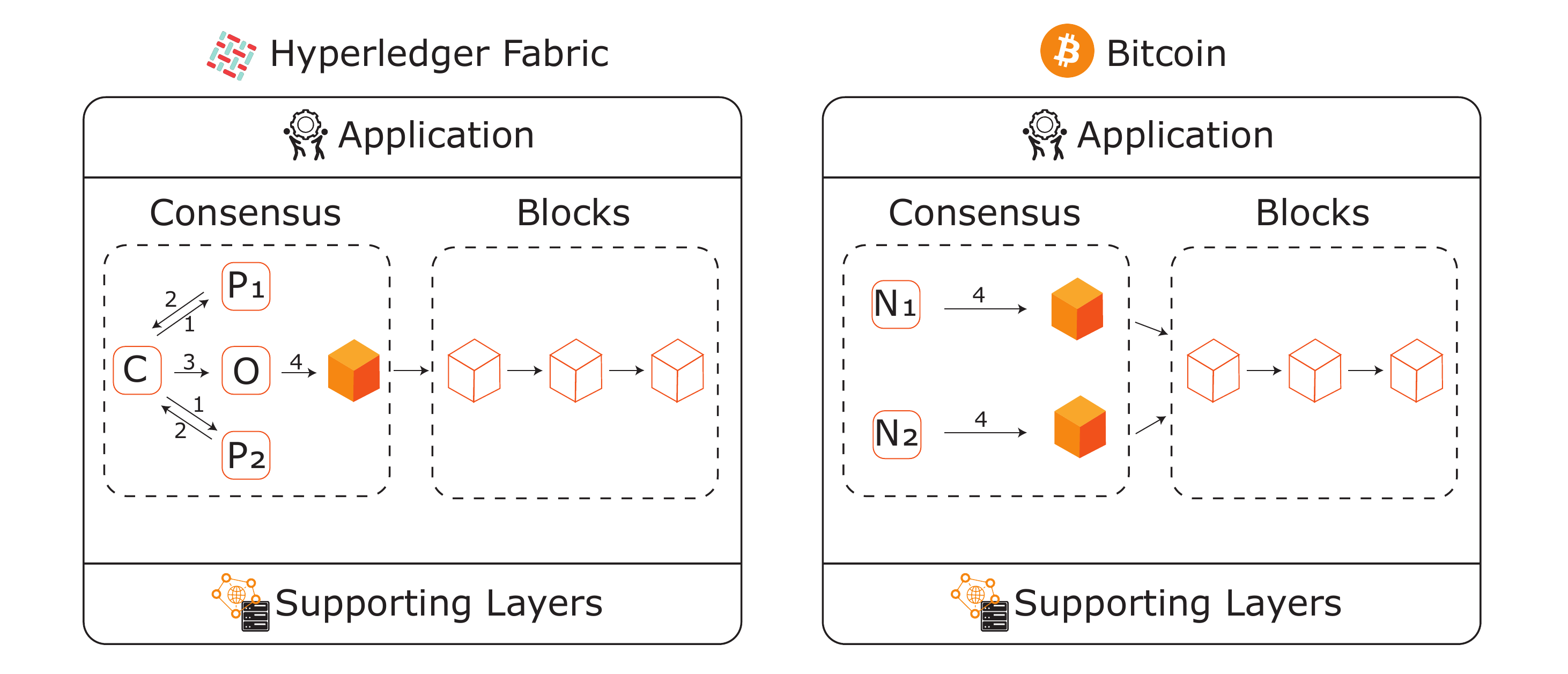}
        \caption{Representation of two blockchains, Hyperledger Fabric \cite{fabric}, and Bitcoin \cite{nakamoto2008}.}
    \label{fig:blockchains}
\end{wrapfigure}

Figure \ref{fig:blockchains} depicts two blockchains: Hyperledger Fabric, a permissioned blockchain; and Bitcoin, a permissionless blockchain. The supporting layers  (e.g., networking, storage, encryption) \cite{Kan2018} provide a basis for the consensus engine, which orders transactions and appends them to the chain of blocks. In Hyperledger Fabric, the consensus is modular, based on endorsement policies. In Fabric, a client (C) sends a transaction proposal to the peer nodes (P), and obtains a signed transaction, called an endorsement (steps 1 and 2). An orderer validates the endorsements and builds a block with valid transactions, appending it to the ledger (steps 3 and 4). In Bitcoin, the consensus is based on the notion of Proof-of-Work (PoW), a cryptographic puzzle that mining nodes need to solve in order to build a valid block. This corresponds roughly to Fabric's steps 1-3. After a node finds a solution to PoW, it then can propose a block of transactions to be appended to the ledger (step 4).

Blockchain trust is based on the incentive models that guide the behavior of the nodes. For instance, in Bitcoin, nodes have the incentive to produce blocks of transactions and support the network because they are rewarded Bitcoins. Conversely, nodes do not have the incentive to disrespect the protocol, as attacks are expensive and nodes can get punished \cite{Conti2018}. In Hyperledger Fabric, where nodes are identified, they have the business incentive to follow the protocol because parties cooperate towards a common goal, and misbehavior can be punished according to the law or applicable governance model. Decentralization, different goals, and incentives support the trust on the blockchain -- parties can share the ledger without relying on a trusted, centralized party.

The ability to distribute trust on a global state fostered the appearance of \emph{decentralized applications} (\emph{dApps}) \cite{antonopoulos2018mastering}. A dApp is a computer program running on a decentralized peer-to-peer network. For example, Steemit\footnote{https://steemit.com/} is a social blogging dApp that rewards content-creators with cryptocurrency. Thus, dApps are based on smart contracts running on a blockchain, but they also have other components that should equally be decentralized.

\subsection{Cross-Blockchain Communication}
\label{sec:ccc}
Cross-blockchain communication involves two blockchains: a \emph{source blockchain}, and a \emph{target blockchain}. The source blockchain is the blockchain in which the transaction is initiated to be executed on a target blockchain. While general-purpose interoperability comes down to a blockchain exposing its internal state to other, cross-chain asset transfers rely on an atomic three-phase procedure: 1) locking (or extinguishing) of an asset on a source blockchain; 2) blockchain transfer commitment, and 3) creation of a representation of the asset on a target blockchain \cite{odap_draft_01,hermes-middleware-2021,scotm}. This procedure, later explained in detail, relies on a \emph{cross-chain communication protocol} (CCCP).

A CCCP defines the process by which a pair of blockchains interact to synchronize cross-chain transactions correctly. Hence, a CCCP allows \emph{homogeneous} blockchains to communicate. For instance, sidechains typically use a CCCP (e.g., Zendoo allows communication between Bitcoin-like blockchains systems \cite{zendo}). Conversely, a \emph{cross-blockchain communication protocol} (CBCP) defines the process by which a pair of blockchains interact to synchronize cross-blockchain transactions correctly. CBCPs allow \emph{heterogeneous} blockchains to communicate (e.g., the Interledger Protocol allows any blockchains that implement the protocol to exchange 
``money packets'' \cite{ILPv4}). The differentiation between CCCPs and CBCPs is important because CCCPs typically can leverage the interoperating blockchains' constructs and functionality (e.g., utilize smart contracts to implement a relay \cite{peacerelay}), whereas CBCPs normally require blockchains to be adapted. However, CBCPs may leverage specific functionalities of both blockchains \cite{btcrelay}. Cross-blockchain, or cross-chain communication, is a requirement for blockchain interoperability. This section provides a few theoretical results regarding cross-blockchain communication, and thus also blockchain interoperability.

Zamyatin et al.~\cite{sok_cdl} prove that ``there exists no asynchronous CCC [cross-chain communication] protocol tolerant against misbehaving nodes''. The authors use a reduction to the fair exchange problem \cite{fair_exchange} to prove that correct cross-chain communication is as hard as the fair exchange problem. As a consequence of the presented theorem, the authors state that ``there exists no CCC protocol tolerant against misbehaving nodes without a trusted third party''. A trusted third party can be centralized or decentralized. Centralized trusted parties are, for example, trusted validators \cite{hyperledger_cactus}. A decentralized trusted party can be another blockchain, in which their participants agree on the global ledger state via a consensus algorithm. However, the trusted party has to ensure that most participants are honest, guaranteeing the correctness of the process is guaranteed. 
Cross-chain protocols, therefore ``use the consensus of the distributed ledgers as an abstraction for a trusted third party.'' \cite{sok_cdl}. Borkowski et al.~\cite{tast_paper7} derive the ``lemma of rooted blockchains'' that states that a source blockchain cannot verify the existence of data on a target blockchain with practical effort. In particular, the source blockchain would need to be able to mimic consensus from the target blockchain, and it would have to store a (potentially large) subset of the target blockchain's block history. On a recent endeavor, Lafourcade and Lombard-Platet \cite{pascal2020} formalize the blockchain interoperability problem, arguing that fully decentralized blockchain interoperability is not possible. More specifically, there is no protocol assuming a full-client that can realize its interoperability functions, such as asset transfer, without a third party's aid. However, a blockchain with two ledgers offers the possibility of interoperability (there is, in fact, the possibility of moving assets from one ledger to the other). This study applies mainly to public blockchains.

The results above are relevant because they lead to an important consideration: \emph{cross-blockchain transactions are not feasible in practice without the participation of a trusted third party}. In other words, although trust assumptions vary greatly from permissionless to permissioned networks, cross-blockchain transactions, as well as cross-chain transactions, require a trusted third party to assure the correctness of the underlying protocol. Most solutions presented throughout this paper present at least one decentralized trust anchor.

\subsection{Blockchain Interoperability Definitions}
\label{subsec:bid}
In this section, we define additional technical terms for an understanding of this study.
    
Vernadat defines interoperability among enterprise systems as \cite{Vernadat2006}: ``a measure of the ability to perform interoperation between [...] entities (software, processes, systems, business units...). 
The challenge relies on facilitating communication, cooperation, and coordination among these processes and units''.
Abebe et al.~propose a general communication protocol as an alternative approach to the ``point-to-point'' blockchain interoperability approach \cite{Abebe2019}. Interoperability is defined as ``the semantic dependence between distinct ledgers to transfer or exchange data or value, with assurances of validity''.
Pillai and Biswas refer that ``cross-communication is not intended to make direct state changes to another blockchain system. Instead, cross-communication should trigger some set of functionalities on the other system that is expected to operate within its own network'' \cite{Pillai2019}. 

A technical report from the National Institute of Standards and Technology (NIST) defines blockchain interoperability as ``a composition of distinguishable blockchain systems, each representing a unique distributed data ledger, where atomic transaction execution may span multiple heterogeneous blockchain systems, and where data recorded in one blockchain are reachable, verifiable, and \emph{referable} by another possibly foreign transaction in a semantically compatible manner'' \cite{NISTIR}.
Hardjono et al.~define blockchain survivability as ``the completion (confirmation) of an application-level transaction [composed of subtransactions] independent of blockchain systems involved in achieving the completion of the transaction.''\cite{Hardjono2019} The concept of transactions and subtransactions relates to ``\emph{best effort delivery}'', that applications must comply to, by ensuring that transactions and their \emph{subtransactions} are completed (i.e., committed) within a certain time frame. 

\begin{wrapfigure}{r}{0.5\textwidth} 
    \centering
    \includegraphics[scale=0.22]{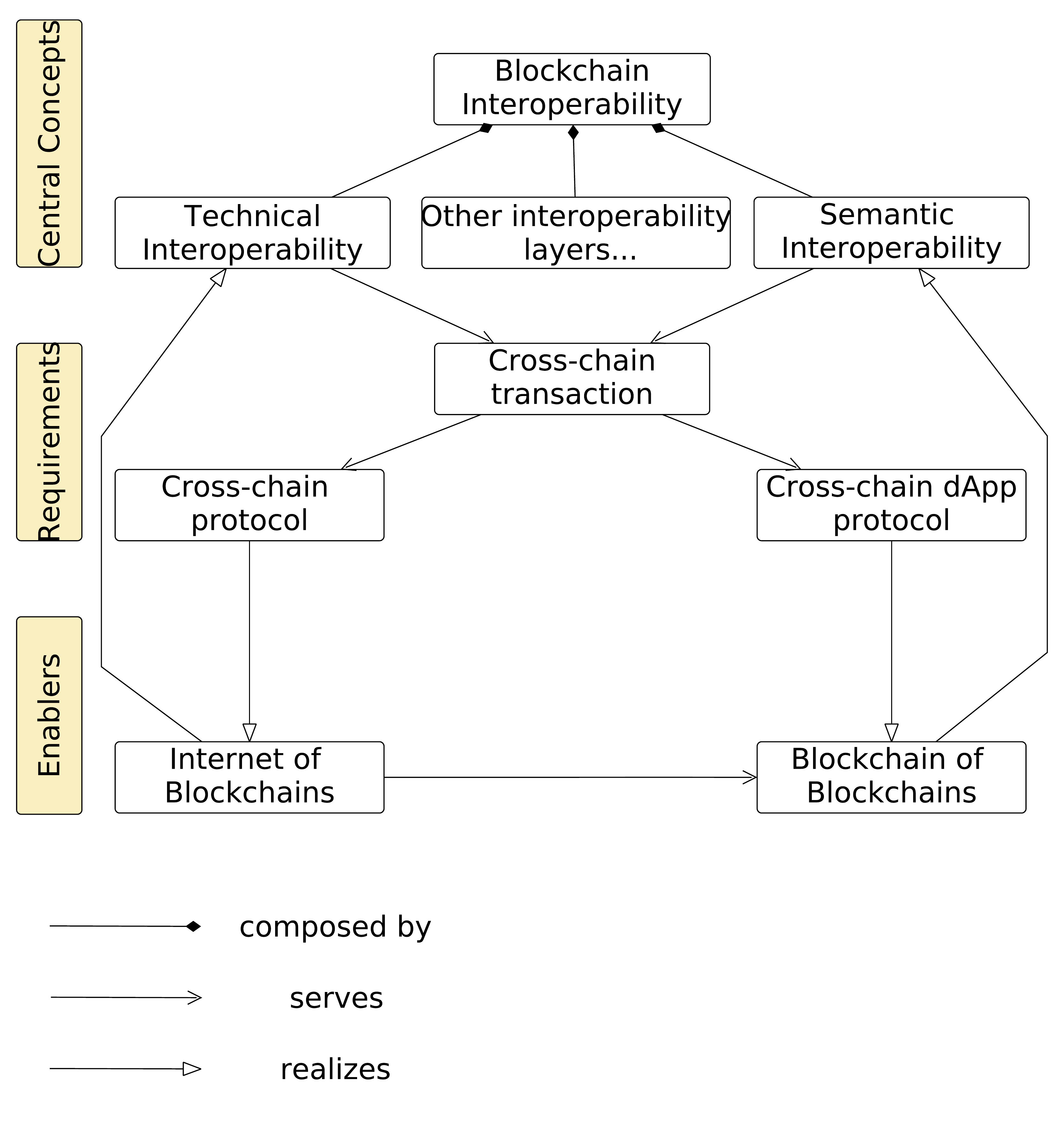}
    \caption{Concept map, illustrating the relationship between different concepts related to blockchain interoperability}
    \label{fig:concept_map}
\end{wrapfigure}

Regarding types of blockchain interoperability, Besançon et al. highlight three \cite{tbig}: interoperability between different blockchains, interoperability between dApps using the same blockchain, and interoperability blockchain and other technologies (such as integration with enterprise systems).
While different definitions tackle different dimensions of interoperability, there is room for improvement. We define several terms that encompass the whole scope of technical interoperability to later provide a holistic definition of technical interoperability (see Figure \ref{fig:concept_map}). 
To recall the definition presented in Section \ref{sec:ccc}, a  {source blockchain} is a blockchain that issues transactions against a {target blockchain}. A \emph{source node} is a node from the source blockchain, and a target node belongs to the target blockchain. When several participants elect a source node and a target node, we achieve decentralization in the context of interoperability \cite{jin2018}.  

A  \ac{CC-Tx}, where ``CC'' stands for \emph{cross-chain}, and ``Tx'' for transaction, is a transaction between different chains, which belong to the same blockchain system (homogeneous blockchains), for example, between EVM-based blockchains. We use the \ac{CC-Tx}, \emph{inter-chain transaction}, and \emph{inter-blockchain transaction} terms interchangeably. 
A \ac{CB-Tx} is a transaction between different blockchains (heterogeneous blockchains), for example, between Hyperledger Fabric and Bitcoin. Note that the terms \ac{CC-Tx} and \ac{CB-Tx} are used as synonyms in the industry, as currently, most solutions connect homogeneous blockchains. 
A \ac{CC-dApp} is a dApp that leverages cross-blockchain transactions to implement its business logic. We use the terms \ac{CC-dApp} and  \emph{cross-blockchain decentralized application} (CB-dApp) interchangeably. Other terms with the same meaning in the literature are 
inter-chain decentralized application and inter-blockchain decentralized application.

A \ac{IoB} is a system ``where homogeneous and heterogeneous decentralized networks communicate to facilitate cross-chain transactions of value'' \cite{vo2018}. We use this definition of IoB throughout this paper.

The term \ac{BoB} is not used consistently 
\cite{Verdian2018, hyperservice}. Verdian et al.~use it to describe the structure that aggregates blocks from different blockchains into ``meta blocks'', organized through a consensus mechanism using \emph{posets} (partially ordered sets) and total order theory \cite{Verdian2018}, thus producing a blockchain of blockchains. A poset consists of a set of elements and their binary relationships that are ordered according to a specific set of rules \cite{lattice}. 


Influenced by those authors, we define a \ac{BoB} \emph{as a system in which a consensus protocol organizes blocks that contain a set of transactions belonging to CC-dApps,  spread across multiple blockchains. Such a system should provide accountability for the parties issuing transactions on the various blockchains and providing a holistic, updated view of each underlying blockchain}. Note that BoB solutions belong to the category with the same name.
Therefore, the notion of \ac{IoB} directly refers to the connection relationships among blockchains, whereas the term \ac{BoB} refers to an architecture made possible by IoB. BoB approaches are concerned with the validation and management of cross-blockchain transactions. 

Figure \ref{fig:concept_map} shows the relationship between the different concepts concerning blockchain interoperability. A \ac{CC-dApp} realizes the blockchain of blockchains approach. This approach can provide the semantic level interoperability (i.e., concerned at transmitting the meaning of the data, which corresponds to the value level interoperability) required by organizations, mappable by the applicational layer. However, it relies on the existence of an IoB -- a network of blockchains. For an IoB to exist, technical interoperability (or mechanical interoperability) is required. In the context of a CC-dApp, cross-chain transactions are ordered by a \emph{cross-chain dApp protocol}. Such protocols should assure transaction atomicity and resolve possible conflicts in transactions spawning across homogeneous and heterogeneous blockchains. 

From the several definitions we encountered during our research, we envision \emph{blockchain interoperability} as:
\emph{the ability of a source blockchain to change the state of a target blockchain (or vice-versa), enabled by cross-chain or cross-blockchain transactions, spanning across a composition of homogeneous and heterogeneous blockchain systems, the IoB.} IoB transactions are delivered via a cross-blockchain communication protocol, thereby granting technical interoperability, enabling \ac{CC-dApp}s. \ac{CC-dApp}s provide semantic interoperability via the BoB. The BoB approach is realized by a cross-blockchain dApp protocol, which provides consensus over a set of cross-chain transactions, thus enabling cross-chain dApps.

\section{Related Literature Reviews}
\label{sec:related_literature_reviews}

Due to the novelty and large-breadth of this research area, few updated surveys cover aspects of blockchain interoperability. We compare existing surveys based on the \emph{criteria} and \emph{sub-criteria} shown in Table \ref{tab:related_survey_criteria}. For example, in the first row, the criteria ``public connector'' evaluates if a study addresses its sub-criteria: work on sidechains, hash-lock time contracts, and notary schemes. On the second row, the criteria Blockchain of Blockchains evaluates if a study describes BoB solutions (1) and if it performs a detailed comparison, including consensus, security, validators, and performance. 
\begin{table}[]
\centering
\resizebox{\linewidth}{!}{%
\begin{tabular}{@{}lllll@{}}
\toprule
\textbf{Criteria}                        & \textbf{Description}                   & \textbf{Sub-criteria 1} & \textbf{Sub-criteria 2} & \textbf{Sub-criteria 3} \\ \midrule
\textbf{Public Connectors (PC)}          & Addresses public connectors            & Sidechains              & Hash lock contracts     & Notary Schemes          \\ \midrule
\textbf{Blockchain of Blockchains (BoB)} & Addresses BoBs                         & Describes solutions     & Detailed comparison     & N/A                     \\ \midrule
\textbf{Hybrid Connectors (HC)}    & Addresses Hybrid Connectors              & Trusted Relays      & Blockchain agnostic protocols           & Blockchain migrators \\ \midrule
\textbf{Architecture (AR)}                & Addresses architectures enabling CCCPs & Proposes architecture   & Presents related work   & N/A                     \\ \midrule
\textbf{Cross-chain Standards (ST)} & Addresses standards for interoperability & Present standards   & Relate standards to solutions           & N/A                  \\ \midrule
\textbf{Cross-analysis (CC)}             & Compares across categories             & Compare categories      & Compare sub-categories  & N/A                     \\ \midrule
\textbf{Use Cases (UC)}                  & Presents use cases using an IoB or BoB & Existing use cases      & Predicted use cases     & N/A                     \\ \midrule
\textbf{Open Issues (OI)}          & Challenges on interoperability           & Research directions & Relate interoperability to other issues & N/A                  \\ \bottomrule
\end{tabular}%
}
\caption{Survey comparison criteria, description, and sub-criteria. }
\label{tab:related_survey_criteria}
\end{table}

Buterin presents a survey on public connector solutions, including notary schemes, sidechains, and hash-time locking techniques \cite{vitalik2016}. Similarly, other surveys focus on public connectors \cite{sok_cdl,tast_paper8,koens2019,singh2020}, with a focus on sidechains and hash lock time contracts. Vo et al. present work mostly on architecture for interoperability, presenting some BoB and HC solutions \cite{vo2018}, while Qasse et al. organize solutions across sidechains, blockchain routers, smart contracts, and industrial solutions \cite{Qasse2019}. Johnson et al. focus on Ethereum as the infrastructure enabling interoperability across several categories of solutions \cite{Johnson2019}. Siris et al.~\cite{inter_approaches}, Kannengieber et al. \cite{niclas2020}, and Bishnoi et al. \cite{Bishnoi2020} tackle a wider range of solutions.

\begin{wraptable}{r}{0.7\textwidth} 
\centering
\footnotesize
\begin{tabular}{@{}lcccccccc@{}}
\toprule
 &
  \multicolumn{3}{c}{\textbf{Solution category}} &
  \multicolumn{5}{c}{\textbf{Detailed Analysis}} \\ \midrule
\multicolumn{1}{c}{} &
   &
   &
  \multicolumn{1}{c|}{} &
   &
   &
   &
   &
   \\
\multicolumn{1}{c}{\multirow{-2}{*}{\textbf{Reference}}} &
  \multirow{-2}{*}{PC} &
  \multirow{-2}{*}{BoB} &
  \multicolumn{1}{c|}{\multirow{-2}{*}{HC}} &
  \multirow{-2}{*}{AR} &
  \multirow{-2}{*}{ST} &
  \multirow{-2}{*}{CC} &
  \multirow{-2}{*}{UC} &
  \multirow{-2}{*}{OI} \\ \midrule
\multicolumn{1}{l|}{Buterin \cite{vitalik2016}, 2016} &
  \cellcolor[HTML]{D8E3BB}+ &
  \cellcolor[HTML]{BF504D}- &
  \multicolumn{1}{c|}{\cellcolor[HTML]{BF504D}-} &
  \cellcolor[HTML]{BF504D}- &
  \cellcolor[HTML]{BF504D}- &
  \cellcolor[HTML]{F79545}$\pm$ &
  \cellcolor[HTML]{D8E3BB}+ &
  \cellcolor[HTML]{D8E3BB}+ \\ \midrule
\multicolumn{1}{l|}{Vo et al.\cite{vo2018}, 2018} &
  \cellcolor[HTML]{BF504D}- &
  \cellcolor[HTML]{F79545}$\pm$ &
  \multicolumn{1}{c|}{\cellcolor[HTML]{F79545}$\pm$} &
  \cellcolor[HTML]{D8E3BB}+ &
  \cellcolor[HTML]{F79545}$\pm$ &
  \cellcolor[HTML]{F79545}$\pm$ &
  \cellcolor[HTML]{F79545}$\pm$ &
  \cellcolor[HTML]{D8E3BB}+ \\ \midrule
\multicolumn{1}{l|}{Borkowski et al. \cite{tast_paper5}, 2018} &
  \cellcolor[HTML]{D8E3BB}+ &
  \cellcolor[HTML]{BF504D}- &
  \multicolumn{1}{c|}{\cellcolor[HTML]{BF504D}-} &
  \cellcolor[HTML]{BF504D}- &
  \cellcolor[HTML]{BF504D}- &
  \cellcolor[HTML]{F79545}$\pm$ &
  \cellcolor[HTML]{BF504D}- &
  \cellcolor[HTML]{D8E3BB}+ \\ \midrule
\multicolumn{1}{l|}{Quasse et al. \cite{Qasse2019}, 2019} &
  \cellcolor[HTML]{F79545}$\pm$ &
  \cellcolor[HTML]{F79545}$\pm$ &
  \multicolumn{1}{c|}{\cellcolor[HTML]{F79545}$\pm$} &
  \cellcolor[HTML]{BF504D}- &
  \cellcolor[HTML]{BF504D}- &
  \cellcolor[HTML]{F79545}$\pm$ &
  \cellcolor[HTML]{F79545}$\pm$ &
  \cellcolor[HTML]{F79545}$\pm$ \\ \midrule
\multicolumn{1}{l|}{Johnson et al. \cite{Johnson2019}, 2019} &
  \cellcolor[HTML]{F79545}$\pm$ &
  \cellcolor[HTML]{F79545}$\pm$ &
  \multicolumn{1}{c|}{\cellcolor[HTML]{F79545}$\pm$} &
  \cellcolor[HTML]{BF504D}- &
  \cellcolor[HTML]{BF504D}- &
  \cellcolor[HTML]{BF504D}- &
  \cellcolor[HTML]{BF504D}- &
  \cellcolor[HTML]{BF504D}- \\ \midrule
\multicolumn{1}{l|}{Zamyatin et al. \cite{sok_cdl}, 2019} &
  \cellcolor[HTML]{D8E3BB}+ &
  \cellcolor[HTML]{BF504D}- &
  \multicolumn{1}{c|}{\cellcolor[HTML]{BF504D}-} &
  \cellcolor[HTML]{BF504D}- &
  \cellcolor[HTML]{BF504D}- &
  \cellcolor[HTML]{F79545}$\pm$ &
  \cellcolor[HTML]{BF504D}- &
  \cellcolor[HTML]{D8E3BB}+ \\ \midrule
\multicolumn{1}{l|}{Siris et al. \cite{inter_approaches}, 2019} &
  \cellcolor[HTML]{F79545}$\pm$ &
  \cellcolor[HTML]{F79545}$\pm$ &
  \multicolumn{1}{c|}{\cellcolor[HTML]{F79545}$\pm$} &
  \cellcolor[HTML]{F79545}$\pm$ &
  \cellcolor[HTML]{BF504D}- &
  \cellcolor[HTML]{D8E3BB}+ &
  \cellcolor[HTML]{BF504D}- &
  \cellcolor[HTML]{BF504D}- \\ \midrule
\multicolumn{1}{l|}{Koens et al. \cite{koens2019}, 2019} &
  \cellcolor[HTML]{D8E3BB}+ &
  \cellcolor[HTML]{D8E3BB}+ &
  \cellcolor[HTML]{BF504D}- &
  \cellcolor[HTML]{BF504D}- &
  \cellcolor[HTML]{BF504D}- &
  \cellcolor[HTML]{F79545}$\pm$ &
  \cellcolor[HTML]{BF504D}- &
  \cellcolor[HTML]{D8E3BB}+ \\ \midrule
\multicolumn{1}{l|}{Singh et al. \cite{singh2020}, 2020} &
  \cellcolor[HTML]{D8E3BB}+ &
  \cellcolor[HTML]{BF504D}- &
  \cellcolor[HTML]{BF504D}- &
  \cellcolor[HTML]{BF504D}- &
  \cellcolor[HTML]{BF504D}- &
  \cellcolor[HTML]{BF504D}- &
  \cellcolor[HTML]{D8E3BB}+ &
  \cellcolor[HTML]{D8E3BB}+ \\ \midrule
\multicolumn{1}{l|}{Kannengießer et al., \cite{niclas2020}, 2020} &
  \cellcolor[HTML]{D8E3BB}+ &
  \cellcolor[HTML]{F79545}$\pm$ &
  \multicolumn{1}{c|}{\cellcolor[HTML]{F79545}$\pm$} &
  \cellcolor[HTML]{BF504D}- &
  \cellcolor[HTML]{BF504D}- &
  \cellcolor[HTML]{F79545}$\pm$ &
  \cellcolor[HTML]{BF504D}- &
  \cellcolor[HTML]{BF504D}- \\ \midrule
\multicolumn{1}{l|}{Bishnoi et al. \cite{Bishnoi2020}, 2020} &
  \cellcolor[HTML]{D8E3BB}+ &
  \cellcolor[HTML]{F79545}$\pm$ &
  \cellcolor[HTML]{F79545}$\pm$ &
  \cellcolor[HTML]{BF504D}- &
  \cellcolor[HTML]{BF504D}- &
  \cellcolor[HTML]{BF504D}- &
  \cellcolor[HTML]{BF504D}- &
  \cellcolor[HTML]{BF504D}- \\ \midrule
\multicolumn{1}{l|}{\emph{this survey}} &
  \cellcolor[HTML]{D8E3BB}+ &
  \cellcolor[HTML]{D8E3BB}+ &
  \multicolumn{1}{c|}{\cellcolor[HTML]{D8E3BB}+} &
  \cellcolor[HTML]{D8E3BB}+ &
  \cellcolor[HTML]{D8E3BB}+ &
  \cellcolor[HTML]{D8E3BB}+ &
  \cellcolor[HTML]{D8E3BB}+ &
  \cellcolor[HTML]{D8E3BB}+ \\ \midrule
 &
  \multicolumn{1}{l}{} &
  \multicolumn{1}{l}{} &
  \multicolumn{1}{l}{} &
  \multicolumn{1}{l}{} &
  \multicolumn{1}{l}{} &
  \multicolumn{1}{l}{} &
  \multicolumn{1}{l}{} &
  \multicolumn{1}{l}{} \\ \bottomrule
\end{tabular}
\caption{Comparison of related literature reviews: PC (Public Connectors), Blockchain of Blockchains (BoB), HC (Hybrid Connectors), AR (architectures for blockchain interoperability), ST (standards), CC (cross-comparison), UC (use cases), OI (open-issues). Each criterion can be ``fulfilled'' (``+'' in green background), ``partially fulfilled'' (``$\pm$'' in orange background) or ``not fulfilled'' (``-`'' in red background), if it addresses all, between one and all, or none of its sub-criteria, respectively. }
\label{tab:related_slr}
\end{wraptable}

We aim at providing a solid, throughout and comprehensive foundation on which researchers can rely upon as a starting point in this field, including a description of the related surveys, which illuminated our research. In contrast to most of the works mentioned above, this paper provides a holistic view of blockchain interoperability by focusing not only on public connectors but also on BoBs and hybrid connectors. By including updated grey literature and focusing on private blockchain interoperability, a comprehensive discussion on standards, use cases, and architecture for interoperability was possible.

\section{Blockchain Interoperability Framework}
\label{sec:bif}
This section presents the Blockchain Interoperability Framework (BIF), a framework classifying solutions collected through our methodology. To drive criteria for assessing the categories (and specific solutions) of blockchain interoperability, we analyzed the solution space using the six ``W'' questions: Who, What, Where, When, Why, and How. The ``Why'' was determined irrelevant to our analysis because its purpose is constant -- connecting different chains (CC-Txs), different blockchains (CB-Txs), or even to arbitrary systems (e.g., enterprise legacy systems). This is instead addressed by the ``where'' question.

\subsection{Deriving Evaluation Criteria}

The ``what'' refers to the \emph{assets} exchanged. An interoperability solution can handle different data objects or assets. Hence it is important to know which data representations a solution supports \cite{wegner96}. Assets can be treated as data (arbitrary payloads), as fungible assets, or non-fungible assets \cite{barnes2020,pilai2020,hyperledger_cactus}. Arbitrary data is often represented via a key-value pair, being the preferred representation of some blockchains \cite{sawtooth,fabric,vukolic_bcpitw}. The key-value is also useful to represent the contents of account-based blockchains \cite{eth2_wiki,algorand,quorum}. Payment tokens are 
fungible tokens \cite{Pillai2019}. 
Utility tokens include tokens used to access a service or application, such as non-fungible tokens (e.g., ERC20 tokens). Finally, asset tokens represent real-world physical or digital instruments, such as blockchain-based promissory notes, regulated by the {Swiss Financial Market Supervisory Authority} \cite{draft-sardon-blockchain-gateways-usecases-00} (see more details in Section \ref{sec:use_cases}), or bonds \cite{barnes2020}. An asset has different maturity levels. In particular, an asset may be standardized, (e.g., ERC tokens\cite{erc20}, standardized schema for utility tokens, ERC1400, a security token \cite{erc1400s,erc1400}) and/or {regulated} \cite{oecd2021,finma,usexchcomm}. Regulated digital assets are backed by legal frameworks. We consider all asset tokens to be regulated. We envision utility tokens as standardized and asset tokens as standardized and regulated (i.e., asset tokens are emitted by legal entities).

The ``who'' question refers to whom controls the CC-Tx process and thus accounts for trust establishment \cite{sidechains_pos,sok_cdl}). It can be the end-user (e.g., \cite{hyperledger_cactus,Frauenthaler2019}), a consortium (e.g., \cite{peggedsidechains, Schulte2019}), or a trusted third party (e.g., cloud services, centralized notary schemes). Some solutions allow different levels of control.

The ``where'' refers to what are the source and target ledgers, as well as what is the support of conducting the CC process. Solutions can support public blockchains (P) or non-public blockchains (NP). We use NP to designate private blockchains, other decentralized ledger technology (DLT) systems, and centralized systems (e.g., VISA payment network).  The supported systems of each solution matter since communication may happen unidirectionally or bi-directionally \cite{hyperledger_cactus}. Blockchain oracles apart, it often is not feasible to have a solution based on a blockchain system connected to a centralized system (e.g., providing insurance data). A smart contract may be the one conducting an asset transfer (on-chain channel, with on-chain CC-Tx validation) versus an off-chain settlement, e.g., techniques using commitment schemes \cite{abebe2020,zendo}, or via (semi-)centralized system (off-chain channel). Typically, on-chain channels offer more resiliency, but off-chain channels are more scalable. Combinations between off-chain and on-chain channels also exist (e.g., payment networks \cite{LN}). Offline channels depend on different proof generation mechanisms \cite{abebe2020, sidechains_pos, zendo}.

The ``when'' refers to the set of processes (e.g., executing CC-Txs) that are defined at \emph{design-time} or \emph{run-time}. \emph{Design-time customization} decisions affect the punctual behavior of a CC-dApp concerning when it is executed. At design-time, a user defines the behavior of the solution \emph{apriori}. If a change is needed, a new instance of the solution needs to be deployed. Conversely, \emph{run-time customization} decisions are flexible, allowing the end-user to adjust the conditions defined by business logic as needed. Solutions in which business logic is changed at run-time are called \emph{flexible approaches}, allowing to adjust business logic and conditions that trigger the execution of a CB-Tx or CC-Tx by a CC-dApp. Most literature reviews focus on design-time approaches and public blockchains, leaving a vast range of recent solutions out of scope. In this survey, we also consider private-private and public-private blockchain interoperability, focusing on flexible approaches.

The ``how'' regards the realization of cross-chain transactions: how are CC-Txs realized on the underlying DLTs? Often, these transactions can be performed using \emph{cross claims}, i.e., by locking/burning an asset on the source blockchain and unlocking/creating its representation on the target blockchain. Cross-claims require two nodes from different blockchains, where one performs one operation in a source blockchain in exchange for its counterparty performing other operations on a target blockchain - each party logs the operation in case a dispute is needed. Typically, cross-claims operate in semi-trusted environments (e.g., private blockchain, regulated blockchain), and can be operated via a (semi) trusted third party \cite{hyperledger_cactus,odap_draft_01,crp_draft_00}. Escrowed cross-claims are the standard mechanism for asset transfers, operating similarly to cross-claims, but in an untrusted environment, leveraging dispute-resolution mechanisms (e.g., via smart contracts requiring inclusion proofs \cite{abebe2020}) or by parties holding custody of assets and collateral, \cite{xclaim,dextt, plasma_vitalik}. Inclusion proofs include applying Merkle tree proofs to block header transfer via a coordinating blockchain, block header transfer, or direct signing \cite{robinson2020}. Collateralization is the process in which a party performing the transfer of assets provides a certain amount of their assets as a guarantee of following the protocol (e.g., not to steal assets from the end-user). If a party misbehaves (e.g., steals assets), the deposit is given to the victim party. Finally, a mediated CC-Tx includes (an offline) trusted party \cite{hyperledger_cactus}. In case of a dispute about an asset transfer between a public blockchain and a private blockchain (P-NP) or a public blockchain and an enterprise system (also P-NP), there needs to be a dispute-resolution mechanism. This is due to NP systems' private nature, although several mechanisms exist to prove internal state belonging to private blockchains. Hence, CC-Txs have a trade-off risk-performance: the less centralization there is on the CC-Tx settlement, the worst the performance, but the lesser the risk.

The ``how'' also relates to the extent to which the implementation of the solution is tested. Solutions might be implemented, tested, and validated (application to a real-world scenario). Testing regards \emph{correctness guarantees}: \emph{behavioral correctness} or \emph{formal correctness}. Behavioral correctness is the ability to guarantee that CC-Txs are issued as intended, without unintended consequences (e.g., asset lock, asset theft). While in practice, behavioral correctness depends on formal correctness, we say a solution has behavioral correctness if it has a suite of test cases \cite{testing}. Formal correctness assures that an algorithm is correct with respect to a specification. Formal verification checks the correctness of algorithms against a specification using, for instance, formal methods. Smart contract verification tools allow developers to reduce the probability of creating bugs, thus incurring penalties, as smart contracts are generally difficult to update once deployed \cite{smartbugs}. Another point of providing trust to the user is the solution to have an open-source implementation, where the code can be peer-reviewed and corrected if needed.

\subsection{Evaluation Criteria}
\label{subsec:criteria_bif}

Having discussed the survey's scope, we next define the set of criteria we use to characterize the interoperability solutions. Similarly to Section \ref{sec:related_literature_reviews}, each criterion can be ``fulfilled'' ``partially fulfilled'' or ``not fulfilled''. If a criterion is a yes/no question (e.g., does the solution support asset type ``data''?), we do not explicitly refer to the fulfillment conditions as they are evident. Next, we detail the criteria type (first-level), criteria sub-type (second level), and criteria from BIF:


\begin{itemize}\small
    \item Asset: this category refers to properties of an asset involved in a CC-Tx.
    \begin{itemize}
        \item Type: what type of assets does the solution support?
            \begin{enumerate}
            \item Data: can the solution manipulate arbitrary data?
            \item Payment tokens: can the solution manipulate cryptocurrencies? This criterion is partially fulfilled if the asset is only used as collateral or to reward a service's operational maintenance.
            \item Utility tokens: can the solution manipulate utility tokens? This criterion is partially fulfilled if the asset is used only as collateral or to reward a service's operational maintenance.
            \item Asset tokens: can the solution manipulate utility tokens? 
            
            \end{enumerate}
                    \item Infrastructure: what are the systems involved?
            \begin{enumerate}
            \item P: This criterion is fully fulfilled if more than two public blockchains are supported. It is partially fulfilled if one or two public blockchains are supported.
            \item NP: This criterion is fully fulfilled if more than two non-public blockchains are supported. It is partially fulfilled if one or two non-public blockchains are supported.
            \end{enumerate}

    \end{itemize}
    
    \item Trust Establishment: this category refers to how a solution provides trust to the users.
    
    \begin{itemize}

        \item Decentralization: who operates the solution instance?
            \begin{enumerate}
            \item End-user
            \item Consortium
            \item Trusted (third) party
            \end{enumerate}
        If multiple criteria are selected, it indicates a solution supports more than one mode of operation.

        \item Channel: where are CC-Tx validated?
            \begin{enumerate}
            \item On-chain: This criteria is partially fulfilled if proofs are created on-chain but validation occurs off-chain.
            \item Off-chain: This criteria is partially fulfilled if proofs are created off-chain but validation occurs on-chain.
            \end{enumerate}
    \end{itemize}    
    
    \item CC-Tx Realization: this category refers to how and where a CC-Tx is settled.
    \begin{itemize}
        \item Mechanism: how are CC-Txs agreed-upon multiple parties?
            \begin{enumerate}
            \item Cross-claim
            \item Escrowed cross-claim
            \item Mediated
            \end{enumerate}

    \end{itemize}
    
    \item Extra-functional: this category refers to the design of the solution itself.
    \begin{enumerate}
    
    \item Tests: the approach provides a set of test cases.
    
    \item Implementation: the approach provides an open-source implementation and is validated in the industry. This criterion is partially fulfilled if the implementation is closed-source. 
    
    \item Validation: the approach is validated in an actual use case scenario. 
    \item Run-time: the business logic of the solution can be changed dynamically, as needed. This criterion is considered not fulfilled if logic is settled when the solution is instantiated, i.e., changing logic requires a new instance.
    \end{enumerate}

\end{itemize}

\section{Overview of Blockchain Interoperability Approaches}
\label{sec:solutions}
We conducted a systematic literature review following the protocol described in Appendix A, yielding 80 relevant documents out of the initial 330. By grouping the publications and grey literature, a pattern arises: these works are either about interoperability across public blockchains holding cryptocurrencies, application-specific blockchain generators with interoperability capabilities, or protocols connecting heterogeneous blockchains. We thus classify each study into one of the following categories: \emph{Public Connectors} (Section \ref{sec_crypto}), \emph{Blockchain of Blockchains} (Section \ref{sec:be}), and \emph{Hybrid Connectors} (Section \ref{subsec:blockchain_connectors}). Each category is further divided into sub-categories. Table \ref{tab:existing_sols} summarizes the work conducted.

\begin{table}[]
\centering
\resizebox{\textwidth}{!}{%
\begin{tabular}{@{}lcccccccccccccl@{}}
\toprule
 &
  \multicolumn{5}{c}{\textbf{Asset}} &
  \multicolumn{8}{c}{\textbf{Trust Establishment}} &
   \\ \midrule
\multicolumn{1}{l|}{} &
  \multicolumn{3}{c|}{\textbf{Type}} &
  \multicolumn{2}{c|}{\textbf{Infra.}} &
  \multicolumn{3}{c|}{\textbf{Decentral.}} &
  \multicolumn{2}{c|}{\textbf{Channel}} &
  \multicolumn{3}{c|}{\textbf{CC-Realization}} &
   \\ \midrule
\multicolumn{1}{l|}{\textbf{Sub-Category}} &
  \multicolumn{1}{c|}{D} &
  \multicolumn{1}{c|}{P} &
  \multicolumn{1}{c|}{U} &
  \multicolumn{1}{c|}{P} &
  \multicolumn{1}{c|}{NP} &
  \multicolumn{1}{c|}{U} &
  \multicolumn{1}{c|}{C} &
  \multicolumn{1}{c|}{TTP} &
  \multicolumn{1}{c|}{OC} &
  \multicolumn{1}{c|}{OF} &
  \multicolumn{1}{c|}{CC} &
  \multicolumn{1}{c|}{ECC} &
  \multicolumn{1}{c|}{M} &
  \multicolumn{1}{c}{\textbf{References}} \\ \midrule
{\color[HTML]{ADD73F} \begin{tabular}[c]{@{}l@{}}Sidechains\\ \& Relays\end{tabular}} &
  \cellcolor[HTML]{D8E3BB}+ &
  \cellcolor[HTML]{F79545}$\pm$ &
  \cellcolor[HTML]{BF504D}- &
  \cellcolor[HTML]{F79545}$\pm$ &
  \cellcolor[HTML]{BF504D}- &
  \cellcolor[HTML]{BF504D}- &
  \cellcolor[HTML]{D8E3BB}+ &
  \cellcolor[HTML]{BF504D}- &
  \cellcolor[HTML]{D8E3BB}+ &
  \cellcolor[HTML]{D8E3BB}+ &
  \cellcolor[HTML]{BF504D}- &
  \cellcolor[HTML]{D8E3BB}+ &
  \cellcolor[HTML]{BF504D}- &
  \cite{plasma,wang-cc_elec-2020} \\
{\color[HTML]{000000} } &
  \cellcolor[HTML]{D8E3BB}+ &
  \cellcolor[HTML]{F79545}$\pm$ &
  \cellcolor[HTML]{BF504D}- &
  \cellcolor[HTML]{F79545}$\pm$ &
  \cellcolor[HTML]{BF504D}- &
  \cellcolor[HTML]{D8E3BB}+ &
  \cellcolor[HTML]{D8E3BB}+ &
  \cellcolor[HTML]{BF504D}- &
  \cellcolor[HTML]{D8E3BB}+ &
  \cellcolor[HTML]{BF504D}- &
  \cellcolor[HTML]{BF504D}- &
  \cellcolor[HTML]{D8E3BB}+ &
  \cellcolor[HTML]{BF504D}- &
  \cite{btcrelay, peacerelay, testimonium,waterloo2019,Frauenthaler2020} \\
{\color[HTML]{000000} } &
  \cellcolor[HTML]{BF504D}- &
  \cellcolor[HTML]{D8E3BB}+ &
  \cellcolor[HTML]{D8E3BB}+ &
  \cellcolor[HTML]{D8E3BB}+ &
  \cellcolor[HTML]{BF504D}- &
  \cellcolor[HTML]{D8E3BB}+ &
  \cellcolor[HTML]{D8E3BB}+ &
  \cellcolor[HTML]{BF504D}- &
  \cellcolor[HTML]{D8E3BB}+ &
  \cellcolor[HTML]{BF504D}- &
  \cellcolor[HTML]{BF504D}- &
  \cellcolor[HTML]{D8E3BB}+ &
  \cellcolor[HTML]{BF504D}- &
  \cite{POA,blockCollider} \\
{\color[HTML]{000000} } &
  \cellcolor[HTML]{BF504D}- &
  \cellcolor[HTML]{D8E3BB}+ &
  \cellcolor[HTML]{D8E3BB}+ &
  \cellcolor[HTML]{F79545}$\pm$ &
  \cellcolor[HTML]{BF504D}- &
  \cellcolor[HTML]{D8E3BB}+ &
  \cellcolor[HTML]{D8E3BB}+ &
  \cellcolor[HTML]{BF504D}- &
  \cellcolor[HTML]{D8E3BB}+ &
  \cellcolor[HTML]{D8E3BB}+ &
  \cellcolor[HTML]{BF504D}- &
  \cellcolor[HTML]{D8E3BB}+ &
  \cellcolor[HTML]{BF504D}- &
  \cite{peggedsidechains,loom,liquid,blockstream,nocust} \\
{\color[HTML]{000000} } &
  \cellcolor[HTML]{D8E3BB}+ &
  \cellcolor[HTML]{D8E3BB}+ &
  \cellcolor[HTML]{BF504D}- &
  \cellcolor[HTML]{F79545}$\pm$ &
  \cellcolor[HTML]{BF504D}- &
  \cellcolor[HTML]{BF504D}- &
  \cellcolor[HTML]{D8E3BB}+ &
  \cellcolor[HTML]{BF504D}- &
  \cellcolor[HTML]{D8E3BB}+ &
  \cellcolor[HTML]{BF504D}- &
  \cellcolor[HTML]{BF504D}- &
  \cellcolor[HTML]{D8E3BB}+ &
  \cellcolor[HTML]{BF504D}- &
  \cite{blocknet,rsk,zendo,horizon2021} \\
 &
  \cellcolor[HTML]{BF504D}- &
  \cellcolor[HTML]{D8E3BB}+ &
  \cellcolor[HTML]{D8E3BB}+ &
  \cellcolor[HTML]{F79545}$\pm$ &
  \cellcolor[HTML]{BF504D}- &
  \cellcolor[HTML]{BF504D}- &
  \cellcolor[HTML]{D8E3BB}+ &
  \cellcolor[HTML]{BF504D}- &
  \cellcolor[HTML]{D8E3BB}+ &
  \cellcolor[HTML]{F79545}$\pm$ &
  \cellcolor[HTML]{BF504D}- &
  \cellcolor[HTML]{D8E3BB}+ &
  \cellcolor[HTML]{BF504D}- &
  \cite{Shlomovits2020,cctxnarges,belotti2020,Robinson2019,Deshpande2020,gugger-bmcc-2020} \\
{\color[HTML]{000000} } &
  \cellcolor[HTML]{BF504D}- &
  \cellcolor[HTML]{D8E3BB}+ &
  \cellcolor[HTML]{BF504D}- &
  \cellcolor[HTML]{D8E3BB}+ &
  \cellcolor[HTML]{BF504D}- &
  \cellcolor[HTML]{D8E3BB}+ &
  \cellcolor[HTML]{D8E3BB}+ &
  \cellcolor[HTML]{D8E3BB}+ &
  \cellcolor[HTML]{BF504D}- &
  \cellcolor[HTML]{D8E3BB}+ &
  \cellcolor[HTML]{D8E3BB}+ &
  \cellcolor[HTML]{BF504D}- &
  \cellcolor[HTML]{D8E3BB}+ &
  \cite{ILPv4,quilt} \\ \midrule
{\color[HTML]{ADD73F} } &
  \cellcolor[HTML]{BF504D}- &
  \cellcolor[HTML]{D8E3BB}+ &
  \cellcolor[HTML]{D8E3BB}+ &
  \cellcolor[HTML]{D8E3BB}+ &
  \cellcolor[HTML]{BF504D}- &
  \cellcolor[HTML]{BF504D}- &
  \cellcolor[HTML]{BF504D}- &
  \cellcolor[HTML]{D8E3BB}+ &
  \cellcolor[HTML]{F79545}$\pm$ &
  \cellcolor[HTML]{BF504D}- &
  \cellcolor[HTML]{BF504D}- &
  \cellcolor[HTML]{BF504D}- &
  \cellcolor[HTML]{D8E3BB}+ &
  See Section \ref{sec:crypto_notaries} \\
\multirow{-2}{*}{{\color[HTML]{ADD73F} \begin{tabular}[c]{@{}l@{}}Notary\\ Scheme\end{tabular}}} &
  \cellcolor[HTML]{BF504D}- &
  \cellcolor[HTML]{D8E3BB}+ &
  \cellcolor[HTML]{D8E3BB}+ &
  \cellcolor[HTML]{D8E3BB}+ &
  \cellcolor[HTML]{BF504D}- &
  \cellcolor[HTML]{D8E3BB}+ &
  \cellcolor[HTML]{D8E3BB}+ &
  \cellcolor[HTML]{BF504D}- &
  \cellcolor[HTML]{D8E3BB}+ &
  \cellcolor[HTML]{BF504D}- &
  \cellcolor[HTML]{BF504D}- &
  \cellcolor[HTML]{D8E3BB}+ &
  \cellcolor[HTML]{BF504D}- &
  \cite{0x,uniswaptokrex,Tian2020} \\ \midrule
{\color[HTML]{ADD73F} HLTC} &
  \cellcolor[HTML]{BF504D}- &
  \cellcolor[HTML]{D8E3BB}+ &
  \cellcolor[HTML]{D8E3BB}+ &
  \cellcolor[HTML]{F79545}$\pm$ &
  \cellcolor[HTML]{BF504D}- &
  \cellcolor[HTML]{BF504D}- &
  \cellcolor[HTML]{D8E3BB}+ &
  \cellcolor[HTML]{BF504D}- &
  \cellcolor[HTML]{D8E3BB}+ &
  \cellcolor[HTML]{BF504D}- &
  \cellcolor[HTML]{BF504D}- &
  \cellcolor[HTML]{D8E3BB}+ &
  \cellcolor[HTML]{BF504D}- &
  \cite{dextt,xclaim,Lu2017,comit,fusion,Dai-cctm-2020,Rueegger-ccas-2020} \\ \midrule
{\color[HTML]{F79545} \begin{tabular}[c]{@{}l@{}}Blockchain\\ of Blockchains\end{tabular}} &
  \cellcolor[HTML]{D8E3BB}+ &
  \cellcolor[HTML]{D8E3BB}+ &
  \cellcolor[HTML]{D8E3BB}+ &
  \cellcolor[HTML]{F79545}$\pm$ &
  \cellcolor[HTML]{BF504D}- &
  \cellcolor[HTML]{D8E3BB}+ &
  \cellcolor[HTML]{D8E3BB}+ &
  \cellcolor[HTML]{BF504D}- &
  \cellcolor[HTML]{D8E3BB}+ &
  \cellcolor[HTML]{BF504D}- &
  \cellcolor[HTML]{BF504D}- &
  \cellcolor[HTML]{D8E3BB}+ &
  \cellcolor[HTML]{BF504D}- &
  \cite{Kwon2016,Wood2017,komodo} \\
{\color[HTML]{000000} } &
  \cellcolor[HTML]{D8E3BB}+ &
  \cellcolor[HTML]{D8E3BB}+ &
  \cellcolor[HTML]{D8E3BB}+ &
  \cellcolor[HTML]{D8E3BB}+ &
  \cellcolor[HTML]{D8E3BB}+ &
  \cellcolor[HTML]{BF504D}- &
  \cellcolor[HTML]{D8E3BB}+ &
  \cellcolor[HTML]{D8E3BB}+ &
  \cellcolor[HTML]{D8E3BB}+ &
  \cellcolor[HTML]{BF504D}- &
  \cellcolor[HTML]{BF504D}- &
  \cellcolor[HTML]{D8E3BB}+ &
  \cellcolor[HTML]{D8E3BB}+ &
  \cite{ark2019,aion2017,overledger03} \\ \midrule
{\color[HTML]{BF504D} \begin{tabular}[c]{@{}l@{}}Trusted \\ Relays\end{tabular}} &
  \cellcolor[HTML]{D8E3BB}+ &
  \cellcolor[HTML]{BF504D}- &
  \cellcolor[HTML]{BF504D}- &
  \cellcolor[HTML]{F79545}$\pm$ &
  \cellcolor[HTML]{F79545}$\pm$ &
  \cellcolor[HTML]{D8E3BB}+ &
  \cellcolor[HTML]{BF504D}- &
  \cellcolor[HTML]{BF504D}- &
  \cellcolor[HTML]{BF504D}- &
  \cellcolor[HTML]{D8E3BB}+ &
  \cellcolor[HTML]{BF504D}- &
  \cellcolor[HTML]{BF504D}- &
  \cellcolor[HTML]{D8E3BB}+ &
  \cite{nissl2020,falazi2020,Kan2018,crossfabric2020} \\
{\color[HTML]{BF504D} } &
  \cellcolor[HTML]{D8E3BB}+ &
  \cellcolor[HTML]{D8E3BB}+ &
  \cellcolor[HTML]{D8E3BB}+ &
  \cellcolor[HTML]{F79545}$\pm$ &
  \cellcolor[HTML]{D8E3BB}+ &
  \cellcolor[HTML]{D8E3BB}+ &
  \cellcolor[HTML]{D8E3BB}+ &
  \cellcolor[HTML]{BF504D}- &
  \cellcolor[HTML]{D8E3BB}+ &
  \cellcolor[HTML]{F79545}$\pm$ &
  \cellcolor[HTML]{D8E3BB}+ &
  \cellcolor[HTML]{BF504D}- &
  \cellcolor[HTML]{D8E3BB}+ &
  \cite{Hardjono2019,hermes-middleware-2021,odap_draft_01,crp_draft_00,zhao2020,wang2020,xiao2020} \\ \midrule
{\color[HTML]{BF504D} B. Agnostic} &
  \cellcolor[HTML]{D8E3BB}+ &
  \cellcolor[HTML]{D8E3BB}+ &
  \cellcolor[HTML]{D8E3BB}+ &
  \cellcolor[HTML]{D8E3BB}+ &
  \cellcolor[HTML]{D8E3BB}+ &
  \cellcolor[HTML]{D8E3BB}+ &
  \cellcolor[HTML]{D8E3BB}+ &
  \cellcolor[HTML]{BF504D}- &
  \cellcolor[HTML]{BF504D}- &
  \cellcolor[HTML]{D8E3BB}+ &
  \cellcolor[HTML]{D8E3BB}+ &
  \cellcolor[HTML]{D8E3BB}+ &
  \cellcolor[HTML]{BF504D}- &
  \cite{hyperledger_cactus,abebe2020,Abebe2019,acctp2020} \\
{\color[HTML]{BF504D} Protocols} &
  \cellcolor[HTML]{D8E3BB}+ &
  \cellcolor[HTML]{D8E3BB}+ &
  \cellcolor[HTML]{D8E3BB}+ &
  \cellcolor[HTML]{F79545}$\pm$ &
  \cellcolor[HTML]{F79545}$\pm$ &
  \cellcolor[HTML]{D8E3BB}+ &
  \cellcolor[HTML]{D8E3BB}+ &
  \cellcolor[HTML]{BF504D}- &
  \cellcolor[HTML]{D8E3BB}+ &
  \cellcolor[HTML]{BF504D}- &
  \cellcolor[HTML]{BF504D}- &
  \cellcolor[HTML]{D8E3BB}+ &
  \cellcolor[HTML]{BF504D}- &
  \cite{hyperservice,ion2018,pilai2020,robinson2020,cantonwhitepaper,pang2020} \\ \midrule
{\color[HTML]{BF504D} } &
  \cellcolor[HTML]{D8E3BB}+ &
  \cellcolor[HTML]{BF504D}- &
  \cellcolor[HTML]{BF504D}- &
  \cellcolor[HTML]{F79545}$\pm$ &
  \cellcolor[HTML]{BF504D}- &
  \cellcolor[HTML]{D8E3BB}+ &
  \cellcolor[HTML]{BF504D}- &
  \cellcolor[HTML]{BF504D}- &
  \cellcolor[HTML]{BF504D}- &
  \cellcolor[HTML]{D8E3BB}+ &
  \cellcolor[HTML]{9B9B9B}N/A &
  \cellcolor[HTML]{9B9B9B}N/A &
  \cellcolor[HTML]{9B9B9B}N/A &
  \cite{Frauenthaler2019,scheid2019,Westerkamp-scmob-2019} \\
\multirow{-2}{*}{{\color[HTML]{BF504D} \begin{tabular}[c]{@{}l@{}}Blockchain\\ Migrators\end{tabular}}} &
  \cellcolor[HTML]{D8E3BB}+ &
  \cellcolor[HTML]{D8E3BB}+ &
  \cellcolor[HTML]{D8E3BB}+ &
  \cellcolor[HTML]{F79545}$\pm$ &
  \cellcolor[HTML]{F79545}$\pm$ &
  \cellcolor[HTML]{D8E3BB}+ &
  \cellcolor[HTML]{D8E3BB}+ &
  \cellcolor[HTML]{BF504D}- &
  \cellcolor[HTML]{D8E3BB}+ &
  \cellcolor[HTML]{BF504D}- &
  \cellcolor[HTML]{BF504D}- &
  \cellcolor[HTML]{D8E3BB}+ &
  \cellcolor[HTML]{BF504D}- &
  \cite{scotm} \\ \bottomrule
\end{tabular}%
}
\caption{ Evaluation of blockchain interoperability solutions by subcategory accordingly to the Blockchain Interoperability Framework. N/A stands for not applicable. Public connectors are represented in green, Blockchain of blockchains in orange, and Hybrid connectors in red. }
\label{tab:existing_sols}
\end{table}

\subsection{Public Connectors}
\label{sec_crypto}

The first family of blockchain interoperability solutions aimed to provide interoperability between cryptocurrency systems, as stated by Vitalik \cite{vitalik2016}. This category identifies and defines different chain interoperability strategies across public blockchains supporting cryptocurrencies, including sidechain approaches, notary schemes, and hash time hash-locks. Some solutions share characteristics of more than one sub-category, and thus they can be considered hybrid. We introduce each sub-category, presenting only two illustrative examples of each one for the sake of space. Appendix \ref{a:crypto} depicts a complete list of Public Connectors approaches. After that, a summarized evaluation table is presented using the BIF. These tables are later discussed in Section \ref{sec:disc_crypto}.

\subsubsection{\underline{Sidechains \& Relays}}
\label{subsec:crypto_side}

A \emph{sidechain} (or \emph{secondary chain}, or \emph{satellite chain}, or \emph{child chain}) is a mechanism for two existing  blockchains to interoperate \cite{peggedsidechains, sidechains_pos},  scale (e.g., via blockchain sharding \cite{omniledger}), and be upgraded \cite{velvet} in which one blockchain (\emph{main chain} or mainchain) considers another blockchain as an extension of itself (the sidechain). The mainchain maintains a ledger of assets and is connected to the sidechain, a separate system attached to the main chain via a cross-chain communication protocol \cite{zendo}. An example is a \emph{two-way peg}, a mechanism for transferring assets between the main chain and the sidechain \cite{singh2020}. Main chains can be sidechains of each other \cite{vitalik2016}, creating each chain's possible to connect to others. Sidechains are considered layer one solutions (built on top of layer 0 solutions - blockchains) to implement layer-2 solutions, such as payment channels \cite{nocust}. The second layer allows off-chain transactions between users through the exchange of messages tethered to a sidechain \cite{sok_layers}. A sidechain is then a construct that allows for offloading transactions from the mainchain, processes it, and can redirect the outcome of such processing back to the main chain. 

For instance, state channels are off-chain sidechains used to implement, for example, payment channels, by offloading transactions of the blockchain \cite{LN}. In a payment channel, participants interact, collecting cryptographically signed messages. Those messages update the current state without publishing it to the mainchain. When the payment channel is closed, the final state is published onto the main chain, where an on-chain dispute/closure phase may occur \cite{nocust}. Payment channels are appropriated for use cases requiring several transactions that can be combined in a single one.

Main chains communicate with sidechains via a CCP, often tightly coupled with the functionality of both chains. The basic components of sidechain design are the mainchain consensus protocol, the sidechain consensus protocol, and the cross-chain communication protocol \cite{zendo}. 
Sidechains allow different interactions between participating blockchains, being the most common the transfer of assets between the main chain and the sidechain (two-way peg) \cite{pow_side,singh2020}. A \emph{two-way peg} works in the following manner: a user, operating on the mainchain, sends X tokens to a custom address that locks assets. Those funds are locked on the mainchain, and a corresponding number of tokens are created on the sidechain. The user can now use the tokens on the sidechain. Eventually, the user can transfer back the tokens to the main chain, which causes assets on the sidechain to be locked or destroyed, depending on the implementation. 
There are three major types of two-way pegs: simplified payment verification, centralized two-way pegs, and federated two-way pegs. \emph{Simplified payment verification} (SPV) \cite{nakamoto2008,spvbwiki} is done by \emph{light clients}, which consist of blockchain clients that can verify transactions on the blockchain without having its entire state. The SPV client only needs the block headers; verifying that a transaction is in a block is to request a Merkle tree proof \cite{merkle} including that transaction. In particular, transactions are represented as Merkle tree leaves. Given a leaf node as a target and a path comprised of nodes and its siblings to the target, verifying a Merkle tree proof of including the target is to reconstruct a partial Merkle tree root.

A relay solution is an SPV client for a source blockchain running on a target blockchain, enabling verification of transactions \cite{testimonium}. This verification enables conditional logic to occur on a target blockchain. Since relays are between blockchains and those blockchains are using behavior from others (bidirectionally or unidirectionally), relays include the presence of sidechains. This is saying, without a sidechain, there are no relay solutions.

%
\emph{Centralized two-way pegs}, on the contrary, trust a central entity, benefiting in terms of efficiency. An example is an \emph{exchange}, an organization, typically a company, that trades cryptocurrencies on behalf of its clients. However, Exchanges are a Notary Scheme, so we defer their explanation to Section \ref{sec:crypto_notaries}. Disadvantages include a single point of failure and centralization. \emph{Federated two-way pegs} try to decentralize the previous solution. In this solution, a group is responsible for locking and unlocking funds instead of just one. Standard implementations rely on multi-signature schemes, in which a quorum of entities must sign transactions to be deemed valid by the network. Although a better option, it does not eliminate centralization.

\begin{wrapfigure}{r}{0.55\textwidth} 
    \centering
    \includegraphics[scale=0.40]{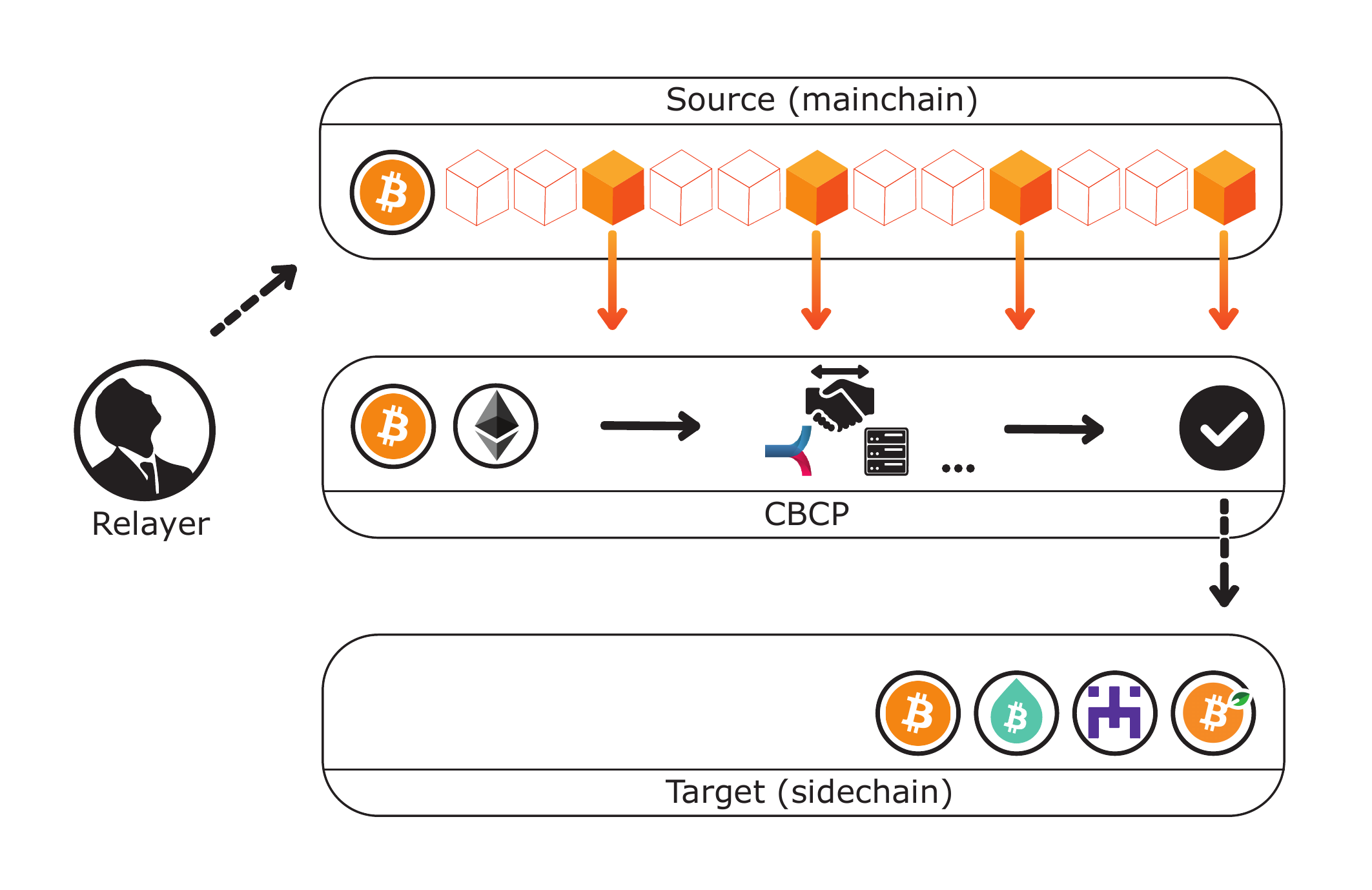}
    \caption{A general sidechain system \cite{btcrelay}}
    \label{fig:sidechains}
\end{wrapfigure}

Figure \ref{fig:sidechains} depicts a system based on the BTC Relay \cite{btcrelay}. In \emph{BTC Relay}, parties called \emph{relayers} keep track of the block headers of the main chain (the Bitcoin network in the figure), and input them to the BTC Relay smart contract, hosted on Ethereum. This procedure builds a pool of Bitcoin headers that can be used (via their stored Merkle trees) to verify on-chain information, including the presence of transactions. This way, any party can request a transaction to be verified by the smart contract that holds the headers' knowledge (via SPV). Transaction validation can be relayed to deployed Ethereum smart contracts, allowing several use cases, for example, the issuance of tokens.

\emph{Zendoo} is a cross-chain transfer protocol that realizes a decentralized, verifiable blockchain system for payments \cite{zendo}. The authors consider a parent-child relationship, where nodes from the sidechain can observe the mainchain's state, but the main chain can only observe the sidechains via cryptographically authenticated certificates. Zk-SNARKSs enable the authentication, validation, and integrity of the information provided by the sidechains via verifiable proofs \cite{zksnarks}. Such proofs are used to generate certificate proofs for the mainchain, enabling a secure verification scheme.

\subsubsection{\underline{Notary Schemes}}
\label{sec:crypto_notaries}
A notary scheme involves a \emph{notary} that is an entity that monitors multiple chains, triggering transactions in a chain upon an event (e.g., a smart contract is deployed) taking place on another chain \cite{vitalik2016}. Notary schemes are, in practice, instantiated as centralized exchanges (EXs) or decentralized exchanges (DEXs). The most popular centralized exchanges, by volume, as of the 8th of February 2021 are Binance\footnote{https://www.binance.com/en}, Coinbase\footnote{https://www.coinbase.com/}, and Huobi Global\footnote{https://www.huobi.com/}. Exchanges facilitate signaling between market participants by managing an order book and matching buyers and sellers. 
If the trust anchor is put on a centralized party, where it holds users' private keys or funds, the notary is a centralized exchange. Otherwise, if exchanges do not execute the trades on behalf of the users, only providing a matching service, they are considered decentralized exchanges. We present the protocols of two decentralized exchanges: 0x \cite{0x}, and Uniswap \cite{uniswap}.

\emph{0x} implements a decentralized exchange as a set of smart contracts (called automated market makers), replacing an on-chain order book with a real-time price-adjustment model. 0x uses a hybrid implementation, ``off-chain order relay with on-chain settlement'', combining the idea of a state channel with settlement smart contracts. Two parties participate: \emph{makers} and \emph{takers}. Makers place orders on the exchange, providing liquidity for the network (a set of decentralized exchanges), while takers place orders matched with the makers' orders. 0x uses the ZRX token and the Ethereum blockchain to incentivize users to host and maintain order books (provide liquidity). In exchange, 0x makers choose the rewards they obtain for each trade - although they have to comply with the DEX policies under the possibility of the order not being disseminated. This approach relies on a smart contract set (smart contract) and several smart contracts representing the different tokens supported. First, a maker creates an order to exchange token A for B, at a given rate, right after it approves a DEX to access its balance of token A. A taker discovers this order and wishes to trade its tokens B for tokens A. The taker grants permission to the DEX to access its tokens, and the DEX performs the exchange after several validations (e.g., the order has not expired, and it has not been filled).

\emph{Uniswap} is a set of smart contracts implementing an automated liquidity pool, serving as a decentralized exchange \cite{uniswap}. Each Uniswap pool provides liquidity for two assets based on the constant set as the reserves' product. Prices for each asset are provided by an on-chain price oracle smart contract. Uniswap can support ERC-20 to ERC-20 trades and even flash loans, a theme explored in the decentralized finance area. A flash loan is a type of loan that does not require collateral, as the debt is repaid within the transaction. Flash loans work because the borrowed asset to be paid within the transaction requesting it \cite{uniswap}.

\subsubsection{\underline{Hashed Time-Lock Contracts}}
\label{subsec:crypto_hashed}

Hashed time-locks contracts (HTLCs) 
initially appeared as an alternative to centralized exchanges, as they enable cross-chain atomic operations \cite{htl}. HTLCs techniques use hashlocks \cite{hashlock} and timelocks \cite{timelock} to enforce atomicity of operations, normally between two parties. A trader commits to make the transaction by providing a cryptographic proof before a timeout to the other. This scheme allows for the creation of multiple outputs (such as multiple payments), depending on solely one hashlock. HTLCs are used in Bitcoin for conditional payments, or cross-chain payments (Bitcoin-Ethereum), i.e.,   \emph{atomic swaps}  \cite{atomicswaps,Black2019,accs2015}. Atomic swaps can be thought as a form of distributed commitment resilient to Byzantine adversaries. Thus, an atomic cross-chain swap is a distributed atomic transaction \cite{Herlihy2018}, settled on-chain.

Several projects implement HTLCs differently, providing different correctness guarantees. However, the general algorithm is quite similar in most of the solutions. Let us consider an HTLC-supported atomic swap between Alice (holding assets of type $a$ in blockchain $\mathcal{B}_a$) and Bob (holding assets of type $b$ in blockchain $\mathcal{B}_b$). An atomic swap can be realized as follows \cite{belotti2020,zapala2020}: 1) Alice generates and hashes a secret $s$, yielding $h$. The protection of a smart contract with hash $h$ is called a hashlock because it will lock a smart contract - only parties with knowledge of secret $s$ can know it since secure hash functions are pre-image resistant (i.e., a hash function cannot be inverted). Alice also creates a timelock $t_b$, corresponding to an upper bound in which the created hashlock can be unlocked, i.e., Bob can unlock the smart contract up to $t_b$, where $t_b$ corresponds to a specified future time or block height; 2) Alice publishes the smart contract in $\mathcal{B}_a$. Bob verifies the deployment, and records $h$ and $t_b$; 3) Bob publishes a smart contract in $\mathcal{B}_b$ locking $b$ with hashlock $h$, but with timelock $ta$ such that $t_a < t_b$, i.e., Alice can claim $b$ before $t_a$. 4) Alice checks that Bob's smart contract has been published and gives as input secret $s$, before $t_a$, acquiring asset $b$. In practice, this triggers a transfer; 5) Bob now sends $s$ to Alice's smart contract in the interval $[t_a,t_b]$, acquiring $a$. Note that if Bob issues the transaction after $t_b$, Bob will not obtain access to $b$. Some solutions utilize the notion of HLTC and enhance it, providing an additional on-chain trust anchor. In particular, two solutions are presented: XCLAIM \cite{xclaim} and the \ac{LN} \cite{LN}.

\emph{XClaim} uses a combination of HLTCs, collateralization, and escrow parties, realizing non-interactive cross-chain atomic swaps \cite{xclaim}. This protocol includes several actors: the requester, the sender, the receiver, the redeemer, the backing vault, and the issuing smart contract. Requesters lock coins to issue tokens, while the redeemer burns tokens to receive coins. The sender sends tokens, while the receiver receives them. After that, the vault smart contract fulfills requests of asset backing and ensures correct redeeming. An issuing smart contract issues and exchanges representations of a token (cryptocurrency-backed assets) and enforces the vault's correct behavior. Considering a transaction between Bitcoin and Ethereum, firstly, the vault locks collateral in Ethereum smart contracts. This collateral defines the amount of CBA that the vault can issue. A user that wants to issue Bitcoin-backed tokens sends Bitcoin to the vault. User A then sends a proof of transaction submitted to the Bitcoin mainchain to a chain relay, e.g., BTC Relay. The chain relay verifies the submitted transaction and alerts the issuing smart contract. The smart contract releases the Bitcoin-backed assets to the user.
On the other hand, a user issues a transaction against the smart contract, locking/burning its backed tokens. The vault releases the Bitcoin to the user, and it submits a proof of the involved operations to the chain relay. The chain relay verifies the proof and only then releases the collateral to the vault. XClaim currently supports exchanges between Bitcoin and Ethereum\footnote{https://github.com/crossclaim/xclaim-sol}. The protocol execution consumes substantially lower Ether than traditional HTLCs.

{\ac{LN}} enables high-volume, low latency micro-payments on the Bitcoin network \cite{LN}.  LN is a payment scheme (i.e., an off-chain sidechain). LN allows several parties to open a payment channel, transact amongst them, and when all the intermediary payments are completed, the final output is sent to the mainchain. LN works as follows: 1) funds are placed into a multi-signature Bitcoin address (two-party multi-signature if only two people are transacting). In order for funds to be changed, two signatures are required. After that, the funds will be managed off-chain via commitment transactions (i.e., a commitment to pay part of the available funds to the other party); 2) Parties can now transact offline under the regime they choose; 3) To settle the payments performed off-chain, both parties sign a new exit transaction. Note that parties can unilaterally close the payment channel in case of conflict. 
LN is considered a precursor of HLTCs because its bi-directional payment channels allow payments to be routed across multiple payment channels using HLTCs.

\subsubsection{\underline{Discussion on Public Connectors}}
\label{sec:disc_crypto}

Public Connectors started emerging as early as 2015 \cite{ZyskindOz2015}, when researchers and practitioners alike saw the potential in cross-chain transactions to support, for instance, atomic swaps \cite{atomicswaps,Black2019,accs2015}, and payment channels \cite{LN}. Sidechains are the solutions increasing the main network's scalability by processing and batching large amounts of transactions before submission on the main blockchain \cite{singh2020, plasma, rsk}. Relays can fetch block headers from sidechains, enabling data verification \cite{btcrelay,peacerelay,waterloorelay}. While sidechains are mainly used on public blockchains, there are also permissioned blockchain sidechains \cite{sidechain_fabric}. We note that some sidechains may have a cross-chain mechanism realization HLTCs, being a solution belonging to multiple categories (e.g., \cite{LN}). 

Most sidechains use Ethereum and have a sidechain consensus mechanism, which is allusive to bidirectional transfers \cite{zendo}. Simple relay schemes, which verify transactions on other chains, such as BTC Relay, have a simple sidechain consensus, as the information flow is unidirectional \cite{btcrelay}. In particular, validators can sign events that happened on the source chain (if validation happens across EVM-based chains) or transfer block headers (via users or aggregation chains)  \cite{robinson2020}. Liquid \cite{liquid}, and POA \cite{POA} rely on a consortium of validators running trusted hardware to execute smart contracts and validate transactions. Other solutions, such as Wanchain \cite{wanchainroadmap} rely on a trusted consortium, but without running trusted hardware.

However, sidechains suffer from several limitations. Safe cross-chain interactions are rooted in the assumption that the main chain is secure, i.e., the network cannot be successfully attacked. Compromising the main chain would invalidate the sidechain logic. Conversely, centralization in sidechains tends to exist to a higher degree than on mainchains, because typically there is a trade-off between decentralization-performance (e.g., lesser validating nodes versus higher throughput). Consequently, if an attacker can obtain control on a (potentially small) set of validators, funds can be stolen from users.  Therefore, it is important to have different stakeholders with different incentives, diminish the likelihood of collusion, and rely on a reasonable quorum of validators to sign each transaction (e.g., 8 out of 11 versus 3 out of 10). If a sidechain has a strong security model, it may lead to a slow transaction settlement, stalling assets, and lowering liquidity. For example, the RSK sidechain \cite{rsk} takes approximately the time to confirm 100 Bitcoin blocks (around 15 hours) to convert BTC to RBTC\footnote{https://developers.rsk.co/rsk/}.  Finally, sidechains typically do not allow for arbitrary code to specify conditions on the pegging mechanism, thus not empowering them to develop more complex applications.

Notaries on the Public Connectors category are cryptocurrency exchanges. 
EXs have the majority of the market share, comparatively to DEXs. While EXs provide services to the end-user, decentralized exchanges tend to provide better exchange fees and security. The trade-off is, therefore, comfort and speed - security. This subcategory provides great flexibility at run-time because EXs and smart contracts that DEXs support triggers (e.g., stop-loss orders).

Notary schemes have to capture the logic of smart contracts in both chains. Although they can capture the full spectrum of interoperability -- both at the value and mechanical levels (see Section \ref{subsec:blockchain_connectors}), practical applications are limited. In summary, notary schemes are intermediaries between blockchains. EXs are notaries because they execute actions on behalf of the end-user (e.g., buy cryptocurrencies conditionally). DEXs are notaries because they provide matching for the end-users by pinning and advertising trade offers encoded in smart contracts.

The HTLCs category was the first one to allow asset exchange in a trustless way. 
HTLCs allow atomic swaps between different blockchains, funding bidirectional payment channels. HTLCs are flexible because they can be chained after each other \cite{zapala2020}, and therefore enable trades even if there is no direct connection between the trading parties. As they serve as programmable escrows, they represent the most trustless and practical approach of the three. However, hashed timelocks might lead to capital retention and unfair trade, as the trader issuing a cross-blockchain asset transfer may only provide the secret on specific conditions (exploring the spread of the cryptocurrency exchange rate) \cite{xclaim}. Many solutions are hybrid, sharing characteristics of HTLCs and sidechains, either exploring collateralization-punishment schemes rooted on smart contracts (\cite{cctxnarges,sai2019,xclaim}, or locking-in and locking-out assets \cite{dextt,fusion,metronome_faq,metronome}. 
HLTCs are practical solutions across public blockchains. HLTCs could also provide asset transfers between private blockchains, but only under the participation of a third party blockchain and a semi-trust environment \cite{hardjono2021}, or if both parties belong to both private blockchains. Current efforts to address these limitations include Hyperledger Cactus \cite{hyperledger_cactus}.

Concluding, Public Connectors are the best approach to perform cryptocurrency trades and moving fungible and non-fungible assets across public blockchains. We encourage the reader to refer to some related surveys focusing on sidechains to complement this survey (see Section \ref{sec:related_literature_reviews}).

\subsection{Blockchain of Blockchains}
\label{sec:be}

\emph{Blockchain of Blockchains} are frameworks that \emph{provide reusable data, network, consensus, incentive, and contract layers for the creation of application-specific blockchains (customized blockchains) that interoperate between each other}. We briefly present Polkadot \cite{Wood2017,polkadot_2} and Cosmos \cite{Kwon2016}, the most widely adopted Blockchain of Blockchains in terms of market capitalization\footnote{USD 22.1B and USD 3.6B respectively, as of February 2021}. A detailed comparison between Polkadot, Cosmos, and Ethereum 2.0 (the baseline) is deferred to Appendix \ref{a:be}. Other Blockchain of Blockchains include Ark \cite{ark2019}, Komodo \cite{komodo}, and AION \cite{aion2017}.

Wood proposes \emph{Polkadot}, a network that aims to connect blockchain networks \cite{Wood2017}. Polkadot provides the foundation for \emph{parachains}, i.e., ``globally-coherent dynamic data structures'' hosted side-by-side. Parachains are, thus, the parallelized chains that participate in the Polkadot network. Specialized parachains called bridges link independent chains  \cite{Wood2017}. Polkadot is based on \emph{Substrate}, a framework for creating cryptocurrencies and other decentralized systems. It guarantees cross-language support with WebAssembly, a light client, and off-chain workers, allowing for integration with other technologies.

Polkadot enables interoperability based on state transition validation, done by the chain-relay validators. Parachains communicate through the Cross-chain Message Passing Protocol (XCMP), a queuing communication mechanism based on a Merkle tree \cite{xcmp}. Communicating state transition proofs from parachain to relay chain is achieved via an erasure-coding scheme. Polkadot scales by connecting up to 100 parachains directly to the relay chain in the short-medium term. A long-term solution is being studied, where second and third-level parachains are added in parallel.

\emph{Cosmos} is a decentralized network of independent parallel blockchains, called \emph{zones} \cite{Kwon2016}. The zones are essentially Tendermint blockchains \cite{tendermint}. Zones can transfer data to other zones directly or via \emph{hubs}. Hubs minimize the number of connections between zones and avoid double spendings. For example, zone A can connect to zone B via Hub C and receive tokens from zone B. Zone A would need to trust the tokens from zone B and Hub C. This scheme allows zones to maintain a reduced number of connections.  Both ways utilize the inter blockchain communication protocol (IBC) \cite{ibc}.

IBC resembles the Internet network layer as it routes arbitrary data packets to a target blockchain. A target blockchain can know that a certain ordered packet with arbitrary data came from another blockchain. By handling transportation and order, the protocol has several steps to achieve cross-zone transactions. First, each chain involved tracks the headers of the others, acting as a light client. When a transfer is initiated, the protocol locks the assets on the origin chain. After that, the proof is sent to the target blockchain, which then represents the locked assets. A similar mechanism is used to recover the original tokens. This scheme allows for interoperability among Tendermint blockchains. Other kinds of blockchains can interoperate with a Cosmos chain via peg zones. Peg zones resemble the pegged sidechain mechanism \cite{peggedsidechains}, in which a representation of the locked token of the source blockchain is created on the target blockchain.

Cosmos abstracts the development of a blockchain into three layers: networking, consensus, and application. Tendermint BFT  realizes the networking and consensus layers. The Tendermint BFT engine is connected to the application layer by a protocol called: the Application Blockchain Interface (ABCI). The Cosmos SDK realizes the applicational layer, allowing developers to develop smart contracts in languages that can be compiled to WASM\footnote{https://blog.cosmos.network/announcing-the-launch-of-cosmwasm-cc426ab88e12}.

\subsubsection{\underline{Discussion on Blockchain of Blockchains}}
\label{sec:block_engine_disc}

Blockchain of Blockchains implementations are similar to relays and sidechains, as there is typically the main chain (often called relay chain) that connects the secondary chains, which can be application-specific blockchains. This scheme allows high throughput and flexibility to the end-users, providing interoperability capabilities between their platform instances. For example, Cosmos's Tendermint-based blockchains interoperate  (instant finality assured), while Polkadot provides interoperability on Substrate-based blockchains (for instance, via Cumulus\footnote{https://wiki.polkadot.network/docs/en/build-cumulus}, a tool for connecting a blockchain to Polkadot). To connect to other chains, Cosmos, Polkadot, AION, and utilize a mechanism similar to pegged sidechains or hashlock time contracts (ARK \cite{ark2019}) to interact with other blockchains, commonly called bridges.

\begin{table*}[]

\caption{Comparison of Blockchain Engine interoperability solutions  \cite{polkadot_comparison,Kwon2016}}
\label{tab:be}
\centering \normalsize
\resizebox{\textwidth}{!}{%
\begin{tabular}{@{}lllllllllll@{}}
\toprule
 &
  \multicolumn{2}{c}{\textbf{Communication}} &
  \multicolumn{6}{c}{\textbf{Properties}} &
  \multicolumn{2}{c}{\textbf{Community}} \\ \midrule
\multicolumn{1}{l|}{} &
  Cross-chain &
  \multicolumn{1}{l|}{Cross-blockchain} &
  Consensus &
  Security &
  Validator &
  Maximum &
  Number of &
  \multicolumn{1}{l|}{Smart} &
  \multicolumn{1}{c}{Launch} &
  \multicolumn{1}{c}{Roadmap} \\
\multicolumn{1}{l|}{} &
  Protocol &
  \multicolumn{1}{l|}{interoperability} &
  Mechanism &
  assumption &
  number &
  Throughput &
  instances &
  \multicolumn{1}{l|}{Contracts} &
   &
   \\ \midrule
\multicolumn{1}{l|}{Polkadot \cite{Wood2017} \mbox{}\hfill \checkmark} &
  XCMP &
  \multicolumn{1}{l|}{$\CIRCLE$} &
  BABE and GRANDPA &
  SM &
  197 &
  $10^{3}$ &
  200 &
  \multicolumn{1}{l|}{WASM} &
  November 2019 &
  Main network launch \\
\multicolumn{1}{l|}{Cosmos \cite{Kwon2016} \mbox{}\hfill \checkmark} &
  IBC Protocol &
  \multicolumn{1}{l|}{$\LEFTcircle$} &
  Tendermint &
  SM &
  125 &
  $10^{3}$ &
  $\textgreater$ 70 &
  \multicolumn{1}{l|}{WASM} &
  March 2019 &
  Governance updates \\
\multicolumn{1}{l|}{ARK \cite{ark2019} \mbox{}\hfill \checkmark} &
  SmartBridge &
  \multicolumn{1}{l|}{$\LEFTcircle$} &
  Delegated proof of stake &
  M &
  51 &
  18.5 &
  Unlimited &
  \multicolumn{1}{l|}{WASM$^{\ast}$} &
  May 2019 &
  ARK Swap Market \\
\multicolumn{1}{l|}{AION \cite{aion2017} \mbox{}\hfill \checkmark} &
  Interchain transactions &
  \multicolumn{1}{l|}{$\Circle$} &
  Proof of intelligence &
  M &
  $\times$ &
  $\times$ &
  $\times$ &
  \multicolumn{1}{l|}{Aion Language} &
  April 2018 &
  Market assimilation \\ \midrule
 &
   &
   &
   &
   &
   &
   &
   &
   &
   &
   \\
\multicolumn{11}{l}{\begin{tabular}[c]{@{}l@{}}\checkmark our description was endorsed by the authors/team  \\ $\times$ not known \\ $\ast$ some languages compilable to WASM, such as Go and .NET, but not all of them\end{tabular}} \\
\multicolumn{11}{l}{\begin{tabular}[c]{@{}l@{}}$\CIRCLE$ can interoperate with instances of the same blockchain engine. Interoperate with more than two heterogeneous blockchains\\ $\LEFTcircle$ can interoperate with instances of the same blockchain engine. Interoperate with up to two heterogeneous blockchains \\ $\Circle$ can interoperate with instances of the same blockchain engine\end{tabular}}
\end{tabular}%
}
\end{table*}

Table \ref{tab:be} maps out the current blockchain engine landscape by extracting and evaluating their main characteristics. Some information was not possible to obtain due to the lack of details on the whitepapers. It is possible to observe that Blockchain of Blockchains is very recent: Polkadot's test network, Kusama \cite{kusama}, was released in November 2019; Cosmos' main network was launched in March 2019. ARK launched in May 2019. AION launched in April 2018. Blockchain of Blockchains has different cross-chain communication protocol, e.g., in Polkadot, {cross-chain message passing}\footnote{https://wiki.polkadot.network/docs/en/learn-crosschain}; in Cosmos, the inter-blockchain communication protocol \cite{Kwon2016}.
Cosmos and Polkadot have some differences regarding their approach: in Cosmos, the idea is to provide blockchains tailored to specific applications. IBC is more generic than XCMP, letting users customize their zones with higher freedom: security and validation are decided per zone. Polkadot restricts this customization but offers a shared security layer, making a trade-off security-customization.

The security assumptions criteria depict the number of nodes assumed to be honest. A supermajority (SM) assumes that at least two-thirds of the nodes are honest, a common condition required by Byzantine fault-tolerant consensus algorithms ($n \textgreater$ $\frac{2}{3}$), while the majority (M) assumes at least half of the nodes are honest ($\textgreater \frac{1}{2}$). The validator number on a network comes with a trade-off: while a higher number is generally better for decentralization and robustness, it comes with an increase of latency towards block production -- and consequently lower throughput. Polkadot currently has around 297 validators, and this number is gradually increasing in the short-term to support up to 100 first-level parachains. 
At the time of writing, Polkadot is developing bridges for Bitcoin \cite{xclaim}, Tendermint, Libra \cite{libra}, and Ethereum. Interoperability between parachains is provided by Substrate.

Currently, Cosmos has 125 validators. The number of validators can rise to 300. Currently, there are around 70 zones, and ``the number is growing''. While Cosmos does hold a limit for zones (as each zone is self-sovereign), there is no limit for how many zones can be attached to a Hub. Cosmos can interoperate with Etheruem. The Cosmos SDK provides interoperability between zones. Cosmos supports multiple peg zone implementations for Bitcoin and one for Ethereum. ARK has 51 validators, which can validate the transactions of a number of blockchains bound to the company's physical resources (instances managed by ARK). ARK can send and receive ERC-20 tokens to the Ethereum blockchain.  We found no information regarding AION's validator number, throughput, or maximum sub-chains \cite{aion2017}. The theoretical throughput of the presented solutions varies: Polkadot's relay chain supports around 1000 transactions per second, considering that a block can include around 7,000 transactions at a 6-second block time (considering current weights, March 2021). Cosmos theoretical throughput can achieve up to dozens of thousands of transactions per second (tps) with two validators. With 64 validators, it falls into several thousand transactions per second. ARK can achieve around 18.5 transactions per second, relying on a proof of work consensus. The number of validators is set to 51. ARK is not a completely decentralized solution, as it manages instances of ARK blockchains. There is no theoretical limit of bridge chains, except the service provider resources. Several optimizations are being done in Cosmos, Polkadot, and ARK, to increase the throughput. The AION The project looks deprecated and stalled. As stated, the ``white paper is both ambitious and experimental'' \cite{aion2017}. AION is now a part of a larger project called the \emph{Open Application Network} (OAN).

Cosmos and Polkadot support smart contracts in languages compilable to WASM (Web Assembly), which means developers can write them in languages such as Go, C++, and JavaScript. AION would support domain-specific languages, Aion language. Blockchain of Blockchains instances achieve inter-chain interoperability by a common point of contact, the ``connector'', analogous with Hyperledger Fabric channels \cite{fabric}. The connectors are the relay chain, the Cosmos Hub, the AION-1 blockchain, and the ARK main net if the technology is Polkadot, Cosmos Network, AION, or ARK, respectively. In Polkadot, the connector provides shared security. The relay-chain (the chain that coordinates consensus and communication between parachains and external blockchains) connects parachains and parachains to bridges. In Cosmos, the connector is loosely coupled to blockchains, providing greater flexibility than Polkadot. We could not extract meaningful considerations about AION's connector. In ARK, it looks like the connector is centralized at the expense of developability and ease of use. Concerning cross-blockchain interoperability, all solutions rely on \emph{bridges} or \emph{adapters} that route transaction from a particular blockchain type to another.

While the provided features can be desirable for end-users, blockchain-engines do not interoperate with each other. In light of this fact, end-users are obligated to choose between existing solutions, leading to sub-optimal leveraging of available resources. Therefore, participant networks have constraints on interoperability, ending at relying on a single blockchain engine solution. Some authors defend that blockchain engine approaches are not universally accepted and cannot eliminate fragmentation \cite{Abebe2019}. Some solutions are even centralized, in the sense that its code is not open-source, and the end-user needs to use an SDK to access core functionalities (e.g., \cite{ark2019,aion2017}). However, ongoing work on building a Tendermint light client for GRANDPA, which would allow Polkadot to interact with Cosmos may allow blockchain engine interoperability in the short-medium term. Thus, in theory, interoperability across Blockchain of Blockchains can also be achieved via the relay chain technique (i.e., a blockchain engine can be a sidechain of other blockchain engines; validation can happen via SPV).

Moreover, Blockchain of Blockchains requires transaction fees to keep the network operating. Given enterprise blockchain systems, a question could be posed: at which point shall an organization pay fees to sustain its business model across several blockchains? While Cosmos can provide flexibility configuring a zone, on Polkadot, this can be harder. Therefore, Blockchain of Blockchains can provide an optimal leveraging for public infrastructures, but that is not necessarily the case for private blockchains.

\subsection{Hybrid Connectors}
\label{subsec:blockchain_connectors}

The \emph{Hybrid Connector} category is composed of interoperability solutions that are not Public Connectors or Blockchain of Blockchains. Directed to both public and private blockchains, Hybrid Connectors attempt at delivering a ``blockchain abstraction layer'' \cite{wef2020}, capable of exposing a set of uniform operations allowing a dApp to interact with blockchains without the need of using different APIs \cite{falazi2020}. 
We derived a set of sub-categories from the studies available: \emph{Trusted Relays}, \emph{Blockchain Agnostic Protocols} (including \emph{Blockchain of Blockchains}), and \emph{Blockchain Migrators}. Trusted relays are directed to environments where a blockchain registry facilitates the discovery of the target blockchains. Typically, such a scheme appears in a permissioned blockchain environment, where trusted escrow parties route cross-blockchain transactions. As the name suggests, Blockchain-agnostic protocols provide technology-agnostic protocols for interoperation between distributed ledger systems but do not guarantee backward compatibility. In other words, to use such protocols, their source code has to be changed to existing blockchains to use such protocols.  Solutions from the blockchain of blockchains category aim to provide mechanisms for developers to build cross-chain dApps. The blockchain migrators sub-category aggregates solutions that perform data migration across blockchains, which resemble the notary schemes discussed in Section \ref{sec:crypto_notaries} (as there is typically a centralized party mediating the migration process).

We introduce each sub-category,
presenting only one illustrative example of each for the sake of space. Appendix \ref{a:connectors} depicts a complete list of Hybrid Connectors. Evaluation tables for each sub-category are discussed in Section \ref{sec:disc_connect}.

\subsubsection{\underline{Trusted Relays}}
\label{sec:block_connector_trusted_relay}
Trusted relays are trusted parties that redirect transactions from a source blockchain to a target blockchain, allowing end-users to define arbitrary business logic. These solutions imply managing different APIs, in which cross-chain consensus may be modular.

\emph{Hyperledger Cactus} (Cactus), previously known as Blockchain Integration Framework, uses an interoperability validator network that validates cross-chain transactions, optionally using a trusted escrow party \cite{hyperledger_cactus}. However, decentralized validators are implemented as well -- making this project move towards a decentralized trusted relay. Cactus allows a party or a set of parties to issue transactions against several ledgers, similarly to some notary scheme solutions \cite{timothesis,timo2019}.
The interoperability is enabled through a set of \emph{interoperability validators}, which are participants from the source and target blockchains. Such validators collect cross-chain transaction requests, sign and deliver them. A CB-Tx is deemed valid, given that a quorum of validators signs them. It is then assumed that the blockchains participating in the network know how to address each other. However, trusted escrows can be replaced by decentralized parties.
Currently, Hyperledger Cactus\footnote{https://github.com/hyperledger/cactus}  supports Hyperledger technologies (e.g., Fabric, Besu), Corda, and Quorum. The roadmap predicts integration with public blockchains and blockchain migration capabilities.

\subsubsection{\underline{Blockchain-Agnostic Protocols}}
\label{sec:block_connector_blockagnosticprotocol}
Blockchain-agnostic protocols enable cross-blockchain or cross-chain communication between arbitrarily distributed ledger technologies by providing a blockchain abstraction layer. These solutions enable BoBs, ``a system in which a consensus protocol organizes blocks that contain a set of transactions belonging to CC-dApps, spread across multiple blockchains. Such system should provide accountability for the parties issuing transactions on the various blockchains and providing a holistic, updated view of each underlying blockchain'' (Section \ref{subsec:bid}). Typically, the cross-chain consensus is fixed, and business logic is more restricted.

The \emph{\ac{ILP}} can be considered a decentralized, peer-to-peer payment network \cite{Thomas2015}. It firstly adopted a generalized hash locking scheme to enable asset transfers, and it was directed to cryptocurrency transfers. Nowadays, ILP is technology-agnostic, defining a ``lowest unit common denominator'' across distributed ledgers, blockchains, fiat payment networks, the ILP packet.

ILP sends payment information in packets by leveraging a network of connectors, which route such packets. At the core of Interledger is the Interledger Protocol (ILPv4) \cite{ILPv4}, which defines how senders, routers (or node, or connector), and receivers interact. Typically, the connector is a money packet router. The root of trust is then the connector, which has to be trusted: companies can settle payments via the routers, given that clearance of such payments is done afterward while being protected by the law. A sender is an entity that initiates a value transfer. A router applies currency exchange and forwards packets of value. The receiver obtains the value transmitted. ILPv4 is a request/response protocol enabled by ILPv4 packets. Each packet contains transaction information, and can be divided into \emph{prepare}, \emph{fulfill}, and \emph{reject} packets. A sender node initiates an exchange of value by sending a \emph{prepare} ILPv4 packet to a receiver. When a receiver obtains the prepared packet, it sends the response back to the sender via routers. The response may be a \emph{fulfill} packet, whereby a transaction has been successfully executed, or a reject packet.

Several specifications for Interledger and related protocols are available\footnote{https://github.com/interledger/rfcs}. 
The Interledger Protocol is discussed by a W3C community group\footnote{https://www.w3.org/community/interledger/} and has a proposal that ``describes data structures and formats, and a simple processing model, to facilitate payments on the Web''\footnote{https://w3c.github.io/webpayments/proposals/interledger/}. The interledger protocol cannot integrate with existing blockchains: each one must be adapted to use ILP.
A disadvantage is that Interledger does not support the transfer of non-fungible tokens (such as  ERC-721\footnote{http://erc721.org/} tokens).



\subsubsection{\underline{Blockchain Migrators}}
\label{sec:block_connector_bm}
Blockchain migrators allow an end-user to migrate the state of a blockchain to another. Currently, it is only possible to migrate data across blockchains, although moving smart contracts is also predicted \cite{hyperledger_cactus}.

\emph{Fynn et al.} present an abstraction for smart contracts to switch to another blockchain consistently, moving the state required by the transaction to the target blockchain and execute it \cite{scotm}. The authors call such abstraction the \emph{Move} operation. The operation works as follows: first, it locks a smart contract on the source blockchain; next, the Move protocol recreates the smart contract in the target blockchain. This method allows arbitrary states to be transferred between blockchains. For example, it allows transferring cryptocurrencies by creating tokens on the target blockchain backed-up by locked tokens on the source blockchain (similarly to pegged sidechains). This method was tested on Ethereum and Hyperledger Burrow (based on Ethereum). The solution assumes the same cross-blockchain smart contracts utilize the same virtual machine, which can be limiting. Furthermore, for such a solution to be deployed, it requires Solidity changes and possibly a soft fork on Ethereum.

\subsubsection{\underline{Discussion on Hybrid Connectors}}
\label{sec:disc_connect}
This section defined the hybrid connector category and its sub-categories: trusted relays, blockchain-agnostic protocols, and blockchain migrators. 

Regarding centralization, almost all adopt a decentralized model. Permissioned blockchain solutions are less flexible, as all involved participants are identified. In particular, trusted relays endorse connections made in a peer-to-peer fashion, upon previous agreement \cite{Abebe2019,Ghaemi2021}. However, Abebe et al.'s work pose some limitations: interoperating networks require a priori knowledge of each other's identities and configurations, hence being static. A discovery service could be implemented using a blockchain registry or a pub-sub mechanism \cite{Ghaemi2021}, in which networks could be added and removed. In trusted relays, it is not completely clear the mechanisms to minimize malicious relay services, apart from replication (whereby the risk of a censorship attack is reduced but not erased).
Hyperledger Cactus could be a true enabler of interoperability, given that a (decentralized) trusted blockchain registry would be deployed, and public escrow parties could replace the overlay of trusted parties. Cactus could, therefore, make the transition between a trusted relay to a semi-trusted relay or even a trustless relay.
 
Blockchain-agnostic protocols will be better positioned to offer interoperability to existing and yet to-exist blockchains, but most do not grant backward compatibility and lack the flexibility to define business logic. This inflexibility is inherent to the provided homogeneous interfaces (containing roles, methods, data, message formats, for instance, \cite{falazi2020}); at least such solutions scale slowly, as adding methods compatible with all the supported blockchains incur in a polynomial effort. However, this category might resemble some of the trusted relay solutions. In particular, both Cactus \cite{hyperledger_cactus}, and SCIP \cite{falazi2020} rely on connectors and validators and gateways, to access the underlying blockchains.  The gateway paradigm implies a (semi) trusted gateway having read/write access to the shared ledger of the blockchain, and often they are expected to participate in the consensus mechanism of the blockchain \cite{hardjono2021}. While there is a higher trust requirement, gateway approaches might be the most suitable to solve interoperability across private blockchains if gateways are framed in a legal and regulatory framework. Proper solutions for enterprises, gateways need infrastructure comprising, for example, public identifiers, a set of connectors, and validators (which Cactus could provide), among others.

From the blockchain of the blockchains category, we highlight Hyperservice, a peer-reviewed paper, and Overledger. Hyperservice tries to achieve full dApp atomicity by introducing the concept of \emph{stateless smart contracts}. Using a stateless smart contract, a \ac{CC-dApp} can load a clean state for a contract, using a valid block. While it can partially solve forks in the underlying blockchains, a \ac{CC-dApp} utilizes, the application of this concept paves a direction to decouple smart contract execution from the consensus layer \cite{hyperservice}. Overledger is a sorted list of messages that are interpreted as the state of a cross-blockchain application. While this is an exciting approach to blockchain interoperability, the solution is proprietary, hindering community efforts for more complex solutions.


Blockchain migrators respond to an enterprise need: migration in case of disaster or performance issues \cite{belchior2020_bpvi,bandara2019}. The two presented solutions can only provide data portability across a small set of public blockchains. It is currently impossible to reproduce the chain of events via smart contracts, as that requires a smart-contract translator functionality.
 

A limitation that we identified in the context of Hybrid Connectors is that most solutions do not support hard forks (i.e., the separation of a blockchain into two different blockchains) nor propose a solution for eventual forks, unlike some public connectors (most HTLCs and notary schemes). 
Forks do not happen regularly, and some solutions offer a quick analysis of the problem and acknowledge their importance \cite{hyperledger_cactus, blockCollider, Verdian2018}. However, this is still a problem that can affect the dependability of cross-chain dApps; dealing with forks is still an open issue. For instance, the protocol used in Hyperservice is unable to revert any state update to smart contracts when a dApp terminates prematurely, i.e., it does not grant atomicity. If one does not have atomicity guarantees, it forces the cross-blockchain application into an inconsistent state when a fork occurs. This can put at risk the purpose of the project:  functional cross-blockchain applications. The same problem applies to, for instance, Overleder \cite{overledger03}.

While one might be tempted to conclude that standardization could improve cross-blockchain API design, some argue that APIs are unlikely to generalize well across radically different technologies. Blockchain-agnostic protocols are more likely to be standardized than APIs, as shown historically by successful standards efforts such as HTTP or the TCP/IP family. Finally, solutions that prove cross-smart contract capabilities are emerging, but still in development \cite{scheid2019,Verdian2018,hyperservice,blockCollider,Abebe2019}.

\section{Discussion, Use Cases and Research Questions}
\label{sec:discussion_sec}

This section presents a comprehensive summary of each blockchain interoperability category we extracted and our considerations about blockchain interoperability. Then it presents use cases and finishes with answers to the research questions we proposed.

\subsection{Discussion}
\label{sec:discussion}

Although blockchain interoperability is a complex technology, connecting blockchains ends up being a manageable approach, despite differences in, for example, data structures, digital signature schemes, transmission protocols, verification mechanisms, consensus mechanisms, token issue mechanisms, and smart contract language. However, ``there is a scant effort today to address the standardization of the various infrastructure building blocks – messages, data formats, and flows – to support the interoperability across blockchains'' \cite{hardjono2021}.

Different categories of solutions approach the interoperability problem differently. Our paper firstly introduced Public Connectors in Section \ref{sec_crypto} and stressed their importance. Token exchange is arguably no longer the whole scope of blockchain interoperability \cite{hyperservice}. Instead, various interoperability approaches emerged in the last years, whereby many of them aimed at generalizing blockchain interoperability. In particular, emerging solutions can be categorized as Hybrid Connectors, which provide cross-blockchain communication, and Blockchain of Blockchains, which allow an end-user to create customized, interoperable blockchains at the expense of vendor lock-in.

Public connectors are the most cited among industry and academia, as they provide practical solutions to real-world problems: asset transfers. As these were the first solutions to emerge, not surprisingly, some may not succeed. It seems that the merge of sidechain and protocols relying on an escrow party (enforced by smart contracts) are the most suitable solutions for interoperability among public blockchains. We argue that the flexibility, decentralization, and security of such proposals can be utilized for secure interoperability. However, creating and maintaining a decentralized application using several blockchains was difficult - and hence the Blockchain of Blockchains solutions appeared. Those can facilitate blockchain adoption while providing built-in interoperability among instances of the same platform, whereas variations of the solutions mentioned above can be used to bridge Blockchain of Blockchains to other blockchains.


While Blockchain of Blockchains, such as Cosmos or Polkadot provide a consensus engine and a security infrastructure to build blockchains, blockchain of blockchains aims at developing solutions using different infrastructures. In particular, Cosmos and Polkadot might progress towards \emph{homogenity}, as they support only the creation of Tendermint-based blockchains and Substrate-based blockchains, respectively. While they provide interoperability capabilities, mainly on the chains relying on their technology and other desirable features (shared layer of security, decentralization, governance, better scalability), the end-users choice will be tied to specific implementations. Paradoxically, such solutions might contribute to data and value silos, as solutions built with them cannot connect with an arbitrary blockchain \cite{Abebe2019}. Despite this fact, one could argue that this problem can be alleviated by building bridges/adapters. These solutions are promising but are challenging to integrate with legacy systems and, generally, private blockchains - and hence the hybrid connectors started appearing.

Hybrid Connectors,  specifically blockchain migrators and blockchain of blockchains, progress towards a user-centric, blockchain-agnostic view, enabling enterprise-connected CC-dApps. 
Arguably, the most suitable solution for connecting private blockchains is the usage of blockchain-agnostic protocols; however, they do not grant backward compatibility (as all previous solutions have to be adapted to integrate the adopted communication protocol). To overcome this fact, the short-medium-term solution would be using trusted relays. An interesting way for trusted relays to venture is by decentralizing the escrow party:  from a set of trusted validators to a network of public nodes. It then follows from this survey that one could perceive trusted relays and blockchain-agnostic protocols to be good solutions to link private blockchains; and sidechain, smart-contract-based protocols suitable to solve interoperability among public blockchains. 

A network of blockchain engine-powered blockchains can be leveraged using Hybrid Connectors. For instance, there is a possible synergy between Cosmos and the Interledger Protocol: when a user wants to make an in-app payment with fiat currency (e.g., dollars) within a Cosmos zone, he or she can rely on the interledger protocol as a payment rail. If using cryptocurrencies to pay (e.g., Bitcoin), the interledger router can route the transactions for a payment channel (e.g., Lightning Network), providing more trustful interaction. To connect this ecosystem to private blockchains, bridges have to be developed. To make such bridges trustable, a possible solution would be to elect a group of validator nodes, via an overlay network, that participates in the consensus of public blockchains and private blockchains. This way, cross-chain, and cross-blockchain transactions can be endorsed.

It is worth mentioning that several cross-chain programming languages are appearing, such as the Hyperservice Language \cite{Liu2019} and DAML \cite{daml}. DAML provides a unified Blockchain programming model by abstracting the underlying blockchains and exposing a higher-level abstract ledger on top, similarly to HSL. DAML has different integration degrees: DAML as an application on the target platform; and integration where the DAML runtime engine validates transactions. Programs compiled on such languages can run on top of a BoB platform.

To conclude this discussion, we recall to the reader that blockchain development has been done in silos since its inception. New solutions for blockchain interoperability started emerging as of 2017, and, perhaps not surprisingly, such solutions are also being adopted in silos. While Public Connectors methods are commonly used nowadays, we focus on Blockchain of Blockchains and Hybrid Connectors. Blockchain of Blockchains and Hybrid Connectors allows interoperability between blockchains and other distributed ledger technologies and enterprise systems in the medium term. This promotes the development of blockchain interoperability standards. While blockchain matures, industries will tend to incorporate this technology into their business processes. Then, we predict that mass adoption will follow.

\subsection{Supporting Technologies and Standards}
\label{sec:technologies_standards}

Besides the presented solutions, there is work towards the support and standardization of blockchain interoperability. Blockchain interoperability standards attempt to create a ``standardized transaction format and syntax'', which is common to all blockchains, and secondly, a ``standardized minimal operations set,'' common to all blockchains \cite{Hardjono2019}. In particular, a standardized format is important because while fungible and non-fungible assets have a single, well-defined representation in each blockchain, arbitrary data can be manipulated freely. First, we introduce indirect contributions that promote blockchain interoperability and then the existing standards. 

Recent efforts are visible in enabling heterogeneous smart contract integration through service-orientation \cite{falazi_scip}, allowing external consumer applications to invoke smart contract functions uniformly. A cross-blockchain data storage solution becomes a feasible solution to achieve application interoperability, whereby applications rely on one blockchain. Some dApps\footnote{https://ethlance.com/} already leverage the InterPlanetary File System (IPFS) \cite{Benet2014} to create a common storage, adjacent to the blockchain. The InterPlanetary File System provides a peer-to-peer network for storing and delivering arbitrary data in a distributed file system, potentially facilitating the transfer of data across blockchains \cite{belchior2021-bungee}. Organizations are working on standardizing digital assets. The Token Taxonomy Initiative\footnote{https://tokentaxonomy.org/} is a consortium dedicated to digital token standardization. It proposes a standard to identify tokens' behavior, properties, and control interfaces according to a token classification hierarchy. This project allows application developers to utilize a standard code set for interacting with tokens regardless of the blockchain platform, thus incentivizing blockchain interoperability. In the context of general interoperability, the Ethereum ERCs are a \emph{de facto} standard\footnote{https://eips.ethereum.org/erc}. 

Oracles are mechanisms that software systems provide an external source of truth for blockchains \cite{oracles2020}, and they can be centralized or decentralized \cite{Fan2018}. Typically, centralized oracles are not as dependable as decentralized oracles, as they constitute a single point of failure. 

Hyperledger Avalon \cite{hyperledger_avalon} defers intensive processing from the main blockchain to an off-chain channel to support centralized yet trustable oracles (by using trusted execution environments). Since multiple blockchains can use the same data, it fosters interoperability. 

Open source projects like Hyperledger Indy\footnote{https://www.hyperledger.org/projects/hyperledger-indy} and Hyperledger Aries\footnote{https://www.hyperledger.org/projects/hyperledger-aries} operate in the field of digital identity and self-sovereign identity. Central concepts of self-sovereign identity are decentralized identifiers (DIDs) \cite{did} and verifiable credentials \cite{vc2017}. Decentralized Identifiers can be created, managed, and shared using Zero-Knowledge Proofs (ZKPs) mechanism, even allowing to create new access control models \cite{ssibac}. Such technologies allow for identity portability, enabling cross-blockchain identities \cite{Hyland-Wood2018}. 

So far, the presented standards are called DLT/Blockchain Enabling Technology Standards because they focus on standardizing elements that blockchains can use, as opposed to DLT/Blockchain Generic Framework Standards \cite{lima2018}. These refer to standardization of blockchain interoperability data and metadata formats, identity, and protocols, namely the IETF, ISO, Enterprise Ethereum Alliance, IEEE, The EU Blockchain Observatory \& Forum, and W3C. 

At the Internet Engineering Task Force (IETF), work is being done defining a set of drafts that guide the implementation of ODAP, a protocol using gateways \cite{odap_draft_01,crp_draft_00,draft-hardjono-blockchain-interop-arch-01}. The ISO Technical Committee 307 works towards the ``standardization of blockchain and distributed ledger technologies''\footnote{https://www.iso.org/committee/6266604.html}, but did not produce any standard yet. Subgroup 7 (ISO/TC/SG7) focuses specifically on interoperability. 
The Enterprise Ethereum Client Specification, currently on its seventh version, ``defines the implementation requirements for Enterprise Ethereum clients, including the interfaces to external-facing components of Enterprise Ethereum and how they are intended to be used'', including cross-chain interoperability \cite{eeav7}. The IEEE Blockchain Initiative\footnote{https://blockchain.ieee.org/standards} and the IEEE Standards Association\footnote{https://standards.ieee.org/}, through the IEEE Standards P3203, P3204, and P3205 \footnote{https://blockchain.ieee.org/standards} work at providing ``interfaces and protocols of data authentication and communication for homogeneous and heterogeneous blockchain interoperability''. The EU Blockchain Observatory \& Forum by the European Commission aims to 1) the monitoring of blockchain activities in Europe, 2) the management of the source of blockchain knowledge, 3) the creation of a forum for sharing information, and 4) the creation of recommendations on the role the EU could play in blockchain \cite{eublock}. The same entity points out the likelihood of an increasing number of standards and adoption within governments \cite{eu_forum}. The W3C, via the Interledger Payments Community Group \footnote{https://www.w3.org/community/interledger/}, is connecting payment networks, including decentralized ledgers. Other organizations working in the area include BIA, BiTA, BRIBA, BSI, CESI, DCSA, EBP, GS1, and MOBI \cite{wef2020}.

Standardization efforts focused on a specific blockchain (DLT/Blockchain Platform-Specific Standards) are, for example, the 0302 Aries Interop Profile \footnote{https://github.com/hyperledger/aries-rfcs/tree/master/concepts/0302-aries-interop-profile} and the Hyperledger Fabric Interoperability working group \footnote{https://wiki.hyperledger.org/display/fabric/Fabric+Interop+Working+Group}.

Multiple standards will likely arise and be used, for each vertical industry, as there is a lack of generalized interoperability standards. Standards are then reused across industries (e.g., IEEE P2418.5). Solving interoperability in a specific sector would then pave the way for standards in other industries because the main requirement is domain expertise (ontologies are good starting points for standardization) \cite{lima2018}. The heterogeneity created by standards will pose a regulation challenge, as blockchains may spread across different jurisdictions \cite{hermes-middleware-2021}. 

\subsection{Use Cases with Multiple Blockchains}
\label{sec:use_cases}
In this Section, we present use cases with multiple blockchains. More use cases can be found in Appendix \ref{a:use_cases}.

The industry is still applying blockchain to use cases using only one blockchain. Consequently, it is expected that use cases with multiple blockchains are rare. Notwithstanding, according to the existing resources, it seems that there is considerable interest in use cases using multiple blockchains. As long as the technologies mature, novel, disruptive use cases may be found. For the sake of space, we present some general use cases involving an IoB \cite{draft-sardon-blockchain-gateways-usecases-00}. We refer readers to Appendix F for more use cases relative to Public Connectors, Hybrid Connectors, and Blockchain of Blockchains.

The first big IoB use case is asset transfers, where users can transfer assets from one blockchain to another. While some approaches implement this use case in an ad-hoc way, the emergence of central bank digital currencies (CBDCs) \cite{mbcb,oecd2021}, requires further efforts and standardization \cite{visa2020}. A CBDC is a digital version of a sovereign currency of a nation. A CBDC is issued by central banks, where each unit represents a claim on the value held by such central bank. Many blockchains features are appealing to implement CBDCs, particularly the offered immutability, transparency, and trust distribution. Some central banks are already experimenting with blockchain, including the Monetary Authority of Singapore and the Bank of Canada \cite{draft-sardon-blockchain-gateways-usecases-00}. As each CBDC can be implemented with a blockchain, and each central bank might choose a different technology, interoperability between them is achieved using an IoB or even a BoB.

Another major use case is interoperability across supply chains \cite{draft-sardon-blockchain-gateways-usecases-00, hyperledger_cactus}. A supply chain is a chain of value transfer between parties, from the raw product (physical or intellectual) to its finalized version. Managing a supply chain is a complex process because it includes many non-trusting stakeholders (e.g., enterprises, regulators). As many markets are open and fluid, enterprises do not take the time to build trust - and instead, rely on a paper trail that logs the state of an object in the supply chain. This paper trail is needed for auditability and typically can be tampered with, leading to blockchain's suitability to address these problems \cite{wef2020}. A key challenge of blockchain-based supply chains is to interoperate with other DLT systems. Interoperability granted each participant of the supply chain (e.g., supplier, manufacturer, retailer) can participate at several supply chains (and thus several blockchains) using a single endpoint, simplifying the interaction process while reducing costs. Other use cases comprise  connecting Hyperledger Fabric and Ethereum with Singapore Exchange and Monetary Authority
of Singapore via node integration and  EVRYTHNG, a product connecting multiple chains via API to
digitize products \cite{wef2020}.

Finally, identity and data portability can be provided by an IoB approach. Identity paradigms like self-sovereign identity \cite{ssibac} can increase identity portability by providing users control of their identities. Typically, this is achieved by rooting user credentials in a blockchain. Hence, if blockchains can communicate with identity providers that are blockchains, one can use the same identity in different blockchains. Data portability complies with blockchains, allowing blockchain users to use their data outside of a blockchain without requiring significant effort. 

\subsection{Answers to the Research Questions}
\label{sec:res_q_a}
In this section, we provide answers to the presented research questions (further elaborated on Section \ref{sec:res_q}).

\begin{enumerate}
    \item \textbf{What is the current landscape concerning blockchain interoperability solutions, both from the industry and the academia? i.e., what is the current state of blockchain interoperability solutions?}

    The first step towards blockchain interoperability has been creating mechanisms allowing the exchange of tokens (e.g., cryptocurrencies). We categorized such solutions as Public Connectors (Section \ref{sec_crypto}). Such category comprises Sidechains and Relays (Section \ref{subsec:crypto_side}), Notary Schemes (Section \ref{sec:crypto_notaries}), and Hash Time Lock Contracts (Section \ref{subsec:crypto_hashed}). This category provides an idea of the emergence of blockchain interoperability - but this area no longer applies solely to token exchanges between homogeneous blockchains.

    Novel blockchain interoperability approaches are Blockchain of Blockchains (see Section \ref{sec:be}) and Hybrid Connectors (Section \ref{subsec:blockchain_connectors}). Hybrid Connectors fall into four sub-categories: trusted relays (Section \ref{sec:block_connector_trusted_relay}), blockchain-agnostic protocols (Section \ref{sec:block_connector_blockagnosticprotocol}), and blockchain migrators (Section \ref{sec:block_connector_bm}. We also analyzed related literature reviews on blockchain interoperability, in Section \ref{sec:related_literature_reviews}.

    \item \textbf{Is the set of technological requirements for blockchain interoperability currently satisfied?}
    
    There are two requirements for realizing technical interoperability \cite{vitalik2016}: a pair of sufficiently mature blockchains to build artifacts that promote interoperability and ``some application or need that cannot be served by implementing it on a single blockchain.'' There are several blockchains that can be considered mature enough to support applications built on top of them \cite{fabric, Gorenflo2019,Wood2017,Kwon2016}. On the other hand, interoperability regarding blockchain needs to have the necessary infrastructure and facilitating technologies. In particular, the production of standards \cite{Hyland-Wood2018,token_taxonomy_initiative} technologies like decentralized identifiers \cite{did}, verifiable credentials \cite{vc2017}, cross-blockchain communication protocols \cite{dextt,xclaim,sok_cdl}, and the representation of blockchain smart contracts \cite{Hyland-Wood2018} can foster the likelihood for blockchain interoperability standards and solutions, as they remove considerable barriers to blockchain interoperability. 
    
    On the other hand, there is a set of cross-blockchain use cases that validate the need for interoperability, which will inevitably foster it \cite{hermes-middleware-2021}. In conclusion, the set of critical requirements for blockchain interoperability is currently satisfied, but there is still work to be done at standardization and interoperability across public-private and private-private blockchains.

    \item \textbf{Are there real use cases enabling a value chain coming from blockchain interoperability?}
    
Regarding the third research question, some authors defend that blockchain interoperability is important and crucial for the survivability of this technology \cite{Hardjono2019,Pillai2019 hermes-middleware-2021,hyperservice}.
Standards are paving the way for blockchain adoption \cite{Hyland-Wood2018, Deshpande2017}. It is likely that ``forward-looking interoperability standards are most likely to result in successful standards creation and facilitate industry growth'' \cite{Hyland-Wood2018}. Conversely, standardization is a requirement for mass adoption that is being developed. Given the multiple blockchain interoperability solutions, both Hybrid Connectors, and Blockchain of Blockchains, some of them with considerable weight on the industry, we believe this is a very likely scenario. In Section \ref{sec:use_cases}, we expose multiple use-cases that may benefit from cross-blockchain technology, which can foster adoption by enthusiasts and enterprises. 
In conclusion, we envision reliable solutions and standards emerging in the following years and a steady increase in blockchain adoption by enterprises and individuals alike.
\end{enumerate}

As a value enhancer and maturing key factor, interoperability will ultimately decide the survival of this technology. Based on the evidence collected and analyzed, we foresee increased attention to this research area, with blockchain interoperability gaining traction among the academia and the industry.

\subsection{Open Issues and Challenges}
\label{sec:issues}
In this section, we present open issues and challenges regarding blockchain interoperability and, in a more general sense, the adoption of blockchain.  

Nowadays, solutions available to build decentralized applications lack interoperability, thwarting scalability  \cite{tbig}. As Liu et al.~note, ``it is very challenging to enforce correct executions in a full trust-free manner where no trusted authority is allowed to coordinate the executions on different blockchains'' \cite{hyperservice}. Although interesting and notorious recent advances in this area make interoperability a reality, there is still a gap between theory and practice, as much of the existing work is mostly conceptual.

Given the vast amount of blockchain platforms, fragmentation regarding solutions and their approach to interoperability is strongly present, for example, in IoT scenarios \cite{Zhu2019}. A combination of multiple platforms tailored for specific purposes, which can be public, private, or consortium, adds an overhead to manage workflows. In particular, this concern is intensified when multiple blockchains are serving a specific application.

Concerning blockchain scalability, the internet of blockchains can be realized upon improvements to current performance, both in public and private blockchains. Techniques such as implicit consensus and data sharding can improve transaction throughput and storage \cite{omniledger}. However, blockchain sharding requires solving cross-blockchain transaction routing and retrieval and asset referencing (also known as the discoverability problem).   

It is challenging to coordinate transactions from different blockchains to support a cross-chain dApp, as different blockchains have different properties (e.g., architecture, protocols \cite{Abebe2019}, service discovery, access control, between others). In particular, reverting a transaction that depended on another can be cumbersome, especially given different transaction finalities from different blockchains). Some solutions have proposed a mechanism to overcome such a challenge (blockchain of blockchains) \cite{Verdian2018,hyperservice}. Although a promising approach, it is still unclear the applicability of these solutions to arbitrarily complex cross-blockchain dApp. More research is required to confirm the feasibility of this approach.

Some authors \cite{vo2018} highlight problems related to the GDPR, such as security, trust, confidentiality, and data privacy issues. In particular, security threats are exacerbated by the presence of multiple blockchains and possible multiple administrators. Regarding privacy, the authors underline problems with the right-to-forget, in which a user can ask his or her data to be deleted from the blockchain. Currently, most blockchains do not provide effective mechanisms that can respond to this request. Blockchain fine-grain access control is appointed as a key requirement to minimize information leakage and confidentiality risk. 

Blockchain interoperability reduces dependencies on a single blockchain, and consequently, risk (e.g., the blockchain is attacked) \cite{dextt}, it does not eliminate the inherent risks. It is worth underscoring that the multiple blockchain approach is more complicated than the sum of its parts, as there is extra complexity underlying the cross-chain communication. This adds challenges to governance: whereas a private consortia can use Hybrid Connectors at will to interoperate systems (decentralized and/or decentralized), the governance model is not straightforward within community projects, supported by public blockchains.

In short, the most relevant open issues towards blockchain interoperability are:
\begin{itemize}
    \item The gap between theory and practice, including the lack of standardization and implementations \cite{hardjono2021,Zhu2019},
        \item Discoverability \cite{Verdian2018,hyperservice, Abebe2019},
    \item Privacy and Security \cite{Thomas2015,vo2018,Wood2017,sok_cdl},
    \item Governance \cite{Hardjono2019, Hardjono2019a,wef2020,Qasse2019}.
\end{itemize}

Notwithstanding, security \cite{surv_sec, sok_crpyto}, privacy \cite{casino2019}, and scalability (e.g., using sharding \cite{surv_shar} or novel blockchain systems \cite{appendableblock}) remain the most prominent areas to be improved in the blockchain space.

\section{Research Directions}
\label{sec:research_directions}
New tools, frameworks, standard proposals, and even programming models are emerging and need further development. Programming models such as Polkadot and Cosmos offer developers a way to create their blockchains effectively and connect them to other blockchains. Protocols such as ILP and UIP allow cross-blockchain transactions. Programming languages such as HSL and DAML aim at embedding different blockchain models, providing an abstraction for cross-blockchain dApps. 

Although one can have good reasons to utilize blockchain interoperability solutions for public or private blockchains, few solutions are available for connecting them. The problem of obtaining state from permissioned blockchains effectively  \cite{abebe2020} makes interoperating with private blockchains a challenge \cite{wef2020,iiconsortium2020}. Thus, connecting public and private blockchains bidirectionally remains an open problem.

One of the problems that bidirectional communication across permissioned and permissionless ledgers poses is semantic compatibility. Technical interoperability does provide the technical foundation that realizes interoperability but does not grant semantic interoperability per se \cite{Hardjono2019}. There is, therefore, a gap: how can we effectively combine both blockchain types to enable new use cases? How to make sure a solution complies with the goals of all involved stakeholders? Disciplines as view integration can help to provide an answer \cite{belchior2020_bpvi}. View integration is the process that combines views of the same business process into a consolidated one by combining the different views of the stakeholders participating in different blockchains.

Another considerable obstacle for blockchain adoption is its fast-paced development. The development of blockchain interoperability standards may provide a way for more flexibility regarding backward compatibility. 


In the light of the present study and the identified open issues and challenges, we propose research directions based on some sections of our survey: research on architecture for enabling blockchain interoperability, Public Connectors, Blockchain of Blockchains, Hybrid Connectors, and supporting technologies, standards, use cases, and others.

\emph{Architecture for Blockchain Interoperability} (Section \ref{a:architecture}):
\begin{itemize}
    \item Define a blockchain interoperability maturity model, modeling interoperability at its various layers (e.g., technological, semantic, organizational).
    
    \item Model the different views on the various types of interoperability, according to different stakeholders (e.g., the provider's technical view on a cross-blockchain dApp \emph{versus} the semantic view of the end-user on the same cross-blockchain dApp).
    
    \item Study blockchain interoperability semantics by exploring, for example, the research area of view integration \cite{view_int}.
    \end{itemize}

Public Connectors (Section \ref{sec_crypto}):    
    \begin{itemize}

    \item Research on how permissioned blockchains can benefit from sidechains to improve scalability and privacy.
    
    \item Develop protocols to allow fiat money exchange, higher liquidity on decentralized exchanges. Conversely, improve the level of privacy and security of centralized exchanges.
    
    \end{itemize}

Blockchain of Blockchains (Section \ref{sec:be}):
\begin{itemize}
    \item Integration of existing blockchain systems with Blockchain of Blockchains.
    \item Study how Blockchain of Blockchains can provide a reliable interoperability scheme bridging permissioned blockchains and permissionless blockchains.
    \item Connect Blockchain of Blockchains to both centralized systems and decentralized ledger systems (e.g., connect Polkadot to Visa).
    \end{itemize}
    
    Hybrid Connectors (Section \ref{subsec:blockchain_connectors}):
\begin{itemize}
    \item Decentralize the trust of trusted relays by integrating them with public blockchains (e.g., by submitting the state periodically to a public blockchain);
        \item Study how blockchain-agnostic protocols can be easily adapted to existing ledgers. 
        \item Explore the blockchain of blockchains approach as an advance in dependable blockchain-based applications.
            \item Improve atomicity and consistency guarantees on cross-blockchain decentralized applications.
        
        \item Explore blockchain migration across public and permissioned ledgers. Such migration schemes can be decentralized and adapt to functional and non-functional requirements imposed by stakeholders.
        \item Explore blockchain migration via non-trusted relays (e.g., using a set of public escrow nodes following a protocol).
                \item Develop frameworks for multiple blockchain management. Such frameworks should respond to multiple stakeholder needs, decentralizing trust.
                \item Model integration abstraction layers that enable the development of universally connected contracts. 
            \item Research on the visualization of CC-Txs.
\end{itemize}

Supporting technologies and standards, use cases, and others (Section \ref{sec:discussion}):

\begin{itemize}
\item Work along with regulators and standardizing bodies to come with blockchain interoperability standards across industries

            \item Research on blockchain interoperability programming languages, supporting tools, and standards, including but not limited to cross-blockchain programming languages and frameworks, decentralized identifiers and verifiable credentials, and blockchain interoperability standards for enterprise blockchains;
        \item Explore new use cases using multiple blockchains, the ``value-level'' interoperability \cite{hyperledger_cactus}.

    \item Research synergies between cryptocurrency-based interoperability approaches, Blockchain of Blockchains, and Hybrid Connectors.
    
    \item Study security aspects of blockchain interoperability.
    \item Understand the implications of the different interoperability layers (value, semantic, organizational, among others).

\end{itemize}

\section{Conclusion}
\label{sec:concl}
In this paper, we performed a systematic literature review on blockchain interoperability. We systematically analyzed, compared, and discussed 80 documents, corresponding to 45 blockchain interoperability solutions. By including grey literature, we expect to thwart intrinsic limitations regarding the blockchain interoperability research area, such as a considerable presence of the industry. By exploring each solution methodologically, this study provides interesting insights, distributed across three categories: Public Connectors, Blockchain of Blockchains, and Hybrid Connectors. 
Despite sidechain and HLTC solutions are gaining traction in the industry, blockchain interoperability are not solely Public Connectors solutions. New approaches started emerging since 2017. Hybrid Connectors provide a varied landscape of solutions, adapted for the majority of the use cases. They are likely to be used to produce cross-blockchain dApps. Blockchain of Blockchains are likely to be adopted by the industry in the short-medium term, by leveraging easy-to-produce, customizable blockchains.


Our findings allow us to conclude that conditions to research on blockchain interoperability are fulfilled, allowing a multitude of new use cases. Thus, we expect interest in this research area to raise considerably. 

This work is towards making the blockchain ecosystem more practical, by easing work for developers and researchers. We expect that this study provides a robust and dependable starting point whereby developers and researchers can work in the blockchain interoperability research area.
 
\begin{acks}

The authors would like to thank to the anonymous reviewers that constructively provided suggestions that significantly improved this paper. Thanks to Peter Somogyvari, Paul DiMarzio, Jag Sidhu, Sergio Lerner, Andy Leung, Travis Walker, Bill Laboon, Josh Lee, Austin King, Oliver Birch, Thomas Hardjono, and Miguel Matos for fruitful discussions regarding blockchain interoperability. We thank Daniel Hardman and Ken Elbert for constructive discussions about DIDs and verifiable credentials. Special thanks go to Iulia Mihaiu, Cláudio Correia, Benedikt Putz, Francisco Braga, Gavin Wood, João Ferreira, Miguel Pardal, Jonas Gehrlein, and Dilum Bandara for constructive  comments and suggestions that greatly contributed to improving the paper. The authors express their gratitude to the Linux Foundation for providing the Diversity \& Inclusion scholarship. 
 This work was partially supported by the EC through project 822404 (QualiChain), and
by national funds through Fundação para a Ciência e a Tecnologia (FCT) with reference
UIDB/50021/2020 (INESC-ID).
\end{acks}


\bibliographystyle{ACM-Reference-Format}
\bibliography{references}
\appendix



\begin{acronym}[ftyp]
\acro{IoB}{\emph{Internet of Blockchains}}
\acro{BoB}{\emph{Blockchain of Blockchains}}
\acro{DLT}{\emph{Distributed Ledger Technology}}
\acro{AS}{\emph{Autonomous System}}


\acro{LN}{{{Ligthning Network}}}
\acro{ILP}{{{Interledger Protocol}}}

\acro{CC-dApp}{\emph{Cross-Chain Decentralized Application}}
\acro{CC-Tx}{\emph{Cross-Chain Transaction}}
\acro{CB-Tx}{\emph{Cross-Blockchain Transaction}}
\acro{AAA}{Authentication, Authorization, and Accounting}
\acro{ABAC}{Attribute-Based Access Control}
\acro{AD}{Access Directory}
\acro{AM}{Attribute Manager}
\acro{AP}{Attribute Provider}
\acro{API}{Application Programming Interface}
\acro{AS}{Autonomous System}
\acro{BRBAC BN}{Blockchain Role-Based Access Control Business Network}
\acro{CC}{Cloud Computing}
\acro{CFT}{crash fault-tolerant}
\acro{DAC}{Discretionary Access Control}
\acro{DLT}{Distributed Ledger Technology}
\acro{EMR}{Electronic Medical Record}
\acro{ID}{Identifier}
\acro{IoT}{Internet of Things}
\acro{JSON}{JavaScript Object Notation}
\acro{MAC}{Mandatory Access Control}
\acro{PAP}{Policy Administration Point}
\acro{PDP}{Policy Decision Point}
\acro{PEP}{Policy Enforcement Point}
\acro{PIP}{Policy Information Point}
\acro{PKI}{Public Key Infrastructure}
\acro{PRP}{Policy Retrieval Point}
\acro{RBAC}{Role-Based Access Control}
\acro{SP}{Smart Policy}
\acro{tps}{transactions per second}
\acro{VM}{Virtual Machine}
\acro{XACML}{eXtensible Access Control Markup Language}
\acro{XML}{Extensible Markup Language}
\acro{ftyp}{File Type}
\acro{EHR}{Electronic Health Record} 
\acro{TBAC}{transaction based access control}
\acro{ZKP}{Zero knowledge proofs}
\end{acronym}

\section{Methodology}
\label{a:method}
This section presents the methodology we followed in conducting the systematic literature review about blockchain interoperability. Our methodology follows several phases, as advised by several authors, \cite{Group2007,frantz}. In the planning phase, we select the research questions, the data sources, the search terms, the practical screening criteria, and the methodological screening criteria. In the review phase, we abstract data from selected papers, identifying the underlying conceptual mechanisms for interoperability. We then correlate approaches \emph{intra-category} and \emph{inter-category} (via the discussion subsections). Finally, we report the review and synthesize the findings. 

We give special attention to grey literature, as some authors defend that it includes ``a broader scope of literature, providing a more comprehensive view of the available evidence'' \cite{greyliterature, Group2007}. In particular, we analyze grey literature as a way to include recent endeavors. In particular, we argue that including grey literature is relevant, as:
(i) blockchain interoperability is in active development, and there is still a reduced number of academic studies, (ii) some research is concentrated on the industry, and (iii) grey literature reduces the publication bias \cite{Group2007}.

Notwithstanding, grey literature 
is not often updated (e.g., whitepapers \cite{blockCollider,aion2017,POA, Wood2017}).
To the best of our knowledge, we picked the most recent whitepaper versions and made the effort of looking through the documentation for updates. Nonetheless, it is possible that a newer version is available, or that we missed out on relevant information. That is why we systematically contacted the authors of the projects (see Section \ref{subsec:search_proc}). This methodology allows us to validate or view of the project at hand while addressing some shortcomings of researching grey literature. 
Hence, we built a list of references and contacts, which we engaged during our research. We indicate when we obtained feedback from authors on their projects, using the ``checkmark'' sign (\checkmark). More specifically, the checkmark typically indicates that we have taken the authors or their respective team's feedback into consideration, regarding a specific project. 
Exceptions occur whenever the legend of a table indicates so (for example, in Table \ref{tab:related_slr}, the checkmark indicates that an author discusses the referenced criteria. A caveat of our approach is that grey literature is not, necessarily, quality scientific work, as it is not peer-reviewed  \cite{blockchainresearchframework}.

Moreover, in order for our grey literature search to be ``systematic, transparent, and reproducible,'' we adopt recommendations from Mahood et al.~\cite{greyliterature}. In particular, they recommend ``that searches include online databases, web search engines and websites, university, and institutional repositories, library catalogs, as well as contacting subject specialists, hand-searching and consulting reference lists of relevant documents''. We then include grey literature, as the result of retrieving references from scientific articles, and consultation with both academics and professionals in the area of blockchain interoperability. We, therefore, define grey literature as: Github documentation, whitepapers, technical and institutional reports, initial coin offer plans, magazine articles, academic dissertations, consultant reports, book chapters, and blog posts. With such sources, we believe that it is possible to construct a reliable, updated, and extensive understanding of blockchain interoperability.

We believe this approach leads to  adequate coverage and transparency in blockchain interoperability research and, consequently, provides accurate information to the reader in a research area evolving so quickly. In a research area on its inception, and given its fragmentation, we acknowledge that we may have missed some advances in this field. We commit to updating our knowledge base in the light of the new information being produced, to yield the most comprehensive results possible.

\subsection{Research Questions}
\label{sec:res_q}
Taken into account the different stakeholders of the blockchain technology, and the previous literature reviews limitations, we propose the following research questions, addressed by this paper:

\begin{enumerate}
    \item \textbf{What is the current landscape concerning blockchain interoperability, both from the industry and the academia?}
Bitcoin and Ethereum fostered hundreds of cryptocurrencies and use cases, shortly after their inception. Heterogeneous solutions appeared to further deliver customization, tailored for enterprise use-case scenarios that benefit with blockchain technology. Soon after this solution proliferation, and in particular, with the vast number of platforms emerging, the blockchain interoperability problem started to be tackled by industry and academia \cite{vitalik2016, vo2018,kaur2018,Kan2018}. Although some attempts of classifying blockchain interoperability solutions have been made \cite{Qasse2019,vitalik2016,tast_paper5}, they are either outdated, or not capturing the whole interoperability spectrum. 
    
\item \textbf{Is the set of technological requirements for blockchain interoperability currently satisfied?}
According to several authors, the prerequisites for blockchain interoperability are: (i) the existence of a cross-blockchain communication protocol that can transfer arbitrary data in a trustless and decentralized way, comparable to the transport layer of the Internet \cite{Hardjono2019}, (ii) a pair of sufficiently mature blockchains that can be bridged through such protocol, and (iii) the need for applications benefiting from a multiple-blockchain approach \cite{vitalik2016}, i.e., IoB-powered BoB applications.
 This research question is particularly important since it gives a perspective if research and focus should be put in the direction of blockchain interoperability.

    \item \textbf{Are there real use cases enabling a value chain coming from blockchain interoperability?} According to some authors \cite{Hardjono2019,Pillai2019,Hileman2017,hyperservice,hardjono2021_gateways}, blockchain interoperability is a core requirement for the survival of the technology.
    Given stable, matured blockchain interoperability mechanisms, one needs to explore which solutions can be built, which sectors it may benefit, and what are the use cases foreseeable in the short and medium-term.
\end{enumerate}

\subsection{Data Sources}
The online repository using for the majority of the research is Google Scholar. Google Scholar is a modern search engine owned by Google, which indexes most major digital libraries, including but not limited to IEEE Xplore, ACM Digital Library, Science Direct (another major search engine for digital libraries), ASCE, Scopus, Web of Science, SpringerLink, and arXiv (known for containing grey literature). According to Google's documentation\footnote{https://scholar.google.com/intl/en/scholar/help.html\#coverage}, ``Google Scholar includes journal and conference papers, theses and dissertations, academic books, pre-prints, abstracts, technical reports, and other scholarly literature from all broad areas of research''. It includes  ``academic publishers, professional societies, and university repositories, as well as scholarly articles available anywhere across the web. Google Scholar also includes court opinions and patents''. It covers grey literature, making it a suitable option to reduce the publication bias \cite{Group2007}. Google Scholar's coverage is arguably the biggest across other academic search engines for Computer Science \cite{fagan2017}, and it meets the criteria recommended in guidelines for conducting systematic literature reviews \cite{fagan2017,brereton2017}. Fagan critiques Google Scholar for giving too much importance to the citation count and therefore suggests the usage of additional search tools to conduct the review \cite{fagan2017}. However, as we are aiming for a bigger coverage, by studying most work concerning blockchain interoperability up to this date, the bias introduced by the citation count does not significantly impair our study. Hence, and to simplify our research process, we rely on Google Scholar.

Furthermore, in order to add resiliency to our study, we compiled a list of appropriate search terms from our knowledge of the literature – previous searches on this topic, well-known projects on the community and suggestions from other researchers, to identify additional references not previously captured. Such references were included in the review. 

\subsection{Search Process}
\label{subsec:search_proc}
We divided the search process into three phases: searching for related literature reviews, searching for relevant peer-reviewed scientific papers, and searching for relevant grey literature.

We aim to find relevant literature directed to blockchain interoperability, which can be synonyms with \emph{chain interoperability}, \emph{interconnected blockchain}, \emph{multiple blockchains}, and \emph{internet of blockchains}. One could consider the concept of \emph{blockchain sharding} a novel solution to address blockchain scalability, which can ultimately foster blockchain interoperability since shards need to communicate with each other. However, due to the extension of the blockchain sharding research area, and because of space constraints, we purposely leave it out of the scope of this research. 

In the first phase of the search process, \emph{identification}, we queried \emph{``interblockchain survey''} OR ``blockchain interoperability survey'' OR \emph{``IoB''}, where we obtained 86 results. From those 86 results, only one was explicitly a literature review concerning blockchain interoperability (i.e., contained the term ``survey'' at the title). 

Next, we performed a keyword-based search. We limited the scope of queries until the present date of writing, i.e., the 14th February 2020, thus covering literature up to the present day. Notwithstanding, we updated this paper with both academic literature and grey literature dated up to the end of May 2020.
Google Scholar treats all terms specified in the search query as an \emph{AND} operator: it yields search results for all the terms. Henceforth, all queries presented in this document assume such quotes. Therefore, we opt by restricting this feature, as querying \emph{blockchain interoperability} yields more than 9,000 results. By using quotes in the search, we limited its range. Hence, a query with the keywords \emph{blockchain} and \emph{interoperability} yields results only if both terms are present. We then searched the terms \emph{interchain communication}, \emph{interconnected blockchain}, and \emph{blockchain interoperability}, as they semantically seem the most suitable terms for our search. We obtain 262 results: and chose not to include terms as \emph{multiple blockchains} or \emph{chain interoperability}, because although related, those terms are too vague and yield too many results not directly related to this study, respectively 494 and 665 results.

In the third phase, we collected relevant work classified as grey literature. We retrieved the collected reference list and used techniques as snowballing to expand our document repository further. We obtain an additional 69 documents.

\subsection{Screening and Eligibility Processes}
\begin{figure}[h]
    \centering
    \includegraphics[scale=0.30]{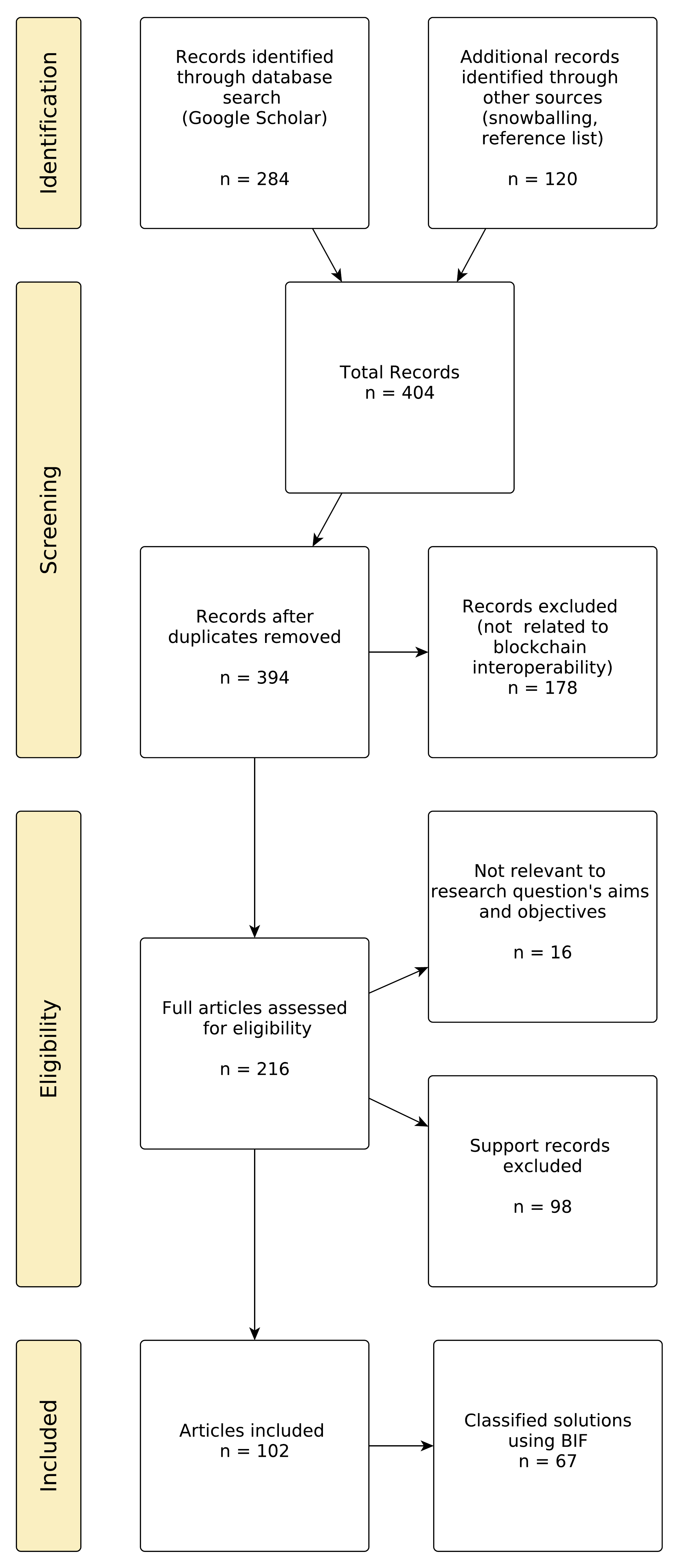}
    \caption{PRISMA diagram specifying our literature research methodology.}
    \label{fig:prism}
\end{figure}

In this section, we define our methodology for the eligibility criteria. Figure \ref{fig:prism} represents an adapted \emph{Preferred Reporting Items for Systematic Reviews and Meta-Analyses (PRISMA)} diagram \cite{prisma}, considering all steps of our literature research methodology.

In terms of the included documents (papers, grey literature), we first examined the title, abstract, and keywords. When these three elements do not provide enough insights to decide on whether include the document on this study, we examined the full-text body of the documents. This first screening aims to conclude about the feasibility of a given document to answer the proposed research questions. 

Due to the small number of available papers, we had a lenient approach regarding the exclusion criteria: we only excluded papers that do not comprehensively tackle blockchain interoperability. For example, papers which focus is state of the art on blockchain applications, security, scalability, consensus mechanisms, and economic models, even if they tackle blockchain interoperability, are excluded. In contrast, papers with at least a section dedicated to blockchain interoperability are taken into consideration. The process above leads to a total number of 404 documents. After excluding 178 non-related papers, 10 duplicates, 16 not relevant papers, and 98 support papers (papers that, although crucial for the understanding of this topic, they are not included in the comparison of solutions), we achieve a total of 102 documents, from which 67 were systematically compared.

\section{An Architecture for Blockchain Interoperability}
\label{a:architecture}
This section discusses existing architectures for interoperable blockchains, the ``internet of blockchains'' approach. We then present a consolidated architecture. 

Zhu et al.~define several layers for a blockchain \cite{Zhu2019}. The \emph{data layer} defines the representation of data in the blockchain (e.g., transactions aggregated into blocks vs transactions represented in a directed acyclic graph). The \emph{network layer} defines the type of nodes in the peer-to-peer network (e.g., full nodes and light nodes \cite{nakamoto2008}). The \emph{consensus layer} represents the consensus algorithm the network uses and its security assumptions. The \emph{contract layer} represents the execution environment for smart contracts, which provide the foundation for the application layer, which include the blockchain-enabled business logic.

Other authors proposed architectures for blockchain interoperability composed of several layers: Jin et al. proposed the data, network, consensus, contract, and application layers \cite{jin2018}, while Kan et al.~proposed the basic, blockchain, multi-chain communication, and application layers \cite{Kan2018}. 

\begin{figure*}[h!]
    \centering
    \includegraphics[scale=0.50]{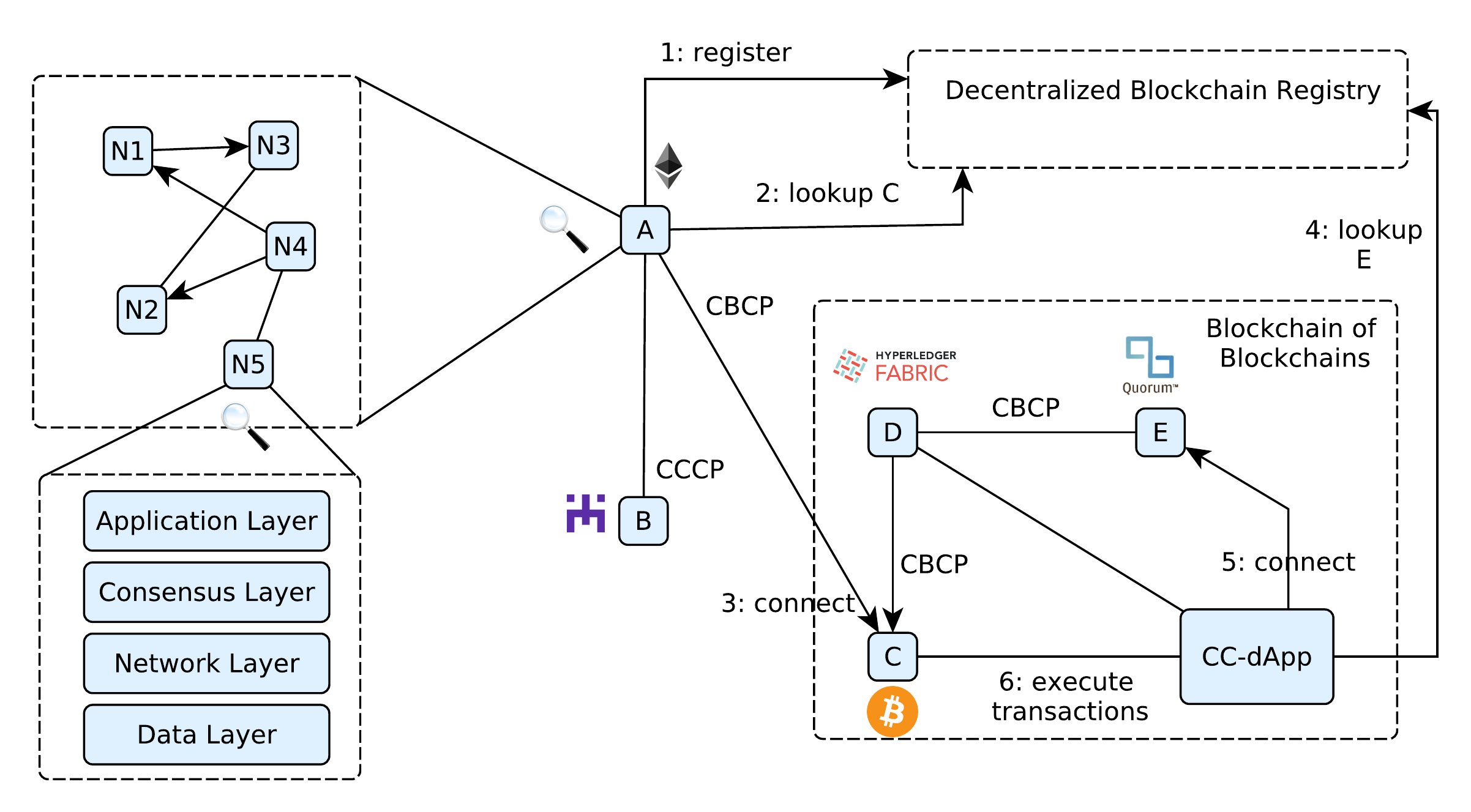}
    \caption{Architecture for Interoperable Blockchains: a network comprised of five blockchains (A to E) and a cross-chain decentralized application (CC-dApp).}
    \label{fig:aib}
\end{figure*}

Hardjono et al.~proposed an architecture inspired by the architecture of the Internet \cite{Hardjono2019}. The proposed architecture has as central concepts the  \ac{AS} (or \emph{routing domain}) and gateway. A routing domain is a network ecosystem operating with specific rules, under an administrative domain. An AS is a set of IP networks that form a single administrative domain, which maps to a blockchain network. A gateway supports cross-domain routing in order to allow communication among networks in different ASs. Gateways are nodes that support interoperability, such as smart contracts or trusted third parties.

Our proposal is influenced by previous work: in particular, we envision each blockchain as an autonomous system, which communicates to others via a cross-blockchain protocol. Most nodes on public and private blockchains can serve as interoperability gateways. To facilitate communication among blockchains, one can rely on decentralized blockchain registries, that can identify and address oracles, blockchains, and their components (e.g., smart contracts, and certificate authorities)  \cite{vo2018}. A registry for both public and private blockchains could be written in a public blockchain with strong security assumptions (e.g., a high degree of decentralization). Alternatively, the contents of the registry can be recorded in a custom public blockchain maintained by the stakeholders of major blockchains, or enforced by trusted hardware \cite{Hardjono2019}. The decentralized registry would act as a (preferably) decentralized domain name system \cite{dns}, but for blockchains instead of domains. A simple implementation would be leveraging a multi-signature Ethereum smart contract where a consortium could manage a registry of gateway nodes. 

We leave further discussions on a decentralized blockchain registry for future work. Note that this registry is optional, and it is not essential for enabling an IoB.

Figure \ref{fig:aib} illustrates our proposal for an architecture for the IoB, the enabler of technical interoperability. Although we represent a BoB in the figure, we do not detail its architecture at this stage. Blockchain$_A$ (A) and Blockchain$_B$ (B) are both public, EVM-based blockchains, namely Ethereum and POA Network. Blockchain$_D$ (D) and Blockchain$_E$ (E) are private blockchains, namely Hyperledger Fabric and Quorum. A blockchain node belonging to the Ethereum network, Blockchain$_A$, registers its communication endpoint (i.e., IP address) on the blockchain registry (step 1). After that, it looks up for the address of a node belonging to Blockchain$_C$ (C), Bitcoin (step 2). CCCP and CBCP  protocols can provide unilateral or bidirectional interoperability. In step 3, a CBCP establishes communication between the Ethereum node and the Bitcoin node, unilaterally, since the Ethereum node can read Bitcoin's blocks headers (e.g., via \cite{btcrelay}), but not the other way around. Blockchain$_D$ and Blockchain$_E$ are heterogeneous, thus connected by a CBCP. A CC-dApp is already connected to blockchain$_C$ and blockchain$_D$, and further connects with blockchain$_E$, after fetching its address on the blockchain registry (steps 4 and 5). Step 4 assumes the necessary credentials to access the private blockchain are held by the CC-dApp user(s) (e.g., private keys, X.509 certificates). A CC dApp protocol allows an end-user to realize the semantic interoperability, by leveraging blockchain$_C$, blockchain$_D$, and blockchain$_E$ (step 6).
These steps accomplish connectivity among blockchains, thus forming an IoB, and therefore enabling a BoB.

CCCPs (e.g., XClaim \cite{xclaim}) and CBCPs (e.g., inter-blockchain protocol \cite{ibc} or the Interledger Protocol \cite{ILPv4}) can be employed to manage the end-to-end communications between blockchain networks, addressable by the blockchain registry. While such protocols can provide seamless interoperability for future blockchains, via standardization, they are not compatible with existing blockchains. Existing blockchains would require to refactor several layers: the network, consensus, contract, and application layers \cite{Zhu2019}, would need to be changed. 
\begin{figure}[htb]
    \centering
    \includegraphics[scale=0.55]{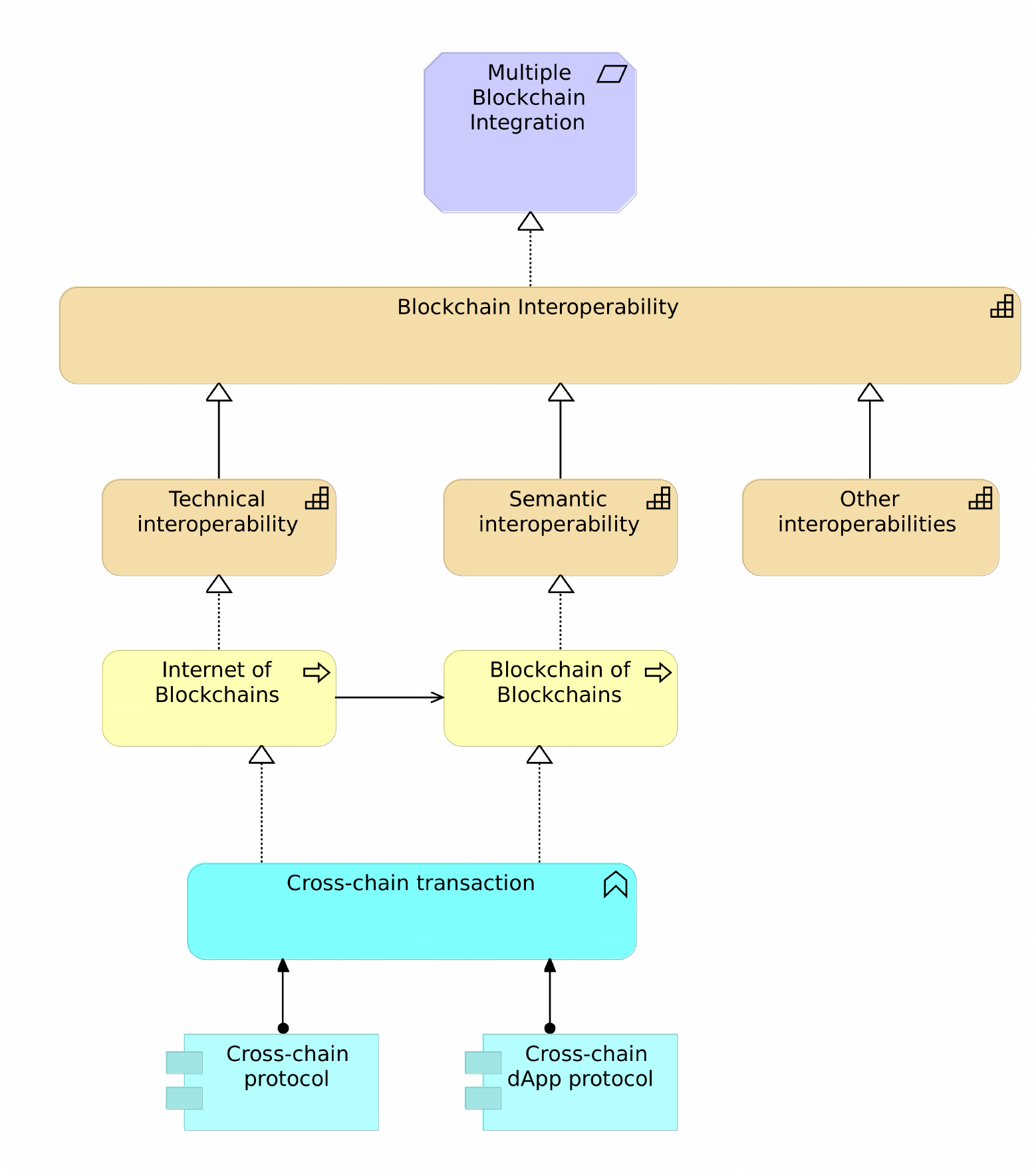}
    \caption{Simplified blockchain interoperability model, represented in Archimate}
    \label{fig:model}
\end{figure}

In Figure \ref{fig:model}, we model the layers of blockchain interoperability that correspond to the proposed architecture, using the Archimate modeling language \cite{group2016archimate}, a standard for enterprise architecture modeling. Blockchain interoperability, technical interoperability and semantic interoperability are capabilities, abilities that the business processes ``Internet of Blockchains'' and ``Blockchain of blockchains'' possesses (as they enable interoperability at different levels). 
``Cross-chain protocols'' and ``cross-chain dApp protocols'' are applicational components that realize the ``cross-chain transaction'' function. Other interoperability layers are left for future work.

Regardless of the interoperability solution employed, it is likely that the network layer has to suffer refactoring, and consequently the consensus layer since there are blockchains with different transaction finalities \cite{Das2019}. Transaction \emph{finality} can be probabilistic or deterministic, and refers to when parties involved in a transaction can consider it committed to the blockchain. For example, Bitcoin needs around 6 confirmed blocks to consider a transaction final with a high probability (probabilistic), whereas Tendermint transactions are final right after their execution (deterministic). Several abstractions that include transactions from other blockchains can be implemented on the contract layer. These changes have repercussions on the application layer, as now it can handle more complex operations. The application can now expose APIs to dispatch cross-blockchain transactions, as illustrated in some works \cite{hyperservice, Verdian2018, hyperledger_cactus}. The data layer would not necessarily have to be changed.  

Although this could be a viable solution, it is logistically cumbersome to adjust all blockchains in production to use a specific set of inter-blockchain protocols and to adapt their different layers. As this solution is not feasible in practice, at least in the short term, blockchain interoperability solutions are typically tailored for a specific blockchain or a set of specific blockchains. Nevertheless, we believe that as the technology matures blockchain interoperability standards will guide technical efforts, leading to convergence towards interoperability within the blockchain space. 

Throughout this paper, blockchain-agnostic solutions, as well as specific solutions will be presented and discussed.

\section{Public Connectors}
\label{a:crypto}

\subsection{\underline{Sidechains}}
\label{subsec:crypto_side_appendix}

We now describe some of sidechain solutions we identified in the literature. Table \ref{tab:crypto_sidechains} summarizes these solutions. An analysis of this table is conducted in the discussion.


The \emph{Peace Relay} is inspired by BTC Relay, allowing communication between EVM-based blockchains \cite{peacerelay}. Peace allows Ethereum contracts to verify account states and transactions from Ethereum Classic, and vice-versa, allowing a two-way peg (given that the Peace relay smart contract is deployed on both chains).

\emph{Testimonium} is a relay solution that follows a validation-on-demand pattern, validating blockchain block headers on-chain \cite{testimonium}. As block headers are accepted optimistically, validation-on-demand locks block headers for a specific lock time, where off-chain clients (disputers) can challenge their validity.

\begin{table*}[]
\centering
\caption{Comparison of \emph{Sidechains} solutions}
\label{tab:crypto_sidechains}
\resizebox{\textwidth}{!}{%
\begin{tabular}{@{}lllllll@{}}
\toprule
\multicolumn{1}{c}{\textbf{Reference}} &
  \multicolumn{1}{c}{\textbf{Mainchain}} &
  \multicolumn{1}{c}{\textbf{\begin{tabular}[c]{@{}c@{}}Sidechain \\ consensus\end{tabular}}} &
  \multicolumn{1}{c}{\textbf{Summary}} &
  \multicolumn{1}{c}{\textbf{Strong points}} &
  \multicolumn{1}{c}{\textbf{Weak points}} &
  \multicolumn{1}{c}{\textbf{Roadmap}} \\ \midrule
\multicolumn{1}{l|}{BTC Relay \cite{btcrelay} \mbox{}\hfill \checkmark} &
  \multicolumn{1}{l|}{Ethereum} &
  \multicolumn{1}{l|}{$\times$} &
  \multicolumn{1}{l|}{\begin{tabular}[c]{@{}l@{}}Ethereum smart contract \\ reading Bitcoin's blockchain\end{tabular}} &
  \multicolumn{1}{l|}{\begin{tabular}[c]{@{}l@{}}Simple solution \\ relying on verifying \\ block headers\end{tabular}} &
  \multicolumn{1}{l|}{Limited functionality} &
  None \\ \midrule
\multicolumn{1}{l|}{Peace Rekay \cite{peacerelay}} &
  \multicolumn{1}{l|}{Ethereum} &
  \multicolumn{1}{l|}{$\times$} &
  \multicolumn{1}{l|}{\begin{tabular}[c]{@{}l@{}}SPV on EVM-based\\ blockchains\end{tabular}} &
  \multicolumn{1}{l|}{Allows two way pegs} &
  \multicolumn{1}{l|}{\begin{tabular}[c]{@{}l@{}}It is expensive to verify\\ Ethereum block headers\end{tabular}} &
  None \\ \midrule
\multicolumn{1}{l|}{Testimonium \cite{testimonium}} &
  \multicolumn{1}{l|}{Ethereum} &
  \multicolumn{1}{l|}{$\times$} &
  \multicolumn{1}{l|}{\begin{tabular}[c]{@{}l@{}}EVM-based blockchains\\ SPV\end{tabular}} &
  \multicolumn{1}{l|}{Effiecient validation} &
  \multicolumn{1}{l|}{\begin{tabular}[c]{@{}l@{}}Mainly support EVM-\\ based blockchains\end{tabular}} &
  \begin{tabular}[c]{@{}l@{}}Batch submission \\ of block headers\end{tabular} \\ \midrule
\multicolumn{1}{l|}{POA Network \cite{POA} \mbox{}\hfill \checkmark} &
  \multicolumn{1}{l|}{Ethereum} &
  \multicolumn{1}{l|}{\begin{tabular}[c]{@{}l@{}}Proof of \\ authority\end{tabular}} &
  \multicolumn{1}{l|}{\begin{tabular}[c]{@{}l@{}}Applicational interoperability\\  to EVM-based dApps\end{tabular}} &
  \multicolumn{1}{l|}{Inexpensive consensus} &
  \multicolumn{1}{l|}{\begin{tabular}[c]{@{}l@{}}Validators confined to \\ one country \\ (geographic concentration)\end{tabular}} &
  \begin{tabular}[c]{@{}l@{}}POA-based \\ stable token\end{tabular} \\ \midrule
\multicolumn{1}{l|}{Liquid \cite{peggedsidechains,confidential_assets} \mbox{}\hfill \checkmark} &
  \multicolumn{1}{l|}{Bitcoin} &
  \multicolumn{1}{l|}{\begin{tabular}[c]{@{}l@{}}Strong \\ federations\end{tabular}} &
  \multicolumn{1}{l|}{\begin{tabular}[c]{@{}l@{}}Strong federation-based \\ settlement network\end{tabular}} &
  \multicolumn{1}{l|}{\begin{tabular}[c]{@{}l@{}}Strong federation of \\ functionaries \\ maintain the network\end{tabular}} &
  \multicolumn{1}{l|}{\begin{tabular}[c]{@{}l@{}}Consensus secured by\\  specialized hardware\end{tabular}} &
  \begin{tabular}[c]{@{}l@{}}Wallet and \\ mining services\end{tabular} \\ \midrule
\multicolumn{1}{l|}{Loom Network \cite{loom} \mbox{}\hfill \checkmark} &
  \multicolumn{1}{l|}{Ethereum} &
  \multicolumn{1}{l|}{\begin{tabular}[c]{@{}l@{}}Delegated \\ proof \\ of stake\end{tabular}} &
  \multicolumn{1}{l|}{\begin{tabular}[c]{@{}l@{}}dApp platform with \\ interoperability capabilities\end{tabular}} &
  \multicolumn{1}{l|}{\begin{tabular}[c]{@{}l@{}}Support for a high number \\ of tokens\end{tabular}} &
  \multicolumn{1}{l|}{Closed source solution} &
  \begin{tabular}[c]{@{}l@{}}Integrations with \\ major blockchains\end{tabular} \\ \midrule
\multicolumn{1}{l|}{Zendoo \cite{zendo}} &
  \multicolumn{1}{l|}{Bitcoin} &
  \multicolumn{1}{l|}{Proof of stake} &
  \multicolumn{1}{l|}{Sidechain creation platform} &
  \multicolumn{1}{l|}{\begin{tabular}[c]{@{}l@{}}zk-Snark solution allows the \\ mainchain to verify the\\  sidechain without disclosing \\ sensitive information\end{tabular}} &
  \multicolumn{1}{l|}{\begin{tabular}[c]{@{}l@{}}zk-Snarks are \\ computationally\\  expensive\end{tabular}} &
  \begin{tabular}[c]{@{}l@{}}Further specification \\ of the protocol\end{tabular} \\ \midrule
\multicolumn{1}{l|}{RSK \cite{rsk} \mbox{}\hfill \checkmark} &
  \multicolumn{1}{l|}{Bitcoin} &
  \multicolumn{1}{l|}{DECOR+} &
  \multicolumn{1}{l|}{\begin{tabular}[c]{@{}l@{}}Federated sidechain, in which \\ RBTC is tethered to BTC\end{tabular}} &
  \multicolumn{1}{l|}{\begin{tabular}[c]{@{}l@{}}Merge mining allows \\ reutilization of work\end{tabular}} &
  \multicolumn{1}{l|}{\begin{tabular}[c]{@{}l@{}}Relies on PoW, \\ energetically inneficient\end{tabular}} &
  \begin{tabular}[c]{@{}l@{}}Decentralized bridge \\ with Ethereum\end{tabular} \\ \midrule
\multicolumn{1}{l|}{Blocknet \cite{blocknet} \mbox{}\hfill \checkmark} &
  \multicolumn{1}{l|}{Ethereum} &
  \multicolumn{1}{l|}{\begin{tabular}[c]{@{}l@{}}Proof \\ of stake\end{tabular}} &
  \multicolumn{1}{l|}{\begin{tabular}[c]{@{}l@{}}EVM-based blockchain \\ with interoperability capabilities\end{tabular}} &
  \multicolumn{1}{l|}{\begin{tabular}[c]{@{}l@{}}Blocknet protocol \\ allows trustless\\ blockchain interoperability\end{tabular}} &
  \multicolumn{1}{l|}{\begin{tabular}[c]{@{}l@{}}Currently limited \\ to digital assets\end{tabular}} &
  \begin{tabular}[c]{@{}l@{}}EOS/NEO/other \\ integrations\end{tabular} \\ \midrule
\multicolumn{7}{l}{} \\
\multicolumn{7}{l}{\begin{tabular}[c]{@{}l@{}}\checkmark our description was endorsed by the authors/team  \\ $\times$ not specified \\ $\ast$ although zk-Snarks are not a consensus algorithm, consensus on which operations were performed at each sidechain is obtained through a process that uses zk-Snarks \\ to generate proofs of sidechain state that, on its turn, generate certificate proofs for the mainchain\end{tabular}} \\
 &
   &
   &
   &
   &
   &
  
\end{tabular}%
}
\end{table*}
 
\emph{POA Network} encompasses an EVM-based blockchain as well as the {POA Bridge} \cite{POA}. The POA Bridge is a component that enables cross-application transactions with Ethereum, providing support for ERC-20 tokens. For instance, the POA20 token represents the POA token available to use on the Ethereum main network. The sidechain achieves consensus through proof of authority.

A newer feature from POA, \emph{Arbitrary Message Bridge},\footnote{https://docs.tokenbridge.net/amb-bridge/about-amb-bridge} allows transferring arbitrary data between  EVM-based chains (e.g., POA, Loom, Ethereum Classic). This feature can be used for cross-chain smart contract invocations. POA implemented a POA-based stable token, through the xDai chain.\footnote{https://www.poa.network/roadmap} POA is an open-source project.\footnote{https://github.com/poanetwork}

\emph{Elements}\footnote{https://elementsproject.org} is a sidechain-capable blockchain platform. \emph{Liquid} is a federated pegged sidechain \cite{peggedsidechains,confidential_assets}, based on Elements, relying on the concept of \emph{strong federations} \cite{liquid}. Strong federations introduce the concepts of a federated two-way peg, in which entities move assets between two chains. In strong federations, a role called block-signers maintains the consensus of the blockchain, while the watchmen realize cross-chain transactions. Software running on hardware security modules achieve consensus. Hardware security modules (HSMs) are physical computing devices that actively hides and protects cryptographic material, e.g., via limited network access and features that provide tamper evidence \cite{Rostami2014}. Moreover, a \emph{k-of-n} multi-signature scheme is also used to endorse block creation. 

Liquid supports several assets, including fiat currencies and cryptocurrencies, such as Bitcoin. When Bitcoins are pegged to the Liquid sidechain, they are backed by an L-BTC token, which represents one Bitcoin. 
The roadmap predicts updates to wallet and mining services\footnote{https://blockstream.com/2020/02/10/en-blockstream-2019-review-building-foundations/}. Liquid is an open-source project\footnote{https://github.com/Blockstream?q=liquid\&type=\&language=}.

\emph{Loom Network} is a dApp platform, which relies on sidechains connected to Ethereum, Binance Chain, and Tron \cite{loom}. Loom is a federated two-way peg, whereby a set of 21 validators and token delegators validate cross-asset transactions. Loom uses  \emph{Delegated Proof of Stake} (DPoS) as the consensus mechanism for transactions happening on the sidechain.

\emph{Proof of Stake} (PoS) is an alternative to PoW that aims to reduce energy consumption \cite{correia2019byzantine}. In PoS, the ability for nodes to append blocks to the ledger depends on their stake, that often depends on the amount of currency they own. In DPoS only a subset of the nodes participate in the consensus, which is based on PoS.

The roadmap predicts integration with more blockchain networks\footnote{https://medium.com/loom-network/5183ce02267}. Loom is open-source components\footnote{https://github.com/loomnetwork}. 

\emph{RSK} is a general-purpose smart contract platform pegged to the Bitcoin network that offers improvements in security and scalability of the latter \cite{rsk}, and the first sidechain solution in production (January 2018). It relies on a combination of a federated sidechain with an SPV. Each smart Bitcoin (RBTC), the native token of RSK, is tethered to one Bitcoin. 

In order to get RBTCs, a user has to send Bitcoin to a specific multi-signature address (an address controlled by several parties, through the several signatures) located at the Bitcoin network. That address is controlled by the RSK Federation, which is composed of several stakeholders. The federation members use hardware security modules. By leveraging HSMs, each validator can protect its private keys, and enforce the transaction validation protocol \cite{rsk}. Moreover, an additional layer of security that prevents any corrupt collaborator from forcing the HSM from each stakeholder to sign a fake peg-out transaction: nodes automatically follow the blockchain with the highest cumulative proof of work. 

After the transaction is finished, a proof of transfer (via SPV) is generated and given as an input to a smart contract on the RSK network, called the bridge contract. The bridge contract then sends a corresponding amount of RBTC tokens to the address present at the RSK network that corresponds to the Bitcoin address sending Bitcoin to the RSK address. RSK has a virtual machine that executes smart contracts in the Bitcoin network. 

RSK uses consensus mechanism designated DECOR+  and a technique called merge-mining, which allows users to mine in both the RSK and Bitcoin networks without performance penalties. RSK introduces shrinking-chain scaling, a technique to compress blocks after they are mined. 

The RSK roadmap predicts the development of a decentralized bridge between RSK and Ethereum\footnote{https://blog.rsk.co/noticia/hawkclient-building-a-fully-decentralized-bridge-between-rsk-and-ethereum/}.
RSK is an open-source project\footnote{https://github.com/rsksmart}.

\emph{{Blocknet}} is blockchain based on  PoS that includes a protocol for interoperability among public and private blockchains \cite{blocknet}. 
At its core, Blocknet has several components: the XBridge, XRouter, and XCloud \cite{blocknet_docs,blocknet_comp}. XBridge allows exchanging digital assets, powered by a set of APIs, and relying on SPV. XRouter actuates as an inter-chain address system, providing lookup capabilities to the network. XCloud, relying on XRouter, provides a decentralized oracle network, that can be used to obtain trusted data.

\subsection{\underline{Notary Schemes}}
\label{sec:crypto_notaries_appendix}

Despite this evolution, commonly used notary schemes are centralized cryptocurrency Exchanges (e.g., Binance, Coinbase, BKEX, LBank, Bilaxy, BitForex). Most exchanges are centralized (237), against 22 decentralized exchanges listed by CryptoCompare, at the time of writing.\footnote{https://www.cryptocompare.com/exchanges/\#/overview}

\begin{figure}[h]
    \centering
    \includegraphics[scale=0.30]{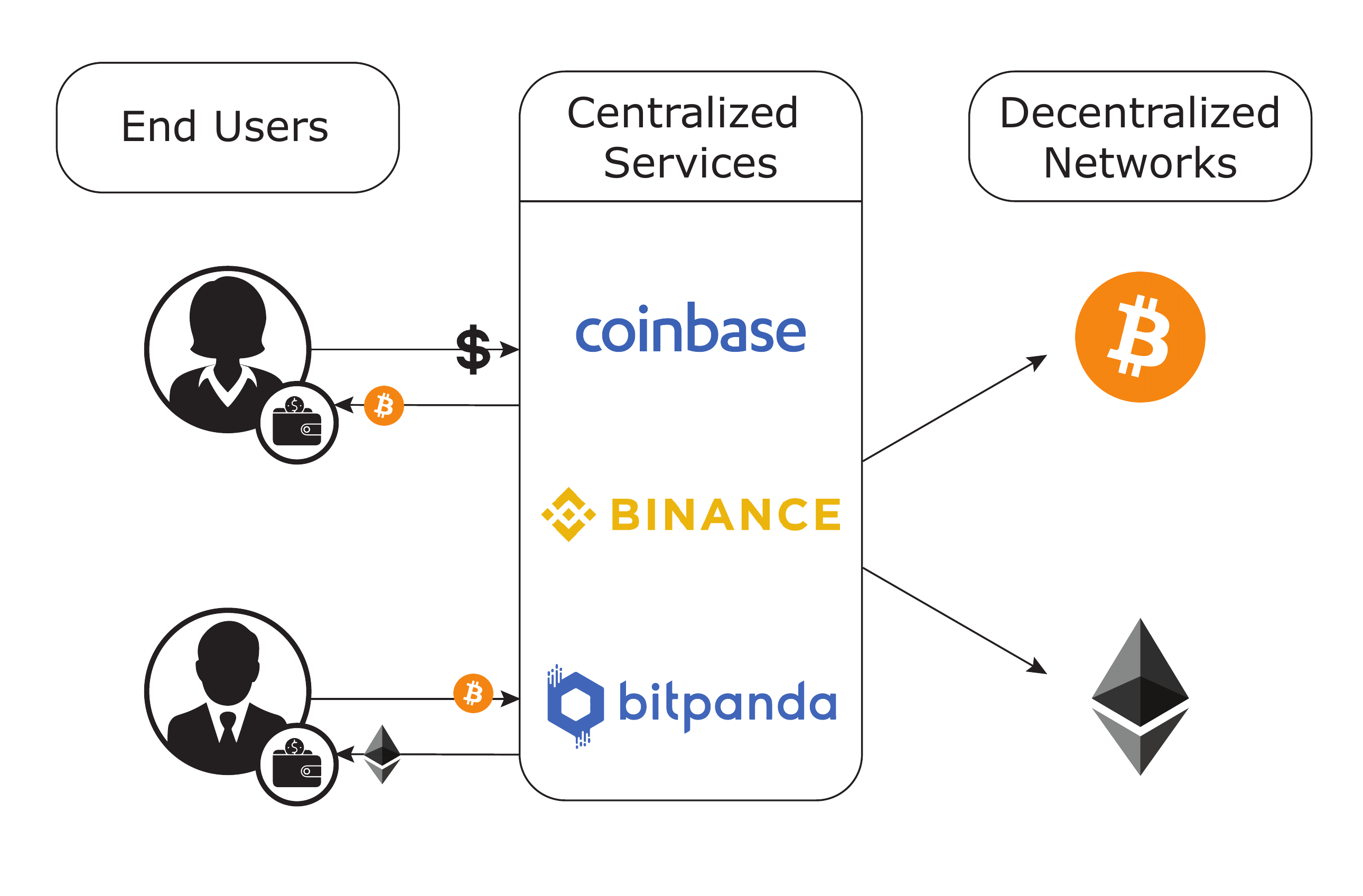}
    \caption{Alice and Bob buy cryptocurrencies via a centralized exchange. The assets are held by a custodial wallet.}
    \label{fig:crypto_cexchange}
\end{figure}

Figure \ref{fig:crypto_cexchange} represents the task of a user acquiring cryptocurrencies via centralized exchanges. Users buy cryptocurrencies with fiat currencies, and are credited the bought assets on their respective wallets, owned by the exchange, i.e., the exchange also known as \emph{custodial wallets}. Exchanges acquire such cryptocurrencies directly on the network, or via an intermediary, and provide arbitrage services. 

Although a simple way to obtain cryptocurrencies, some attacks have been conducted to exchanges, leading to loss of very large cryptocurrency sums \cite{ddosex}.

\begin{figure}
    \centering
    \includegraphics[scale=0.28]{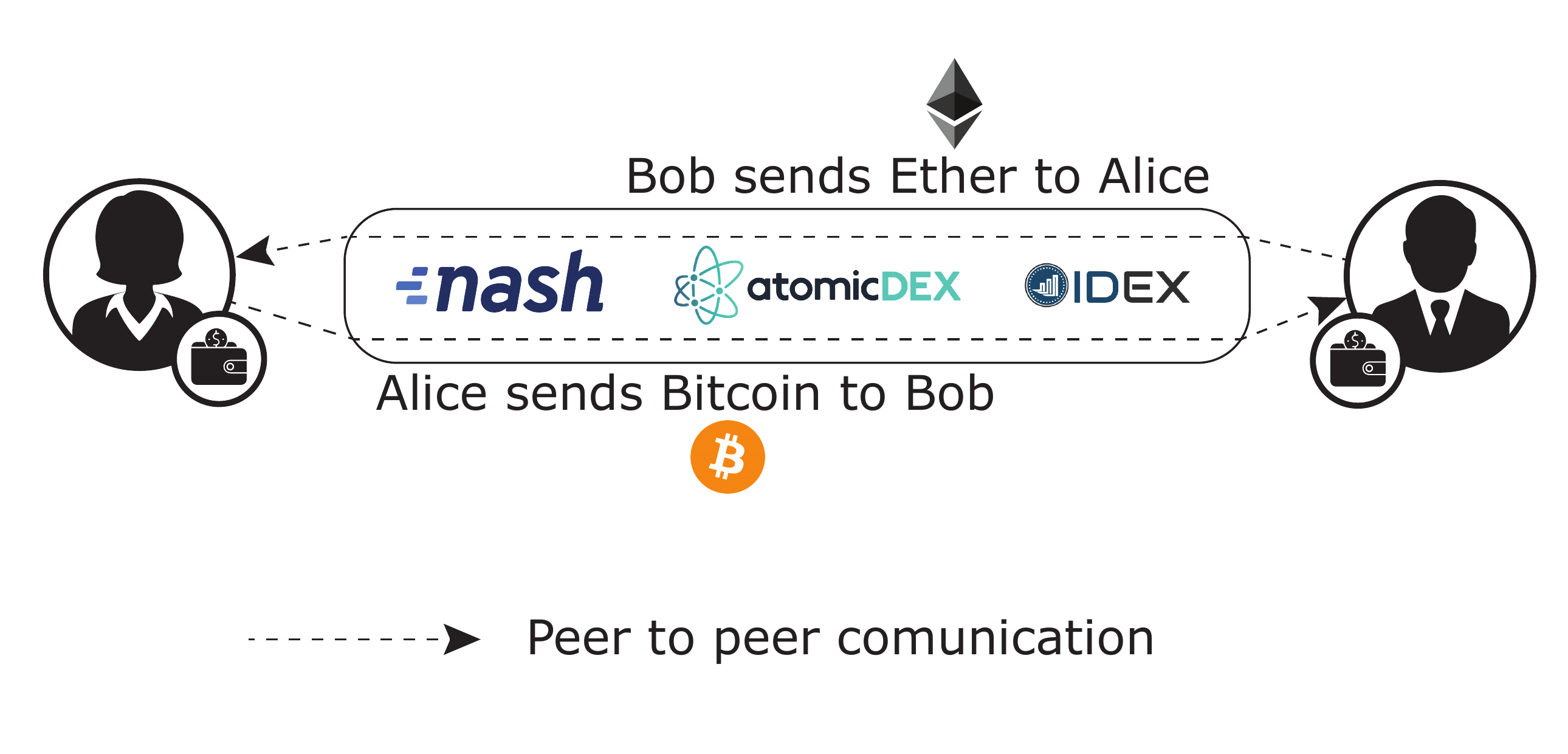}
    \caption{Alice can send cryptocurrencies directly to Bob, and vice-versa. Each user holds their private keys. The exchange is a facilitator of the transactions.}
    \label{fig:crypto_dexexchange}
\end{figure}

Decentralized exchanges can be implemented with hashed timelocks (see Section \ref{subsec:crypto_hashed}), or other technologies (see Section \ref{sec:use_cases}). Figure \ref{fig:crypto_dexexchange} depicts users exchanging assets via a decentralized exchange (e.g., Nash, AtomicDEX, IDEX). When trading via a decentralized exchange, users typically do not disclose their private keys, eliminating the single point of failure inherent with centralized exchanges.

\emph{Agent Chain} is a project aiming to exchange assets between blockchains using a multi-signature scheme \cite{agentchain}. A trader maps the possessed assets to AgentChain, which combines several trading operators in a trading group. Members of that group generate an account using a multi-signature, to serve as a deposit pool, containing the assets. Tokens are then locked. An arbitration mechanism is introduced in case of a malicious trading group. 

\subsection{\underline{Hashed Time-Locks}}
\label{subsec:crypto_hashed_appendix}

\emph{Black et al.} propose the concept of \emph{atomic loans}, based on atomic swaps \cite{Black2019}. Atomic loans allow market participants to create loans in a trustless manner, enabling liquidity. The process of  atomic loans is rooted in the foundations of HTLCs and has several phases: the loan period, in which the loan withdrawal and repayment process is handled; the bidding period; the seizure period; and the refund period. The last four phases happen in case the loan is not repaid in due time during the bidding period phase.

\begin{table*}[]
\centering
\caption{Comparison of \emph{Hash Lock Time Contract} solutions}
\label{tab:crypto_htlc}
\resizebox{\textwidth}{!}{%
\begin{tabular}{@{}lllllll@{}}
\toprule
\multicolumn{1}{c}{\textbf{Reference}} &
  \multicolumn{1}{c}{\textbf{\begin{tabular}[c]{@{}c@{}}Supported\\ Chains\end{tabular}}} &
  \multicolumn{1}{c}{\textbf{Architecture}} &
  \multicolumn{1}{c}{\textbf{Summary}} &
  \multicolumn{1}{c}{\textbf{Strong points}} &
  \multicolumn{1}{c}{\textbf{Weak points}} &
  \multicolumn{1}{c}{\textbf{Roadmap}} \\ \midrule
\multicolumn{1}{l|}{Black et al.~\cite{Black2019}} &
  \multicolumn{1}{l|}{$\times$} &
  \multicolumn{1}{l|}{Lender, borrower} &
  \multicolumn{1}{l|}{\begin{tabular}[c]{@{}l@{}}Leverage HTLC to provide \\ fiat/stablecoinaccess for \\ cryptocurrency holders\end{tabular}} &
  \multicolumn{1}{l|}{Decentralized solutions} &
  \multicolumn{1}{l|}{\begin{tabular}[c]{@{}l@{}}Inneficient (atomic swaps);\\ requires over-collateralization\end{tabular}} &
  $\times$ \\ \midrule
\multicolumn{1}{l|}{Wanchain \cite{Lu2017} \mbox{}\hfill \checkmark} &
  \multicolumn{1}{l|}{\begin{tabular}[c]{@{}l@{}}Bitcoin, \\ Ethereum\end{tabular}} &
  \multicolumn{1}{l|}{\begin{tabular}[c]{@{}l@{}}Vouchers, validators,\\ storemen (Wan protocol)\end{tabular}} &
  \multicolumn{1}{l|}{\begin{tabular}[c]{@{}l@{}}Connects major \\ currency exchanges\end{tabular}} &
  \multicolumn{1}{l|}{\begin{tabular}[c]{@{}l@{}}Cross-Chain Bridge \\ Node Staking Rewards\end{tabular}} &
  \multicolumn{1}{l|}{\begin{tabular}[c]{@{}l@{}}Storemen are not \\ completely decentralized\end{tabular}} &
  \begin{tabular}[c]{@{}l@{}}General \\ interoperability\end{tabular} \\ \midrule
\multicolumn{1}{l|}{LN \cite{LN}} &
  \multicolumn{1}{l|}{Bitcoin} &
  \multicolumn{1}{l|}{\begin{tabular}[c]{@{}l@{}}Relies on \\ multi-signature \\ channel addresses\end{tabular}} &
  \multicolumn{1}{l|}{\begin{tabular}[c]{@{}l@{}}High volume, low latency\\  micropayment enabler\end{tabular}} &
  \multicolumn{1}{l|}{\begin{tabular}[c]{@{}l@{}}Increases Bitcoin \\ performance, \\ solution in production\end{tabular}} &
  \multicolumn{1}{l|}{Timelock expiration exploits} &
  $\times$ \\ \midrule
\multicolumn{1}{l|}{Komodo \cite{komodo}\mbox{}\hfill \checkmark} &
  \multicolumn{1}{l|}{\begin{tabular}[c]{@{}l@{}}Bitcoin, \\ Ethereum\end{tabular}} &
  \multicolumn{1}{l|}{\begin{tabular}[c]{@{}l@{}}Liquidity provider nodes,\\  buyers, sellers\end{tabular}} &
  \multicolumn{1}{l|}{\begin{tabular}[c]{@{}l@{}}Atomic swap \\ decentralized exchange\end{tabular}} &
  \multicolumn{1}{l|}{\begin{tabular}[c]{@{}l@{}}Provides a framework \\ for cross-chain atomic swaps\end{tabular}} &
  \multicolumn{1}{l|}{\begin{tabular}[c]{@{}l@{}}All products are \\ ``highly experimental"\end{tabular}} &
  \begin{tabular}[c]{@{}l@{}}Derivative tokens\\ on the decentralized \\ exchange\end{tabular} \\ \midrule
\multicolumn{1}{l|}{COMIT \cite{comit}} &
  \multicolumn{1}{l|}{\begin{tabular}[c]{@{}l@{}}Bitcoin, \\ Ethereum\end{tabular}} &
  \multicolumn{1}{l|}{Traders, COMIT protocol} &
  \multicolumn{1}{l|}{\begin{tabular}[c]{@{}l@{}}Open protocol facilitating \\ trustless cross-blockchain \\ applications\end{tabular}} &
  \multicolumn{1}{l|}{\begin{tabular}[c]{@{}l@{}}Adds negotiation phase \\ to the atomic swap\end{tabular}} &
  \multicolumn{1}{l|}{\begin{tabular}[c]{@{}l@{}}Does not support \\ negotiation protocols\end{tabular}} &
  \begin{tabular}[c]{@{}l@{}}Protocol for privacy\\  preserving swaps\end{tabular} \\ \midrule
 &
   &
   &
   &
   &
   &
   \\
\multicolumn{7}{l}{\checkmark our description was endorsed by the authors/team} \\
$\times$ not specified &
   &
   &
   &
   &
   &
  
\end{tabular}%
}
\end{table*}

\emph{Wanchain} aims to provide deposit and loan services with cryptocurrencies \cite{Lu2017}. When a transfer request is sent to Wanchain, it issues the corresponding tokens in the existing smart contract that locks them on the target blockchain. Wanchain's validator nodes receive such request, verify that a transaction has been placed into the target blockchain, and creates a representation of the tokens to be transferred (a new smart contract token, analogous to the original currency). 

When a party that has a representation of the original tokens wants to send them to a third party, the locked assets in a smart contract are released to the beneficiary of the transaction. As Wanchain creates a representation of tokens as a means of exchanging assets, we can consider that such a solution is a notary scheme, although decentralized (several validator nodes operate the network). Wanchain's architecture includes the following nodes: vouchers, the cross-chain transaction proof nodes; validators, the verification nodes; and storeman, the locked account management nodes. Vouchers check whether a transaction has been confirmed on a source blockchain. Validators verify the asset registry from the source blockchain: in case it is a new asset, it is registered and added into the registry. Storeman manages locked accounts, facilitating cross-chain transactions. An incentive mechanism rewards the participants to perform their functions. 
More recently, Wanchain is working towards more general interoperability, by promoting cross-chain integration with enterprise blockchains and supporting Web Assembly (WASM) smart contracts \cite{wanchainroadmap}.

\emph{{COMIT}} is a protocol allowing for atomic swaps, based on HLTCs \cite{comit}. COMIT defines several atomic swap protocols that support different cryptocurrencies and tokens, such as HAN (HTLCs for Assets that are Native to the ledger), HErc20 (HTLCs for the Erc20 asset), and HALight (HTLCs for Assets on the Lightning ledger). COMIT nodes can trade Bitcoin for Ether or ERC-20 tokens. The COMIT protocol\footnote{https://github.com/comit-network/comit-rs/} allows one to exchange assets directly with another user (e.g., Bitcoin for Ether).

Apart from HLTCs and sidenchains, there is a set of approaches that share characteristics from several subcategories presented, for instnace, using distributed private key schemes or collateralization with HLTCs. \emph{Distribute private key approaches} rely on the distribution of users' and organizations' private keys, i.e., in splitting each private key in a set of parts \cite{Deng2018}. This leads to distributing the control of assets among several parties. Such schemes can be used to implement decentralized two-way pegs, as well as decentralized notaries. Other approaches combine sidechains and protocols based on escrow parties, relying on smart contracts. An \emph{escrow} is an arrangement in which a third party regulates a transaction or group of transactions between two parties. An escrow typically holds assets (e.g., cryptocurrency) from one of the parties that serves as the collateral of a transaction (assets pledged by a borrower to protect the interests of the lender). Some of those solutions include:

\emph{Tokrex} enables the exchange of cryptocurrencies between different blockchains in a decentralized way, by leveraging the concept of \emph{meta-swap} \cite{tokrex}. A meta swap happens when a sender transmits his private key instead of signing an on-chain transaction. For that, a domain-specific language, Tokrex TLQ, allows developers to write cross-chain applications that run on a decentralized network infrastructure. Tokrex relies on escrow nodes distributing the generated keys, a modularized distributed key generator, cross-chain swaps, and an Incentivization scheme to keep the escrow and validator nodes honest.

\emph{Fusion} is an interoperable blockchain, focused on financial use cases \cite{fusion}. Fusion owns a proprietary technology, DCRMS (Distributed Control Rights Management System), which allows users to lock-in and lock-out assets across blockchains. DCRMS is a decentralized custodian model, which tries to prevent private keys from being a single point of failure: asset control is decentralized along network nodes, instead of them relying on individuals and centralized organizations. The distributed storage and generation of a private key keeps a single entity of obtaining full control of an asset. Fusion supports any chain that uses EcDSA signatures, which includes Bitcoin, Ethereum and other EVM-based blockchains.

\begin{table*}[]
\centering
\caption{Comparison of \emph{Alternative} solutions}
\label{tab:crypto_others}
\resizebox{\textwidth}{!}{%
\begin{tabular}{@{}lllllll@{}}
\toprule
\multicolumn{1}{c}{\textbf{Reference}} &
  \multicolumn{1}{c}{\textbf{\begin{tabular}[c]{@{}c@{}}Main Supported \\ Chains\end{tabular}}} &
  \multicolumn{1}{c}{\textbf{Architecture}} &
  \multicolumn{1}{c}{\textbf{Summary}} &
  \multicolumn{1}{c}{\textbf{Strong points}} &
  \multicolumn{1}{c}{\textbf{Weak points}} &
  \multicolumn{1}{c}{\textbf{Roadmap}} \\ \midrule
\multicolumn{1}{l|}{\multirow{3}{*}{Tokrex \cite{tokrex} \mbox{}\hfill \checkmark}} &
  \multicolumn{1}{l|}{\multirow{3}{*}{$\times$}} &
  \multicolumn{1}{l|}{\multirow{3}{*}{\begin{tabular}[c]{@{}l@{}}Validation and escrow nodes, \\ distributed key generation\end{tabular}}} &
  \multicolumn{1}{l|}{\multirow{3}{*}{\begin{tabular}[c]{@{}l@{}}Cryptocurrency exchange \\ enabling meta-swaps\end{tabular}}} &
  \multicolumn{1}{l|}{\multirow{3}{*}{\begin{tabular}[c]{@{}l@{}}Allows ``real time" \\ value exchange\end{tabular}}} &
  \multicolumn{1}{l|}{\multirow{3}{*}{\begin{tabular}[c]{@{}l@{}}Both sender and receiver \\ know the private key used \\ for asset transfer\end{tabular}}} &
  \multirow{3}{*}{$\times$} \\
\multicolumn{1}{l|}{} & \multicolumn{1}{l|}{} & \multicolumn{1}{l|}{} & \multicolumn{1}{l|}{} & \multicolumn{1}{l|}{} & \multicolumn{1}{l|}{} &  \\
\multicolumn{1}{l|}{} & \multicolumn{1}{l|}{} & \multicolumn{1}{l|}{} & \multicolumn{1}{l|}{} & \multicolumn{1}{l|}{} & \multicolumn{1}{l|}{} &  \\ \midrule
\multicolumn{1}{l|}{\multirow{3}{*}{Fusion \cite{fusion} \mbox{}\hfill \checkmark}} &
  \multicolumn{1}{l|}{\multirow{3}{*}{Ethereum}} &
  \multicolumn{1}{l|}{\multirow{3}{*}{\begin{tabular}[c]{@{}l@{}}FUSION distributed control\\  rights services\end{tabular}}} &
  \multicolumn{1}{l|}{\multirow{3}{*}{\begin{tabular}[c]{@{}l@{}}Distributed storage of a\\  private key and \\ cryptoasset mapping\end{tabular}}} &
  \multicolumn{1}{l|}{\multirow{3}{*}{\begin{tabular}[c]{@{}l@{}}Distributes trust and \\ responsability of\\ managing private keys\end{tabular}}} &
  \multicolumn{1}{l|}{\multirow{3}{*}{\begin{tabular}[c]{@{}l@{}}Does not provide\\  instant atomic swaps\end{tabular}}} &
  \multirow{3}{*}{\begin{tabular}[c]{@{}l@{}}Decentralized oracle \\ services\end{tabular}} \\
\multicolumn{1}{l|}{} & \multicolumn{1}{l|}{} & \multicolumn{1}{l|}{} & \multicolumn{1}{l|}{} & \multicolumn{1}{l|}{} & \multicolumn{1}{l|}{} &  \\
\multicolumn{1}{l|}{} & \multicolumn{1}{l|}{} & \multicolumn{1}{l|}{} & \multicolumn{1}{l|}{} & \multicolumn{1}{l|}{} & \multicolumn{1}{l|}{} &  \\ \midrule
\multicolumn{1}{l|}{\multirow{3}{*}{Sai et al. \cite{sai2019}}} &
  \multicolumn{1}{l|}{\multirow{3}{*}{Ethereum}} &
  \multicolumn{1}{l|}{\multirow{3}{*}{Neutral observers}} &
  \multicolumn{1}{l|}{\multirow{3}{*}{\begin{tabular}[c]{@{}l@{}}Neutral observers monitor \\ transactions to avoid \\ double spending\end{tabular}}} &
  \multicolumn{1}{l|}{\multirow{3}{*}{\begin{tabular}[c]{@{}l@{}}Trustees can choose \\ any node to be an \\ observer\end{tabular}}} &
  \multicolumn{1}{l|}{\multirow{3}{*}{\begin{tabular}[c]{@{}l@{}}Trustees that choose \\ observers are assumed \\ to be honest\end{tabular}}} &
  \multirow{3}{*}{\begin{tabular}[c]{@{}l@{}}Behaviour of malicious \\ trustee\end{tabular}} \\
\multicolumn{1}{l|}{} & \multicolumn{1}{l|}{} & \multicolumn{1}{l|}{} & \multicolumn{1}{l|}{} & \multicolumn{1}{l|}{} & \multicolumn{1}{l|}{} &  \\
\multicolumn{1}{l|}{} & \multicolumn{1}{l|}{} & \multicolumn{1}{l|}{} & \multicolumn{1}{l|}{} & \multicolumn{1}{l|}{} & \multicolumn{1}{l|}{} &  \\ \midrule
\multicolumn{1}{l|}{XClaim \cite{xclaim}} &
  \multicolumn{1}{l|}{\begin{tabular}[c]{@{}l@{}}Bitcoin, \\ Ethereum\end{tabular}} &
  \multicolumn{1}{l|}{\begin{tabular}[c]{@{}l@{}}Requester, sender, \\ receiver, redeemer, \\ the backing vault, \\ issuing smart contract\end{tabular}} &
  \multicolumn{1}{l|}{\begin{tabular}[c]{@{}l@{}}HTLC-based trustless \\ protocol that manages \\ crpytocurrency-backed assets\end{tabular}} &
  \multicolumn{1}{l|}{\begin{tabular}[c]{@{}l@{}}Good performance \\ compared to \\ traditional HLTCs\end{tabular}} &
  \multicolumn{1}{l|}{\begin{tabular}[c]{@{}l@{}}Over-collateralization \\ can lead to locked funds\end{tabular}} &
  \begin{tabular}[c]{@{}l@{}}Asymmetric and \\ non-fungible \\ cryptocurrency-backed \\ assets\end{tabular} \\ \midrule
\multicolumn{1}{l|}{\begin{tabular}[c]{@{}l@{}}DeXTT\\ \cite{dextt, tast_paper2, tast_paper7, tast_paper6}\end{tabular}} &
  \multicolumn{1}{l|}{Ethereum} &
  \multicolumn{1}{l|}{\begin{tabular}[c]{@{}l@{}}PBTs, claim-first transactions,\\ deterministic witnesses\end{tabular}} &
  \multicolumn{1}{l|}{\begin{tabular}[c]{@{}l@{}}A protocol implementing \\ eventual consistency for \\ cross-blockchain \\ token transfers.\end{tabular}} &
  \multicolumn{1}{l|}{\begin{tabular}[c]{@{}l@{}}Ensures eventual \\ consistency of balances\\ across blockchains\end{tabular}} &
  \multicolumn{1}{l|}{\begin{tabular}[c]{@{}l@{}}Veto contest poses strict\\ requirements towards \\ signed PoIs\end{tabular}} &
  \begin{tabular}[c]{@{}l@{}}DeXTT  implementation\\ on OmniLayer\end{tabular} \\ \midrule
\multicolumn{1}{l|}{XChain \cite{cctxnarges}} &
  \multicolumn{1}{l|}{Ethereum} &
  \multicolumn{1}{l|}{\begin{tabular}[c]{@{}l@{}}Directed graph, 3PP: \\ contract creation, secret \\ release, and secret relay\end{tabular}} &
  \multicolumn{1}{l|}{\begin{tabular}[c]{@{}l@{}}A 3PP for general \\ cross-chain transactions\end{tabular}} &
  \multicolumn{1}{l|}{\begin{tabular}[c]{@{}l@{}}Generates custom smart\\ contracts for performing\\ cross atomic swaps\end{tabular}} &
  \multicolumn{1}{l|}{\begin{tabular}[c]{@{}l@{}}Only applicable to \\ Ethereum\end{tabular}} &
  $\times$ \\ \midrule
\multicolumn{7}{l}{}                                                                                                                             \\
\multicolumn{7}{l}{\begin{tabular}[c]{@{}l@{}}\checkmark our description was endorsed\\ $\times$ not defined\end{tabular}}                      
\end{tabular}%
}
\end{table*}

\emph{TAST (Token Atomic Swap Technology)} is a project\footnote{https://dsg.tuwien.ac.at/projects/tast/} that aims 
to create the first multi-blockchain token system \cite{pantos}. TAST includes several components explained in a set of documents. 

In one of these documents, the authors present \emph{claim-first transactions}, a protocol for decentralized blockchain asset transfers.~\cite{tast_paper7}. The protocol includes the role of {witness}, who verifies cross-blockchain transactions and is rewarded for that.
Another document presents the notion of \emph{Proof of Intent} (PoI) \cite{tast_paper6}, a cryptographic construction that implements claim first transactions. The notion of deterministic witnesses is introduced as the mechanism for assigning rewards to parties observing claim-first transactions.  

In \cite{tast_paper5}, the authors present
the design of a blockchain interoperability solution based on an atomic cross-chain token transfer protocol.
Other documents summarize the work developed \cite{tast_paper3,tast_paper4} and discuss the requirements for more efficient cross-blockchain token transfers. 

In \cite{tast_paper2}, the authors propose an incentive structure for blockchain relays, presenting an enhanced prototype based on SPV. The presented solution showed that the solution incurred in high operation costs. The most recent whitepaper, \cite{tast_paper1}, introduces optimizations that reduce such costs. This paper shows the applicability of a cross-blockchain token, relying on token incentives and simplified payment verification.

\emph{DeXTT} is an atomic cross-chain token  transfer protocol that migrates assets --  Pan-Blockchain Tokens (PBTs) -- that can exist in different blockchains simultaneously  \cite{dextt}. DeXTT is part of the TAST project.

DeXTT provides eventual consistency of asset balances across blockchains. Eventual consistency, guarantees that eventually all accesses to an item that has not been updated after the access request will return the latest value.
To achieve eventual consistency, the authors use a technique called \emph{claim first transactions} \cite{tast_paper7}, and observers. The \emph{claim transaction}, immediately claims the asset before it is marked as spent, through a  \emph{SPEND transaction}. The party creating a SPEND transaction is called a \emph{witness}, the rewarded party. Observers observe a transfer and propagate such information across blockchains. 
As several observers might compete for a reward, a solution called \emph{deterministic witnesses} is proposed \cite{tast_paper6,dextt}. Deterministic witnesses solve the problem of assigning witness awards by defining a witness context, whereby observers participate.

A cross-blockchain asset transfer starts with a \emph{transfer initiation}. In a transfer initiation, a wallet$_{a}$ expresses the intent of transferring an asset to a wallet$_{b}$, by signing a transaction with its private key. Wallet$_{b}$ then countersigns the transaction, using its private key, (creating a PoI). A PoI proves that a transfer is authorized by both the sender and the receiver. After that, the receiver can then publish the PoI using a \emph{CLAIM} transaction, used to redeem the assets. Only one PoI from a source wallet is valid at each time, eliminating double-spends. 

Right after a PoI is published on a blockchain$_{a}$, the balance of both wallets has not been updated. In order to propagate this information to the other blockchains, in particular blockchain$_{b}$, the protocol follows the \emph{witness contest} phase. Here, observers become contestants that propagate the PoI to other blockchains, through a \emph{CONTEST} transaction. After that, in the \emph{deterministic witness selection} phase, the destination wallet, wallet$_{b}$, posts a \emph{FINALIZE} transaction on each blockchain, finalizing the contest and awarding an observer. The double-spending problem is eliminated via \emph{VETO} transactions, which can be called by any party, and discloses conflicting PoI (e.g., a source wallet tries to send more assets than it owns to several destination wallets).  

DeXTT tolerates blockchain failures, as long as at least one blockchain remains functional. It is meant to be a blockchain agnostic solution, but the most straightforward framing is within public blockchains. The authors presented a proof of concept using Solidity\footnote{https://github.com/pantos-io/dextt-prototype}.

\emph{XChain} includes a 
three-phase-protocol that generalizes atomic cross-chain swaps, in which two entities, the leaders and the followers exchange assets \cite{cctxnarges}. Hashed timelock contracts are leveraged to resolve the order of issuing contracts and reedeming locked funds from smart contracts. Nodes that create the HLTCs are called leaders, which first release the secrets; followers execute transactions that react to the leaders' actions (i.e., when a leader shares the secret of the HTLC to a follower, the follower unlocks its smart contract, and receives funds from other entity, by sharing the received secret). This solution is based on HLTCs and a protocol that guarantees end-to-end and uniformity properties.

\section{Blockchain of Blockchains}
\label{a:be}
We now describe some of sidechain solutions we identified in the literature. Research on Blockchain of Blockchains required substantial \emph{ad-hoc} research, including blog posts, roadmaps, and update announcements, for us to build an updated understanding regarding the latest capabilities of each blockchain engine. 


The \emph{Polkadot} network has several entities engaged in handling transactions: \emph{collator}, \emph{validator}, \emph{nominator}, and \emph{fisherman}. 
Collators produce proofs  for the validators. Transactions are then executed and aggregated in blocks. There is the possibility of collators \emph{to pool}, to coordinate and share the rewards coming from creating blocks on the parachains they actuate.
Validators produce and finalize blocks on the relay chain. The validator role is contingent on a stake that is put on hold to foment good behavior. Validators who misbehave can have their block rewards denied or, in case of recurrence, have their security bond confiscated. Validators are the equivalent to groups of cooperating miners that share block rewards proportionally to their contribution (mining pools) on PoW systems (e.g., Bitcoin). 
Nominators provide their own stake to validators, whereby sharing the rewards and incurring in potential slashing, in case of misbehaving.
Fishermen get bounties for reporting validators' misbehavior, such as helping to ratify an invalid block.
 
\begin{figure}[h]
    \centering
    \includegraphics[scale=0.30]{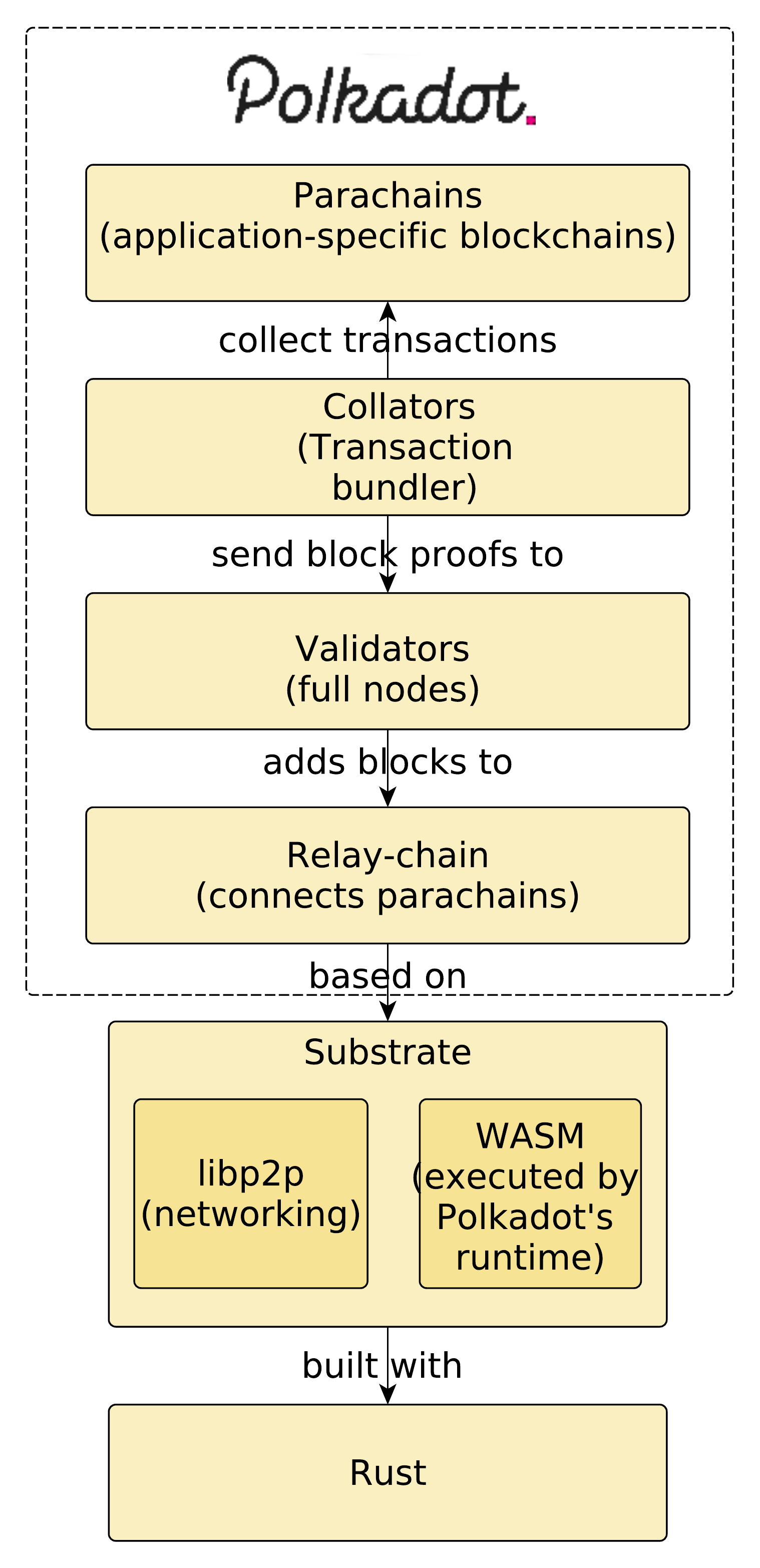}
    \caption{Polkadot's stack \cite{Wood2017,polkadot_research}}
    \label{fig:polkadot_stack}
\end{figure}

Figure \ref{fig:polkadot_stack} depicts the several components constituting Polkadot. Polkadot's relay chain uses Substrate. Polkadot's state machine is compiled to WASM, a virtual environment that can execute the state transition functions \cite{polkadot_research}. Libp2p is a network library for peer-to-peer applications, written in the Rust programming language. Parachains run the application logic, creating transactions as needed. Collators group those transactions and redirect them to Validators, who then deem blocks as valid or invalid. After that, the valid ones are added to the relay-chain.

Polkadot uses the DOT token as an incentive for nodes to behave correctly. DOT has several purposes: (i) decentralize governance (i.e., protocol updates), (ii) operation (i.e., rewarding good actors), and (iii) bonding (i.e., adding new parachains).  

Polkadot's relay chain achieves consensus using BABE and GRANDPA \cite{polkadot_consensus}. BABE is the block production algorithm, and GRANDPA is the finalizing algorithm. To determine a set of validators, Polkadot uses  selection based on PoS, designated Nominated Proof-of-Stake (\emph{NPoS}). Allying NPoS with the rewarding mechanism helps to diminish the impact of attacks such as short-range attack (when a validator attempts to ratify both branches of a fork) or the nothing-at-stake attack (where the risk of simultaneously validating several forks is exploited).
The roadmap comprises the launch of the main network\footnote{https://wiki.polkadot.network/docs/en/learn-roadmap}.

\begin{figure}[h]
    \centering
    \includegraphics[scale=0.30]{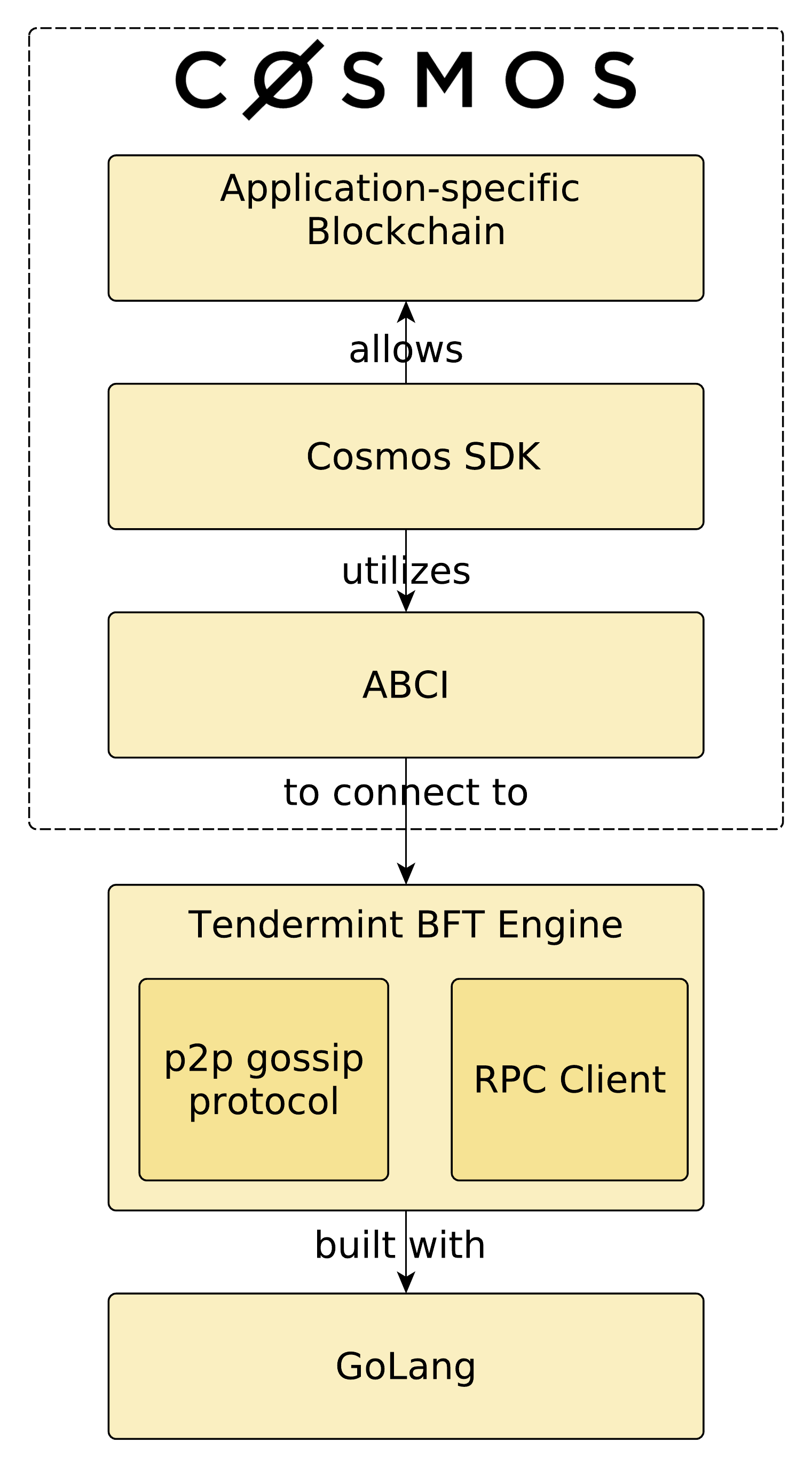}
    \caption{Cosmos Network's stack \cite{Kwon2016}}
    \label{fig:cosmos_stack}
\end{figure}

\emph{Cosmos} is another popular Blockchain of Blockchains. Figure \ref{fig:cosmos_stack} gives a general overview on the Cosmos Network stack. Wrappers can be developed to allow the usage of other programming languages. The applicational layer can be developed with the Cosmos SDK, a framework. This layer connects to the Tendermint BFT Engine (the component responsible for consensus).

Cosmos was limited to asset token on its original inception, now it supports arbitrary data transfers. For CC-Txs, the \emph{relayer} pays a transaction fee on behalf of the transaction sender. The relayer can whitelist any type of financial incentives to keep CC-Txs free.

In Cosmos, validators process blocks of transactions. Validators need to stake ATOM tokens to process blocks and earn transaction fees. Delegators can offload transaction processing to validators, and earn transaction fees.
As a way to promote an open-governance model, participants (e.g., validators and delegators) can hold the ATOM token and vote on  proposals that can change the parameters of the system. Decisions about the network governance, to vote, validate, or delegate transaction validation to other validators are made as a function of how many Atoms are held, similarly to a PoS view. Atoms can also be used to pay transaction fees. 

In Cosmos each zone is sovereign, i.e., it can define, for instance, authentication of accounts and transactions, on-chain governance proposals and voting, validator punishment mechanisms, fee distribution and staking token provision distribution, and creation of new units of staking token.

\emph{ARK} utilizes smart bridges to make instances of its platform interoperable \cite{ark2019}. A smart bridge has two components. The first, \emph{Protocol-Specific SmartBridge} (or \emph{bridgechain}), achieves inter-blockchain communication, by interconnecting the various chains based on ARK. The \emph{Protocol-Agnostic SmartBridge} achieves communication between blockchains that use different consensus mechanisms.

ARK's public network (or the ARK main blockchain) provides the foundation for other blockchains to issue and read transactions. Forging delegates are the entities that create blocks of transactions, analogous to miners in the Bitcoin blockchain. 

The consensus mechanism is a modified version of \emph{Delegated Proof-of-Stake} (DPoS). Holders of the ARK token vote to elect the top 51 delegates, who are randomly chosen to secure the network by validating transactions. By fixing the number of delegators (or forging nodes) at 51, the ``ARK main net strikes a balance between decentralization and performance''. The ARK token is also used to pay cross-chain transaction fees, which can be triggered by smart contracts, coded languages such as JavaScript, Go, Java, and C\#.

The ARK Contract Execution Services (ACES) has ``demonstrated two-way transfers between ARK and Bitcoin, Litecoin, and Ethereum, including issuing smart contracts from ARK to Ethereum, regardless of the underlying protocols''. While the ARK project defends cross-blockchain interoperability, ACES is on its inception. ACES can only provide interoperability on an \emph{ad hoc} basis. Connectors have to be implemented to connect ARK to other blockchains. Furthermore, ACES is that it is not entirely decentralized, as intermediary nodes are necessary to achieve interoperability. ARK plans to add several features to its platform\footnote{https://ark.io/roadmap}, such as integrating HLTCs to provide ARK bridgechains atomic swap capabilities. ARK is a proprietary solution -- it is not open-source. All ARK blockchains are powered by the ARK platform.

\emph{AION} was originally an ERC-20 token implemented on Ethereum \cite{aion2017}. Later, it evolved to a PoS blockchain system designed to provide the foundation for ``custom blockchain architectures''. A token bridge was built to swap tokens from the Ethereum blockchain to the AION blockchain. AION-compliant blockchains communicate through CC-Txs, issued by participating networks and routed by connecting networks. CC-Txs are created and processed on a source blockchain and routed by bridges. Bridges connect participating networks with connecting networks. 

Bridges would sign and broadcast CC-Txs upon payment of a fee and the validation by the source network. They would act as observers, reporting state changes via Merkle tree hashes to the communicating network.

AION’s Transwarp Conduit\footnote{https://github.com/aionnetwork/transwarp\_conduit\/tree/master/aion} is a smart-contract based solution that enables developers to create interchain smart contracts, by listening to the source blockchain contract adapter, and calling the corresponding target blockchain. Users can call such contract, triggering a transwarp conduit node to validate the request. After that, the request is processed by the contract.

The AION project was divided into two distinct brands: the Open Application Network (The OAN)\footnote{https://developer.theoan.com/community} and AION itself. The OAN network is no longer focusing on interoperability; it is an open source public infrastructure for the creation and hosting of ``open apps''.  AION is now the digital asset powering such apps. AION plans to develop the OAN tech stack, as stated by the roadmap\footnote{https://medium.com/theoan/2019-q4-foundation-report-b3a38a28d2b1}.

\emph{Komodo} is a blockchain infrastructure that allows one to create chains pegged to the Komodo blockchain, which is pegged to Bitcoin. Komodo uses delayed Proof of Work to create checkpoints of the Komodo's state that are added to Bitcoin from time to time (a process called notarization). Among other use cases, Komodo-based infrastructure allows atomic swaps, via the AtomicDEX feature \cite{komodo}. To foster adoption, Komodo promotes liquidity provider nodes, which are trading parties that act as market-makers, by buying and selling cryptocurrencies. Komodo is an open-source composable smart chain platform\footnote{https://github.com/KomodoPlatform/komodo}, built on top of Bitcoin and ZCash, which take Merkle tree roots from a smart chain set of blocks and merge them with other Merkle roots, that represent other smart chains. This generates a single Merkle root out of the various Merkle roots, referring to blocks of all smart chains. The mainchain, the KMD ledger, then synchronizes the state of each smart chain, providing interoperability capabilities. This mechanism works similarly to \emph{Delayed Proof of Work} (dPoW). dPoW allows securing a chain with another chain by leveraging a high hash rate (like KMD or even Bitcoin itself). This way, the risk of 51\% attacks is reduced. 

We now compare the Blockchain of Blockchains with highest adoption, Polkadot, and Cosmos. As a baseline, we use \emph{Ethereum 2.0} \cite{ethereum2,eth2_spec,eth2_wiki}, a major upgrade to the current Ethereum public mainnet, to be launched in three phases across 2020-2023. Ethereum 2.0 is an advance in blockchain inteoperability, as it will be composed by shards that interoperate with each other. It features a new execution environment for smart contracts, running on a new virtual machine, eWASM. We compare Polkadot, Ethereum 2.0, and Cosmos in Table \ref{tab:polka_cosmos_eth}.

In phase 0, the beacon chain of the Ethereum 2.0 network will be launched, implementing PoS and managing the validator registry. The beacon chain is meant for testing purposes and does not have functionality: Ethereum 1.0 will continue to operate. In phase 1, the old main chain and the beacon chain are merged, resulting in a single consolidated chain. Blockchain sharding techniques are used to raise Ethereum 2.0 throughput. Phase 2 focuses on enabling ether accounts, transactions, smart contract execution, and possibly further interoperability features \cite{ethhub2020}.

Ethereum 2.0 is suitable to serve as a baseline, as its performance in terms of throughput will be close to Blockchain of Blockchains; and furthermore, Ethereum is one of the most popular blockchains regarding dApps and industrial use cases.

\begin{table}[]
\centering \scriptsize
\caption{Comparison between Polkadot, Ethereum 2.0, and Cosmos \cite{polkadot_research,eth2_spec,c_blog}}
\label{tab:polka_cosmos_eth}
\begin{tabular}{@{}llll@{}}
\toprule
 &
  \textbf{Polkadot} &
  \textbf{Ethereum 2.0} &
  \textbf{Cosmos} \\ \midrule
\multicolumn{1}{l|}{\textbf{Model}} &
  \begin{tabular}[c]{@{}l@{}}Sharded,\\ pure-abstract STF\end{tabular} &
  \begin{tabular}[c]{@{}l@{}}Sharded,\\ fixed-function STF\end{tabular} &
  Bridge-hub \\ \midrule
\multicolumn{1}{l|}{\textbf{\begin{tabular}[c]{@{}l@{}}Consensus \\ protocol\end{tabular}}} &
  GRANDPA/BABE &
  Serenity &
  Tendermint \\ \midrule
\multicolumn{1}{l|}{\textbf{Main Chain}} &
  Relay-chain &
  Beacon Chain &
  Cosmos Hub \\ \midrule
\multicolumn{1}{l|}{\textbf{\begin{tabular}[c]{@{}l@{}}Main Chain State \\ Transition Function\end{tabular}}} &
  \begin{tabular}[c]{@{}l@{}}Abstract\\ meta-protocol\end{tabular} &
  Fixed-function &
  Fixed-function \\ \midrule
\multicolumn{1}{l|}{\textbf{\begin{tabular}[c]{@{}l@{}}Finality fault\\ tolerance\end{tabular}}} &
  33\% &
  33\% &
  33\% \\ \midrule
\multicolumn{1}{l|}{\textbf{\begin{tabular}[c]{@{}l@{}}Finalization expected \\ latency\end{tabular}}} &
  6-60 seconds &
  6-12 minutes &
  Instant \\ \midrule
\multicolumn{1}{l|}{\textbf{\begin{tabular}[c]{@{}l@{}}Horizontal Scaling \\ (sharding)\end{tabular}}} &
  Yes &
  Yes &
  Not available \\ \midrule
\multicolumn{1}{l|}{\textbf{Governance}} &
  \begin{tabular}[c]{@{}l@{}}Lock-vote; Committees;\\ council\end{tabular} &
  Forks &
  Coin-vote \\ \midrule
\multicolumn{1}{l|}{\textbf{BTC Token Support}} &
  Two-way peg &
  Not available &
  Two-way peg \\ \midrule
\multicolumn{1}{l|}{\textbf{ETH Token Support}} &
  Two-way peg &
  One-way-peg &
  Two-way peg \\ \midrule
\multicolumn{1}{l|}{\textbf{EVM Sidechain bridging}} &
  \begin{tabular}[c]{@{}l@{}}Parity PoA \end{tabular} &
  Not available &
  Two-way peg \\
 &
   &
   &
   \\

\end{tabular}
\end{table}

Polkadot and Ethereum 2.0 have a different approach to interoperability than Cosmos. Cosmos relies on a bridge-hub architecture, making it challenging to scale; Polkadot and Ethereum 2.0 have a shared-security/sharded approach, thus providing better scalability. 

Polkadot and Ethereum 2.0 have block production protocols, BABE and RanDAO + LMD Casper, respectively. Moreover, Polkadot and Ethereum 2.0 have finality sub-protocols, GRANDPA, and Casper FFG. Those protocols have to be implemented to provide sharding functionalities. Polkadot can achieve up to 100 shards while Ethereum 2.0 can support 64 shards. Cosmos do not support horizontal scalability via sharding. However, a shared security layer, similar to Polkadot's, is being idealized. In particular, it would allow a zone to inherit the validator set from another zone, allowing for transaction offload. 

On Polkadot, the main chain is the relay-chain, relying on the DOT token. Ethereum's 2.0 main chain is the Beacon chain, using Ether. Cosmos' main chain is the Cosmos Hub, and the token used is ATOM. The main chain state transition function in Polkadot is an abstract meta protocol relying on web assembly. Cosmos and Ethereum 2.0 utilize fixed functions.

The finality fault tolerance, i.e., the minimum required number of faulty nodes to compromise the network, is one third of the nodes less one) for all solutions, with different latencies. Although those solutions have different finality times, one should note that Polkadot and Ethereum 2.0 rely on a sharding strategy.

Polkadot and Cosmos utilize smart contracts and state transaction functions (provide an interface for smart contract execution \cite{polkadot_research}). Ethereum 2.0 only supports smart contracts. All solutions have robust governance mechanisms, namely decision making and decision enactment mechanisms (e.g., multicameral governance mechanism with conviction voting in Polkadot, coin-vote signaling in Cosmos). Polkadot has enhanced governance with a tech committee and an on-chain treasury. In Cosmos, validators can vote on behalf of the ATOMs staked to them, although it is possible to ATOM holders directly vote, canceling the staked validators' vote.

Regarding compatibility and bridging, Polkadot and Cosmos have two-way pegs to the Bitcoin and Ethereum networks. Ethereum 2.0 has a one-way peg with Ethereum, in which only Ethereum users can send Ether to Ethereum 2.0. Both Polkadot and Cosmos can communicate with sidechains. Polkadot further implements bridging capabilities, by leveraging substrate, achieving shard compatibility.

\section{Hybrid Connectors}
\label{a:connectors}

We now describe some of sidechain solutions we identified in the literature. Table \ref{tab:blockchain_connectors} summarizes each solution and aggregates them into the corresponding subcategory. One can assert that from the 14 solutions identified, 3 are trusted relays, 4 are blockchain-agnostic protocols, 4  blockchain of blockchains, and 3  blockchain migrators.

\begin{table}[h]
\centering
\caption{Comparison of \emph{hybrid connector} solutions}
\label{tab:blockchain_connectors}
\resizebox{\textwidth}{!}{%
\begin{tabular}{@{}llllll@{}}
\toprule
 &
  \multicolumn{1}{c}{\textbf{Reference}} &
  \multicolumn{1}{c}{\textbf{\begin{tabular}[c]{@{}c@{}}Transaction \\ validation\end{tabular}}} &
  \multicolumn{1}{c}{\textbf{Protocol}} &
  \multicolumn{1}{c}{\textbf{\begin{tabular}[c]{@{}c@{}}Supported \\ Blockchains\end{tabular}}} &
  \multicolumn{1}{c}{\textbf{\begin{tabular}[c]{@{}c@{}}Public\\  PoC\end{tabular}}} \\ \midrule
\multicolumn{1}{l|}{\multirow{4}{*}{\textbf{\begin{tabular}[c]{@{}l@{}}Trusted \\ Relays\end{tabular}}}} &
  \multicolumn{1}{l|}{Montgomery et al., \cite{block_integ_framework}$^{\ast}$ \mbox{}\hfill \checkmark} &
  \multicolumn{1}{l|}{Trusted escrow party} &
  \multicolumn{1}{l|}{\begin{tabular}[c]{@{}l@{}}Cross-blockchain transactions \\ signed by validator quorum\end{tabular}} &
  \multicolumn{1}{l|}{Private} &
  \checkmark \\ \cmidrule(l){2-6} 
\multicolumn{1}{l|}{} &
  \multicolumn{1}{l|}{Kan et al., \cite{Kan2018}} &
  \multicolumn{1}{l|}{Trusted escrow party} &
  \multicolumn{1}{l|}{\begin{tabular}[c]{@{}l@{}}3-phase-commit\\ protocol\end{tabular}} &
  \multicolumn{1}{l|}{--} &
  $\times$ \\ \cmidrule(l){2-6} 
\multicolumn{1}{l|}{} &
  \multicolumn{1}{l|}{Abebe et al., \cite{Abebe2019}} &
  \multicolumn{1}{l|}{\begin{tabular}[c]{@{}l@{}}Relay service, \\ verifiable proofs, \\ system smart contracts\end{tabular}} &
  \multicolumn{1}{l|}{\begin{tabular}[c]{@{}l@{}}System contracts, \\ communication protocol.\\ protocol buffers\end{tabular}} &
  \multicolumn{1}{l|}{Private} &
  $\times$ \\ \cmidrule(l){2-6} 
\multicolumn{1}{l|}{} &
  \multicolumn{1}{l|}{Falazi et al., \cite{falazi2020}} &
  \multicolumn{1}{l|}{\begin{tabular}[c]{@{}l@{}}Centralized \\ Gateway\end{tabular}} &
  \multicolumn{1}{l|}{\begin{tabular}[c]{@{}l@{}}Smart Contract \\ Invocation Protocol\end{tabular}} &
  \multicolumn{1}{l|}{Private, Public} &
  $\times$ \\ \midrule
\multicolumn{1}{l|}{\multirow{4}{*}{\textbf{\begin{tabular}[c]{@{}l@{}}Blockchain-\\ Agnostic\\ Protocols\end{tabular}}}} &
  \multicolumn{1}{l|}{Hardjono et al. \cite{Hardjono2019}} &
  \multicolumn{1}{l|}{\begin{tabular}[c]{@{}l@{}}Blockchain\\ Gateways\end{tabular}} &
  \multicolumn{1}{l|}{--} &
  \multicolumn{1}{l|}{--} &
  $\times$ \\ \cmidrule(l){2-6} 
\multicolumn{1}{l|}{} &
  \multicolumn{1}{l|}{Vo et al., \cite{vo2018}} &
  \multicolumn{1}{l|}{--} &
  \multicolumn{1}{l|}{\begin{tabular}[c]{@{}l@{}}$\times$ - but Multi-Protocol\\ Communication is referred\end{tabular}} &
  \multicolumn{1}{l|}{--} &
  $\times$ \\ \cmidrule(l){2-6} 
\multicolumn{1}{l|}{} &
  \multicolumn{1}{l|}{Interledger Protocol \cite{Thomas2015}$^{\ast}$ \mbox{}\hfill  \checkmark} &
  \multicolumn{1}{l|}{\begin{tabular}[c]{@{}l@{}}(Trusted)\\ Router\end{tabular}} &
  \multicolumn{1}{l|}{Packet Switching (ILPv4)} &
  \multicolumn{1}{l|}{Private, Public} &
  \checkmark \\ \cmidrule(l){2-6} 
\multicolumn{1}{l|}{} &
  \multicolumn{1}{l|}{Hyperledger Quilt \cite{quilt}$^{\ast}$} &
  \multicolumn{1}{l|}{\begin{tabular}[c]{@{}l@{}}(Trusted)\\ Router\end{tabular}} &
  \multicolumn{1}{l|}{Packet Switching (ILPv4)} &
  \multicolumn{1}{l|}{Private, Public} &
  \checkmark \\ \midrule
\multicolumn{1}{l|}{\multirow{4}{*}{\textbf{\begin{tabular}[c]{@{}l@{}}Blockchain\\  of\\  Blockchains\end{tabular}}}} &
  \multicolumn{1}{l|}{Verdian et al, \cite{Verdian2018}$^{\ast}$ \mbox{}\hfill \checkmark} &
  \multicolumn{1}{l|}{\begin{tabular}[c]{@{}l@{}}BPI, Messaging, \\ Filetering and \\ Ordering layers\end{tabular}} &
  \multicolumn{1}{l|}{\begin{tabular}[c]{@{}l@{}}Based on\\ posets and order theory\end{tabular}} &
  \multicolumn{1}{l|}{Public} &
  $\times$ \\ \cmidrule(l){2-6} 
\multicolumn{1}{l|}{} &
  \multicolumn{1}{l|}{Liu et al., \cite {hyperservice}} &
  \multicolumn{1}{l|}{\begin{tabular}[c]{@{}l@{}}NSB, \\ ISC\end{tabular}} &
  \multicolumn{1}{l|}{UIP protocol} &
  \multicolumn{1}{l|}{Public} &
  \checkmark \\ \cmidrule(l){2-6} 
\multicolumn{1}{l|}{} &
  \multicolumn{1}{l|}{Block Collider \cite{blockCollider}$^{\ast}$ \mbox{}\hfill \checkmark} &
  \multicolumn{1}{l|}{Base tuples} &
  \multicolumn{1}{l|}{\begin{tabular}[c]{@{}l@{}}Proof of Distance\\ (PoD)\end{tabular}} &
  \multicolumn{1}{l|}{Public} &
  \checkmark \\ \cmidrule(l){2-6} 
\multicolumn{1}{l|}{} &
  \multicolumn{1}{l|}{Amiri et al., \cite{Amiri2019}} &
  \multicolumn{1}{l|}{\begin{tabular}[c]{@{}l@{}}Blockchain views, \\ internal and \\ external transactions\end{tabular}} &
  \multicolumn{1}{l|}{\begin{tabular}[c]{@{}l@{}}Hierarchical consensus\\ and one-level consensus\end{tabular}} &
  \multicolumn{1}{l|}{--\textsuperscript{1}} &
  $\times$ \\ \midrule
\multicolumn{1}{l|}{\multirow{3}{*}{\textbf{\begin{tabular}[c]{@{}l@{}}Blockchain \\ Migrators\end{tabular}}}} &
  \multicolumn{1}{l|}{Frauenthaler et al., \cite{Frauenthaler2019}} &
  \multicolumn{1}{l|}{\begin{tabular}[c]{@{}l@{}}Enforced by\\ smart contracts\end{tabular}} &
  \multicolumn{1}{l|}{Adapters} &
  \multicolumn{1}{l|}{Public} &
  \checkmark \\ \cmidrule(l){2-6} 
\multicolumn{1}{l|}{} &
  \multicolumn{1}{l|}{Scheid et al., \cite{scheid2019}} &
  \multicolumn{1}{l|}{\begin{tabular}[c]{@{}l@{}}Enforced by\\ smart contracts\end{tabular}} &
  \multicolumn{1}{l|}{Adapters} &
  \multicolumn{1}{l|}{--} &
  $\times$ \\ \cmidrule(l){2-6} 
\multicolumn{1}{l|}{} &
  \multicolumn{1}{l|}{Fynn et al., \cite{scotm}} &
  \multicolumn{1}{l|}{\begin{tabular}[c]{@{}l@{}}Enforced by\\ smart contracts\end{tabular}} &
  \multicolumn{1}{l|}{Move Operation} &
  \multicolumn{1}{l|}{Public} &
  $\times$ \\ \midrule
 &
   &
   &
   &
   &
   \\
\multicolumn{6}{l}{\begin{tabular}[c]{@{}l@{}}\checkmark our description was endorsed\\ $\ast$ considered grey literature\\ $\times$ lacks implementation or implementation is not public\\ -- Not defined or not applicable\\ \textsuperscript{1} CAPER instance enables cross-aplication transactions\end{tabular}}
\end{tabular}%
}
\end{table}

\subsection{\underline{Trusted Relays}}
\label{sec:block_connector_trusted_relay_appendix}
Trusted relays are trusted parties that redirect transactions from a source blockchain to a target blockchain.

\emph{Kan et al.}  introduce a protocol that delivers atomicity and consistency through asset escrow (third-party releasing locked assets under specific conditions) and a three-phase commit \cite{Kan2018}. This scheme assumes a trusted party. The authors provide a superficial evaluation, consisting of custom-made blockchains.

\emph{Abebe et al.} propose a  generalized protocol for data transfer, with a particular focus on permissioned networks \cite{Abebe2019}. 
They introduce system contracts, a \emph{relay service}, and a communication protocol.   

The conceptual mechanisms that achieve interoperability are the relay service and system contracts. The relay service acts on behalf of each blockchain, serving requests from applications using the blockchains. Relay services communicate with each other using protocol buffers, a method of serializing structured data, and require \emph{verification policies} to be satisfied by the requester (by verifying a proof). They are also responsible for translating the network-neutral protocol messages into blockchain-specific transactions on the target blockchain. Although the authors defend that relayers operate with ``minimal trust'' (as they require verifiable proofs coupled with every request), they are trusted in the sense that they follow the protocol, i.e., do not suffer from Byzantine faults.

System contracts are smart contracts that manage data exposure, such as identity and disclosure of network information. One can consider system contracts to be smart contracts handling infrastructural aspects, being an extension to the business logic encoded in most smart contracts. Moreover, such contracts use access control request policy rules against incoming cross-network requests, and if such information is valid (given an attached verifiable proof), according to a specific verification policy. The generation of proofs based on verification policies, and its subsequent validation, allow for trust distribution regarding cross-network transactions.

\emph{Falazi et al.} \cite{falazi2020} propose an abstraction layer that provides a uniform interface for external client applications to communicate with blockchains and smart contracts. The proposed protocol, Smart Contract Invocation Protocol (SCIP), exposes a interface with several elements (roles, methods, data, and message format), which can be used by applications to issue transactions against different ledgers. The available request messages include (i) the invocation of a smart contract function, (ii) the subscription to notifications regarding function invocations or event occurrences, (iii) the unsubscription from live monitoring, and (iv) the querying of past invocations or events.

\subsection{\underline{Blockchain-Agnostic Protocols}}
\label{sec:block_connector_blockagnosticprotocol}
Blockchain-agnostic protocols enable cross-blockchain or cross-chain communication between arbitrary distributed ledger technologies.

\emph{Hardjono et al.} proposed a model for blockchain interoperability, in the context of the Tradecoin\footnote{https://tradecoin.mit.edu/} project \cite{Hardjono2019}.

Each blockchain is seen as an autonomous system (or routing domain), as a connectivity unit that can scale. Such autonomous systems have a domain-centered control with distributed topology. Entities that execute and validate cross-blockchain transactions are called gateways.

Generally, the conceptual mechanism that underlies the interoperability scheme is the ability of gateways to be autonomous and discoverable. Gateways can then redirect transactions to the corresponding blockchain. 

\emph{Kan et al.} presented a theoretical work on how blockchains can execute cross-chain transactions, via several actors: \emph{validators}, \emph{nominators}, \emph{surveillants}, and \emph{connectors} \cite{Kan2018}. Validators verify and forward blocks to the correct destination. Nominators elect validators. Surveillants monitor the blockchain router's behavior. The proposed protocol aims participants to achieve a dynamic equilibrium state, using incentivization (fees awarded to the parties following the protocol). No implementation details are provided.

\emph{Hyperledger Quilt} is a Java implementation of the Interledger protocol \cite{quilt}. While Interledger implements connectors, Quilt implements several primitives of the Interledger protocol, namely: interledger addresses, ILPv4\footnote{https://github.com/interledger/rfcs/blob/master/0027-interledger-protocol-4/0027-interledger-protocol-4.md}, payment pointers, ILP-over-HTTP, simple payment setup protocol, and STREAM.

Quilt is an open-source project\footnote{https://github.com/hyperledger/quilt}, and it is interoperable with other implementations, such as Interledger Rust\footnote{http://interledger.rs/} and  InterledgerJS\footnote{https://github.com/interledgerjs/ilp-connector}.

Other systems are focused on building cross-blockchain dApps, by organizing blocks that contain a set of transactions belonging to CC-dApps, spread across multiple blockchains. Such system should provide accountability for the parties issuing transactions on the various blockchains, as well as providing a holistic, updated view of each underlying blockchain'' (Section \ref{subsec:bid}).

\emph{Overledger} aims to ease the development of decentralized apps on top of different blockchain infrastructures \cite{Verdian2018,overledger03}. Interoperability is achieved by using a common interface among ledgers.

Overledger proposes a four-layer approach. The \emph{transaction layer} contains different blockchains, and stores transactions coming from them. While the \emph{messaging layer} retrieves relevant information from the transaction layer, coming from heterogeneous blockchains: transactions from a pool of transactions, metadata, or smart contracts. The \emph{filtering layer and the ordering layer} create connections between messages from the messaging layer. Messages are ordered and filtered according to a specific set of rules (e.g., respecting a schema, containing specific cryptographic signatures). In particular, the filtering layer requires knowledge about all the different blockchains included in Overledger.

Overledger requires a block ordering mechanism to ensure the total ordering of cross-blockchain transactions: the application scans the compatible ledgers' transaction hashes and places them into a \emph{verification block}. Transactions in a verification block are modeled as a total poset, in which a binary relationship is used to compare the order of transactions within a block \cite{Verdian2018}. 

Overledger achieves blockchain interoperability using a protocol for message-oriented middleware that implements a protocol similar to 2-phase-commit scheme, instead of relying on adapters between a central blockchain and external blockchains, but no details are given.

\emph{Block Collider} enables smart contract communication among smart contracts located in different chains \cite{blockCollider}. The goal is to alleviate the developer's work while building decentralized apps that use several blockchains. 

Block Collider unifies the latest blocks on each bridged chain via blocks' \emph{base tuples}: every block references the header of the block from each of the bridged chains. This allows Block Collider to be a decentralized \emph{unifying chain}. 

The consensus mechanism for determining the following block head is the proof of distance, a variation of proof of work. Proof of distance uses an algorithm in which a string edit distance scheme is used. In this scheme, the idea is to hash to be filtered within some distance of a reference set. Block Collider is an open-source project\footnote{https://github.com/blockcollider}, and supports various cryptocurrencies, including BTC, ETH, USDT, WAV, LSK, NEO, DAI, and Tether Gold.



\subsection{\underline{Blockchain Migrators}}
\label{sec:block_connector_bm_appendix}
Blockchain migrators allow an end user to migrate the state of a blockchain to another. Currently, it is only possible to migrate data across blockchains, although moving smart contracts is also predicted \cite{hyperledger_cactus}.

\emph{Frauenthaler et al.} propose a framework for blockchain interoperability and runtime selection \cite{Frauenthaler2019}. The framework supports Bitcoin, Ethereum, Ethereum Classic, and Expanse. This framework is app-centric since the user can parameterize the app with functional and nonfunctional requirements. The framework can choose a blockchain at runtime, allowing a blockchain to route transactions to other blockchain, depending on weighted metrics.

Some metrics include the price of writing and reading from a blockchain, the exchange rate between the cryptocurrency supporting a blockchain and the dollar, the average time to mine a block and the degree of decentralization.

Based on such metrics, and their weight, specified by the end-user, the blockchain selection algorithm computes the most appropriate blockchain. According to the authors, switching to another blockchain can help users to save costs and make them benefit from a better infrastructure (e.g., better performance, higher decentralization, better reputation). This solution does not tackle the migration of smart contracts. However, data transfers are possible (i.e., data is copied from the source to the target blockchain).
This project is a centralized application ran by the end user. It is open-source\footnote{https://github.com/pf92/blockchain-interop}.

\emph{Scheid et al.} propose a policy-based agnostic framework that connects, manages, and operates different blockchains \cite{scheid2019, timo2019}.

Policies can be defined to optimize costs or performance. If one chooses to minimize costs associated with data storing, the framework chooses the blockchain which has the cheapest cost of writing. Conversely, performance policies can configure the framework to minimize a transaction's confirmation waiting time. 
The authors include AAA access control, as defined by the OASIS consortium \cite{XACML}, to manage policies.

The platform is blockchain agnostic, but details on supported blockchains are not provided. Although this work is not a functional blockchain migration tool, it allows the flexibility needed for blockchain migrations (see Section \ref{sec:block_connector_bm}).

\section{Use Cases}
\label{a:use_cases}

Example use cases related to cryptocurrency-related techniques are cross-chain payment channels \cite{peggedsidechains,zdag,Lu2017,blockCollider}, efficient multi-party swaps \cite{xclaim, testimonium, dextt}, point of sales and utility tokens \cite{zdag}, and decentralized exchanges \cite{vitalik2016,xclaim}. As a notable use case, we highlight decentralized exchanges \cite{0xproject}, leveraging HLTC techniques to allow users to exchange assets from different blockchains directly with other users. 

Blockchain of Blockchains \cite{Wood2017,Kwon2016,ark2019,aion2017} do implement decentralized exchanges, and predict decentralized banking as use cases. For example, the decentralized exchange Binance \cite{binancesmart} utilizes the Cosmos SDK. Blockchain gaming platforms\footnote{https://xaya.io/}, and stablecoins\footnote{http://bandot.io/} have been implemented with Polkadot. Moreover, Blockchain of Blockchains can stimulate blockchain adoption by enterprises. By using Cosmos, zones can serve as blockchain-backed versions of enterprise systems, whereby services that are traditionally run by an organization or a consortium are instead run as an application blockchain interface on a particular zone. Some authors proposed an IoB approach for a central bank digital currency \cite{mbcb}, which could be realized with a blockchain engine solution. 

Regarding Hybrid Connectors, we highlight blockchain migrators, as solutions that can reduce the risk for enterprises and individuals when investing in blockchain. By reducing risks, investors can expect a higher return on investment \cite{Warren2019}. Hyperledger Cactus, a blockchain interoperability project includes a blockchain migration feature, which allows a consortium of stakeholders operating a blockchain to migrate their assets (data, smart contracts) to another blockchain \cite{hyperledger_cactus}. Other use cases can be realized: cross-blockchain asset transfer, escrowed sale of data for coins, pegging stable coins to fiat currency or cryptocurrencies, healthcare data sharing with access control lists, integration of existing food traceability solutions, and end-user wallet access control. 

More generally, a blockchain of blockchains approach can be leveraged to solve current problems. In \cite{correia2014,depsky}, the authors argue that accidental failures and security events (in particular internal data breaches) is a problem for the end-user. This problem can be alleviated by creating a ``cloud-of-clouds'' for extra security and dependability, on top of individual cloud providers that do not offer enough trust. One could argue that one can use a blockchain of blockchains approach to increase the dependability of services, as well as their security.

\begin{table}[h]
\centering
\setlength{\tabcolsep}{15pt}
\normalsize
\caption{IoB and BoB use cases}
\label{tab:use_cases}
\begin{tabular}{@{}llll@{}}
\backslashbox{Use Case}{Categories}                             & \rot{Public Connectors}  & \rot{Blockchain of Blockchains}   & \rot{Hybrid Connectors} \\ \midrule
\begin{tabular}[c]{@{}l@{}}Decentralized\\ Finance\end{tabular} & \cellcolor[HTML]{D8E3BB}+        & \cellcolor[HTML]{D8E3BB}+  & \cellcolor[HTML]{D8E3BB}+   \\ \midrule
\begin{tabular}[c]{@{}l@{}}Cross-blockchain\\ dApps\end{tabular} &
  \cellcolor[HTML]{BF504D} - &
  \cellcolor[HTML]{F79545}$\pm$ &
  \cellcolor[HTML]{D8E3BB}+ \\ \midrule
\begin{tabular}[c]{@{}l@{}}Blockchain\\ Migration\end{tabular}  & \cellcolor[HTML]{BF504D}- & \cellcolor[HTML]{F79545}$\pm$ & \cellcolor[HTML]{D8E3BB}+   \\ \midrule
\begin{tabular}[c]{@{}l@{}}Enabling Enterprise\\ Business Processes\end{tabular} &
  \cellcolor[HTML]{D8E3BB}+ &
  \cellcolor[HTML]{F79545}$\pm$ &
  \cellcolor[HTML]{F79545}$\pm$ \\ \midrule
                                                                &                                  &                            &                             \\
\multicolumn{4}{l}{\begin{tabular}[c]{@{}l@{}}+  Use case already implemented\\ $\pm$ Use case being developed\\ -- Use case not planned\end{tabular}}    \\
                                                                &                                  &                            &                            
\end{tabular}%
\end{table}

Collecting, storing, accessing, and processing data is not only a common practice across industries but also essential to their thriving. 
Often, a use-case has several stakeholders with different needs, who belong to different organizational boundaries. Those stakeholders might have different access rights to data \cite{Belchior2019, Belchior2019_Audits}. Thus, developers adapt the features of the blockchain they are using to the (sometimes conflicting) needs of their stakeholders. It is important to underline that developers want flexibility regarding their blockchain choice, as they might want to change it in the future \cite{Frauenthaler2019}. This particular need is related to the possibility of vendor lock-in, which also happens in cloud environments \cite{Kaur2017}. The need for this flexibility can be achieved by leveraging blockchain migration or multiple blockchains.


\end{document}